\documentclass[10pt, aps, pra, twocolumn, superscriptaddress, floatfix, showpacs, longbibliography]{revtex4-2}

\usepackage{graphicx}
\usepackage{amsfonts}
\usepackage{amssymb}
\usepackage{xcolor}
\usepackage{amsmath}
\usepackage{txfonts}
\usepackage{lipsum}
\usepackage{color}
\usepackage{wasysym}
\usepackage[colorlinks=true, allcolors=blue]{hyperref}
\usepackage{bbold}
\usepackage{etoolbox} 
\usepackage[normalem]{ulem}
\usepackage{mathtools} 
\usepackage{multirow} 
\usepackage{hhline}
\usepackage{mathrsfs} 
\usepackage{soul}
\usepackage{comment}

\usepackage{letltxmacro}
\LetLtxMacro{\oldsqrt}{\sqrt}
\renewcommand{\sqrt}[2][\mkern8mu]{\mkern-6mu\mathop{}\oldsqrt[#1]{#2}}

\usepackage{url}
\definecolor{indigo(dye)}{rgb}{0.0, 0.25, 0.42}
\usepackage{placeins}

\begin{document}

\title{Dual-space cluster-diagrammatic approach to nonlocal electronic correlations}

\author{Félix Fossati}
\affiliation{CPHT, CNRS, {\'E}cole polytechnique, Institut Polytechnique de Paris, 91120 Palaiseau, France}

\author{Evgeny~A.~Stepanov}
\affiliation{CPHT, CNRS, {\'E}cole polytechnique, Institut Polytechnique de Paris, 91120 Palaiseau, France}
\affiliation{Coll\`ege de France, Universit\'e PSL, 11 place Marcelin Berthelot, 75005 Paris, France}

\begin{abstract}
In this work, we extend the dual triply irreducible local expansion (\mbox{D-TRILEX}) approach for correlated electronic systems by introducing a cluster reference system for the diagrammatic expansion.
This framework allows us to consistently combine the exact treatment of short-range correlation effects within the cluster, with an efficient diagrammatic description of the long-range charge and spin collective fluctuations beyond the cluster.
We demonstrate the effectiveness of our approach by applying it to the one-dimensional nano-ring Hubbard model, where the low dimensionality enhances non-local correlations. Our results show that the cluster extension of \mbox{D-TRILEX} accurately reproduces the electronic self-energy at momenta corresponding to the Fermi energy, in good agreement with the reliable Hirsch-Fye quantum Monte Carlo solution of the problem.
We further compare this method with the more computationally demanding parquet dynamical vertex approximation and find that, our method yields substantially more accurate results at momenta associated with the Fermi surface.
We show that the \mbox{D-TRILEX} diagrammatic extension drastically reduces the periodization ambiguity of cluster quantities when mapping back to the original lattice, compared to cluster dynamical mean-field theory (CDMFT).
Furthermore, we identify the CDMFT impurity problem as the main source of the translational-symmetry breaking and propose a computational scheme for improving the starting point for the cluster-diagrammatic expansion.
\end{abstract}

\maketitle

\section{Introduction}
\label{sec:intro}
The theoretical description of strongly correlated materials relies on solving interacting electronic model Hamiltonians. 
Accounting for electron-electron interaction can significantly modify the characteristics of a non-interacting system. 
In the weak-coupling regime, these interactions renormalize single-particle properties such as the effective electron mass and the spectral function. 
This behavior is well described by Landau Fermi liquid theory~\cite{abrikosov1959, nozieres2018}, and perturbative diagrammatic methods such as the $GW$~\cite{GW1, GW2, GW3} or the fluctuation-exchange (FLEX)~\cite{bickers1989, PhysRevLett.62.961, PhysRevB.57.6884} approximations yield reasonable results. 
However, these methods break down as the interaction strength increases. In the strong-coupling regime, non-perturbative approaches such as dynamical mean-field theory (DMFT)~\cite{RevModPhys.68.13} are required to capture emergent strongly correlated phenomena, including the Mott metal-insulator transition~\cite{Mott, RevModPhys.70.1039} and Hund's metal behavior~\cite{Hunds_metals1, Hunds_metals2}, which lie beyond a weak-coupling-like theoretical description.

The single-site DMFT formulation assumes a purely local self-energy and thus neglects spatial electronic correlations. 
This approximation, while exact in the limit of infinite dimensions~\cite{PhysRevLett.62.324, muller-hartmann1989}, leads to a number of limitations in realistic low-dimensional systems. 
For instance, DMFT cannot capture the momentum-dependent renormalization of the electronic spectral function induced, for example, by strong magnetic fluctuations. In particular, such fluctuations may lead to a momentum-dependent suppression of the density of states at the Fermi surface, known as the nodal-antinodal dichotomy (see, e.g., Ref.~\cite{PhysRevLett.132.236504}).
In addition, DMFT is not designed to describe small-scale systems, such as dimer or one-dimensional models, where accounting for inter-site correlations is essential (see, e.g., Refs.~\cite{10.21468/SciPostPhys.13.2.036, PhysRevB.91.115115}).

Cluster dynamical mean-field theory (CDMFT)~\cite{PhysRevB.58.R7475, PhysRevB.62.R9283, PhysRevLett.87.186401, RevModPhys.77.1027, doi:10.1063/1.2199446, RevModPhys.78.865, PhysRevLett.101.186403} is the most natural non-perturbative extension that incorporates non-local correlations beyond the single-site DMFT. 
By embedding a finite cluster into a self-consistent bath, CDMFT explicitly accounts for short-range correlations within the cluster, which are treated numerically exactly. This construction captures singlet formation and local-moment physics, thereby accounting for the effects that are believed to be responsible for strong-coupling antiferromagnetism, pseudogap formation, and high-temperature superconductivity~\cite{PhysRevB.62.R9283, PhysRevB.76.104509, PhysRevLett.100.046402, PhysRevB.94.125133, PhysRevB.101.045119, danilov2022, dong2022, PhysRevX.8.021048, Cuprates}. 

Despite many successes, CDMFT has several conceptual and practical difficulties.  
First, the Hilbert space of the impurity problem, which has to be solved at every iteration in the DMFT self-consistent loop, grows exponentially with the number of cluster sites \(N_{\mathrm c}\), so the computational cost scales as \(d^{N_{\mathrm c}}\), where $d$ is the number of degree of freedom for one site. 
Quantum Monte Carlo (QMC)~\cite{PhysRevLett.56.2521} and continuous-time quantum Monte Carlo (CT-QMC) methods~\cite{PhysRevLett.77.5130, prokofev1996, rubtsov2004, PhysRevB.72.035122, PhysRevLett.97.076405, Gull2008, RevModPhys.83.349} eliminate the need for exact diagonalization and allow one to avoid the exponential scaling with $N_{\rm c}$. 
However, QMC and \mbox{CT-QMC} approaches are affected by the infamous fermionic sign problem~\cite{PhysRevB.41.9301, PhysRevB.80.155132, PhysRevB.81.045106}, which worsens with the number of lattice sites and is particularly severe for hybridization-expansion (\mbox{CT-HYB}) based methods~\cite{PhysRevLett.97.076405, PhysRevB.74.155107, RevModPhys.83.349}. 
Second, CDMFT breaks the translational symmetry of the original lattice, as it treats bonds within and between clusters in fundamentally different ways: the intra-cluster bonds are handled non-perturbatively within the exactly solved impurity problem, while the inter-cluster bonds are effectively treated as non-interacting.
Reinstating translational symmetry requires a periodization step, in which the cluster outputs are re-expressed as lattice quantities under the assumption of the original lattice symmetry. 
The outcome of this procedure depends sensitively on which quantity is periodized -- Green's function, self-energy, or cumulant -- and can lead to noticeably different single-particle spectra and even changes in the inferred Fermi surface (FS) topology~\cite{PhysRevLett.95.106402, PhysRevB.74.125110}.
Third, CDMFT calculations cannot provide a good resolution in momentum space and thus are unable to resolve fine momentum-space structures, such as small Fermi pockets, shadow bands, momentum-dependent pseudogaps, or sharp dispersions of collective modes~\cite{PhysRevLett.95.106402, PhysRevB.74.125110, PhysRevB.85.035102, PhysRevLett.132.236504}, without moving to much larger (and hence computationally prohibitive) clusters.
Finally, CDMFT neglects correlations that extend beyond the cluster. 
As a result, CDMFT with relatively small clusters tends to overestimate the critical temperatures for magnetic transitions~\cite{PhysRevB.62.R9283, PhysRevB.76.104509, PhysRevB.80.075104}, underestimate the critical interaction strength for the Mott transition~\cite{PhysRevLett.101.186403}, and cannot capture the momentum-resolved spectral functions of long-range collective electronic fluctuations such as plasmons and magnons.

Alternatively, non-local correlations beyond the single-site DMFT can be introduced diagrammatically~\cite{RevModPhys.90.025003, Lyakhova_review}.
In rare cases, the diagrammatic extensions of DMFT enable resummation of all Feynman diagrams up to a given order~\cite{PhysRevB.94.035102, PhysRevB.96.035152, PhysRevB.102.195109} using the diagrammatic Monte Carlo technique~\cite{PhysRevLett.81.2514, Kozik_2010}. 
However, in most cases, this procedure is computationally expensive, and the diagrammatic expansion is typically restricted to a subset of leading contributions, as in the $GW$ extension of DMFT~\cite{PhysRevLett.90.086402, PhysRevLett.92.196402, PhysRevLett.109.226401, PhysRevB.87.125149, PhysRevB.90.195114, PhysRevB.94.201106, PhysRevB.95.245130}, the two-particle self-consistent (TPSC) extension of DMFT~\cite{PhysRevB.107.075158, PhysRevB.107.235101, PhysRevB.107.245137, PhysRevB.111.115143, tpscdmft}, the dual fermion (DF)~\cite{PhysRevB.77.033101, PhysRevB.79.045133, PhysRevLett.102.206401, Brener2020}, the dual boson (DB)~\cite{rubtsov2012, PhysRevB.90.235135, PhysRevB.93.045107, PhysRevB.94.205110, stepanov2018, PhysRevB.100.165128}, the dynamical vertex approximation (D$\Gamma$A)~\cite{PhysRevB.75.045118, PhysRevB.80.075104, PhysRevB.95.115107, doi:10.7566/JPSJ.87.041004, PhysRevB.103.035120}, the triply irreducible local expansion (TRILEX)~\cite{PhysRevB.92.115109, PhysRevB.93.235124, PhysRevB.96.104504}, and the dual TRILEX (D-TRILEX)~\cite{PhysRevB.100.205115, PhysRevB.103.245123, 10.21468/SciPostPhys.13.2.036} methods.
The diagrammatic approaches have produced a series of notable successes, precisely because they self-consistently augment the local physics with spatial correlations at all length scales, including the genuinely long-range fluctuations that remain out of reach for cluster approaches.
As a result, the diagrammatic methods enable accurate description of the N\'eel transition~\cite{PhysRevLett.129.107202, PhysRevB.94.125144, PhysRevX.11.011058, PhysRevLett.102.206401, vandelli2022quantum}, especially in the regime where the magnetic fluctuations have very large correlation length~\cite{PhysRevB.91.125109, PhysRevLett.124.017003, PhysRevLett.124.117602, PhysRevResearch.4.043201}.
The diagrammatic extensions of DMFT are also successful in addressing the superconducting state~\cite{PhysRevB.90.235132, PhysRevB.92.085104, PhysRevB.99.041115, PhysRevB.96.104504, kitatani2020nickelate, Cuprates}, even when compared with the already reliable results obtained from cluster methods~\cite{PhysRevB.62.R9283, PhysRevLett.85.1524, PhysRevLett.110.216405}.
This improvement is often traced back to the ability of the diagrammatic methods to account for the long-range strong spin fluctuations~\cite{PhysRevLett.124.017003, PhysRevX.11.011058, Cuprates}.
Furthermore, the diagrammatic extensions of DMFT are also capable of addressing the charge-ordered state~\cite{PhysRevB.87.125149, Kamil2018, PhysRevB.90.235135, PhysRevB.93.045107, PhysRevB.94.205110, stepanov2021coexisting, stepanov2024} and describing momentum-resolved spectral functions for collective charge~\cite{vilk1996, PhysRevLett.109.226401, PhysRevLett.113.246407} and spin~\cite{PhysRevB.92.115109, stepanov2018, Boehnke_2018, acharya2019evening, suzuki2023distinct} fluctuations.
These methods, however, remain perturbative beyond the local level, so their accuracy strongly depends on the subset of diagrams and varies between different physical regimes.
It is still possible to go beyond these approximations using more sophisticated diagrammatic techniques. However, the downside is that the resulting diagrammatic structure become too complex to be applied efficiently to multi-orbital systems. 

Below, we turn to the \mbox{D-TRILEX} scheme~\cite{PhysRevB.100.205115}, which offers a good compromise between accuracy and cost of numerical calculations~\cite{PhysRevB.103.245123, 10.21468/SciPostPhys.13.2.036}. 
\mbox{D-TRILEX} simultaneously treats collective electronic fluctuations in the charge and spin channels on equal footing without the Fierz-ambiguity problem.
The single-particle and collective fluctuations are treated self-consistently within the functional formulation of the method~\cite{10.21468/SciPostPhys.13.2.036}, so that non-local electronic correlations are coherently incorporated beyond DMFT~\cite{stepanov2021coexisting, Vandelli2024_PbSi, PhysRevB.110.L161106, PhysRevLett.132.236504, Cuprates, stepanov2024signatures}. 
The \mbox{D-TRILEX} approach can be viewed as a simplified version of the DB theory: it retains much of the accuracy of the parent theory~\cite{PhysRevB.103.245123}, yet avoids the computational cost of the four-point vertex function by relying only on a numerically less expensive three-point object. 
As a result of this simplification, the diagrammatic structure of the self-energy and polarization operator in \mbox{D-TRILEX} reduces to a $GW$-like form~\cite{PhysRevB.100.205115, 10.21468/SciPostPhys.13.2.036}, which is significantly more computationally efficient than in DF, DB, or D$\Gamma$A methods, as it avoids solving the Bethe–Salpeter equation in frequency space. 
A rather simple form of the \mbox{D-TRILEX} diagrams allows one to perform feasible multi-orbital~\cite{PhysRevLett.127.207205, PhysRevLett.129.096404, PhysRevResearch.5.L022016, PhysRevLett.132.226501, stepanov2024, Ruthenates} and time-dependent~\cite{vglv-2rmv} calculations.

So far, \mbox{D-TRILEX} calculations have been performed only for the multi-orbital or multi-impurity reference system~\cite{10.21468/SciPostPhys.13.2.036, Vandelli2024_PbSi}. 
Implementing the cluster version of the method would allow us to combine the advantages of both approaches: the exact treatment of short-range correlations via the cluster DMFT reference problem and the efficient diagrammatic description of the long-range collective electronic fluctuations with a good momentum resolution.
This can be of particular importance for the low-dimensional systems, where the non-perturbative short-range correlations coexist with the long-range charge and spin fluctuations.
Additionally, using the cluster reference problem would allow one to address phase transitions associated with non-local order parameters, such as the $d$-wave superconducting state. 
The cluster extensions of the DF~\cite{hafermann2008, Brener2020, PhysRevB.84.155106, PhysRevB.97.125114, danilov2022, DFQMC} and TRILEX~\cite{PhysRevLett.119.166401} methods have already been implemented.
We believe that the cluster \mbox{D-TRILEX} approach would serve as a good alternative to these computational schemes due to a much more efficient diagrammatic structure and the absence of the Fierz ambiguity problem, respectively.

In this work, we present the cluster extension of the \mbox{D-TRILEX} method.
In Sec.~\ref{sec:method} we discuss the general formulation of the method and present the computational scheme. 
In Sec.~\ref{sec:results} we apply our method to the one-dimensional nano-ring Hubbard model. 
The results are benchmarked against the Hirsch–Fye QMC solution of the problem from Ref.~\cite{PhysRevB.91.115115}.
In particular, we demonstrate that for metallic systems, the implemented cluster \mbox{D-TRILEX} approach yields more accurate results for the self-energy at the Fermi energy than the much more computationally demanding parquet D$\Gamma$A approximation.
Further, we discuss several periodization schemes based on the self-energy and Green's function, and analyze how the incorporation of the inter-cluster correlations in the CDMFT scheme results in a partial restoration of translational symmetry that is absent in conventional CDMFT solutions. 
A summary and outlook are given in Sec.~\ref{sec:conclusion}.

\section{Cluster extension of the \mbox{D-TRILEX} method \label{sec:method}}

We start with a general action of the Hubbard model:
\begin{align}
{\cal S} =& - \sum_{{\rm k},\sigma,\{l\}} c^{*}_{{\rm k} \sigma l} \left[(i\nu+\mu)\delta^{\phantom{*}}_{ll'} - \varepsilon^{ll'}_{{\bf k}}\right] c^{\phantom{*}}_{{\rm k} \sigma l'} \notag\\
&+ \frac12 \sum_{\{{\rm k}\},{\rm q}}\sum_{\{l\},\{\sigma\}} U^{\phantom{*}}_{l_1 l_2 l_3 l_4} c^{*}_{{\rm k} \sigma l_1} c^{\phantom{*}}_{{\rm k+q}, \sigma l_2} c^{*}_{{\rm k'+q}, \sigma' l_4} c^{\phantom{*}}_{{\rm k'}\sigma' l_3}.
\label{eq:actionlatt}
\end{align}
In this expression, $c^{(*)}_{{\rm k}\sigma{}l}$ is the Grassmann variable which describes the annihilation (creation) of an electron with momentum ${\bf k}$, fermionic Matsubara frequency $\nu$, and spin projection ${\sigma \in \left\{\uparrow, \downarrow \right\}}$.
The label $l$ numerates the orbital and the site within the unit cell.
To simplify notations, we use a combined index ${{\rm k}\in\{{\bf k},\nu\}}$.
The momentum ${\bf k}$ arises from the translational invariance of the unit cell.
$\varepsilon^{ll}_{\bf k}$ is a Fourier transform of the hopping matrix and $\mu$ is the chemical potential.
$U^{\phantom{*}}_{l_1 l_2 l_3 l_4}$ is the electron-electron interaction within the unit cell. The bosonic momentum ${\bf q}$ and Matsubara frequency $\omega$ dependence is also  depicted by a combined index ${{\rm q}\in\{{\bf q},\omega\}}$.

The diagrammatic expansion in dual techniques can be formulated on the basis of an arbitrary interacting reference problem~\cite{Brener2020}. 
The standard choice is the impurity problem obtained from the DMFT mapping of the lattice problem.
Besides the single-band DMFT impurity problem, the \mbox{D-TRILEX} calculations have been performed only for the multi-orbital or multi-impurity reference systems~\cite{10.21468/SciPostPhys.13.2.036, Vandelli2024_PbSi}. 
Cluster problem has the same algebraic structure and can, in fact, be regarded as multi-orbital system in which the orbital indices correspond to cluster sites.  
However, the principal difficulty is that ordinary multi-orbital single-site impurity problems employ an orthogonal local basis, so both the on-site Hamiltonian and the hybridization function are diagonal.  
In a cluster formulation, the inter-site hopping necessarily introduces off-diagonal matrix elements in both objects. 
The presence of these terms drastically complicates the numerical solution of the impurity problem and frequently results in the fermionic sign problem~\cite{PhysRevB.41.9301,PhysRevB.85.201101, PhysRevB.92.195126}.

\subsection{Diagonalization of the reference system}

CDMFT maps the original lattice problem onto an impurity model of a cluster embedded in a bath described by the hybridization function $\Delta^{ll'}_{\nu}$:
\begin{align}
{\cal S}_{\text{imp}}  = & \sum_{\nu,\sigma,\{l\}} 
c_{\nu\sigma{}l}^{\ast} \left[(i\nu + \mu) \delta^{\phantom{*}}_{ll'} - \Delta^{ll'}_{\nu}\right] c_{\nu\sigma{}l'} \notag\\
+ & \sum_{\{\nu\}, \omega} \sum_{\{l\},\{\sigma\}} U_{l_1 l_2 l_3 l_4}
c_{\nu \sigma l_1}^{\ast} c_{\nu+\omega, \sigma l_2} 
c_{\nu'+\omega, \sigma' l_4}^{\ast} c_{\nu', \sigma' l_3}.
\label{Seff-freq}
\end{align}
In CDMFT, the hybridization function $\Delta^{ll'}_{\nu}$ becomes a matrix in the space of cluster sites $\{l,l'\}$ that in general has off-diagonal terms. 
In order to avoid dealing with the sign problem, 
let us perform a basis transformation, $\mathcal{R}$, to diagonalize the Hermitian matrix $\Delta^{ll'}_{\nu}$:
\begin{align}
\Delta_{\nu} =\mathcal{R}^{\dagger}_{\nu} \, D_{\nu} \,\mathcal{R}^{\phantom{\dagger}}_{\nu},
\label{diag-hybr}
\end{align}
where $D_{\nu}$ is a diagonal matrix. 
We emphasize, that this procedure is not required by the \mbox{D-TRILEX} formalism itself, can handle hybridization functions $\Delta^{ll'}_\nu$ with full off-diagonal structure. 
This diagonalization is required for the numerical solution of the impurity problem and, in particular, for the evaluation of two-particle quantities such as charge and spin susceptibilities and three-point vertex functions. 
Within QMC–based solvers used in the current implementation of the \mbox{D-TRILEX} method, these quantities can be computed only when the hybridization function is diagonal in the orbital basis.
The basis transformation \eqref{diag-hybr} motivates the definition of new Grassmann variables:
\begin{align}
\overline{c}_{\nu\sigma{}l}^{\ast} = \sum_{l'} c_{\nu\sigma{}l'}^{\ast} \big[\mathcal{R}^{\dagger}_{\nu}\big]_{l'l}\,, ~~~~
\overline{c}_{\nu\sigma{}l} = \sum_{l'}\big[\mathcal{R}^{\phantom{\dagger}}_{\nu}\big]_{ll'} c_{\nu\sigma{}l'}\,.
\end{align}
The rotation matrix is unitary, since the hybridization function is Hermitian, ensuring that anti-commutation relations are preserved. The impurity problem~\eqref{Seff-freq} becomes
\begin{align}
{\cal S}_{\text{imp}}  = & \sum_{\nu,\sigma,l} 
\overline{c}_{\nu\sigma{}l}^{\ast} \left[i\nu + \mu - D^{ll}_{\nu}\right] \overline{c}_{\nu\sigma{}l} \notag\\
+ & \sum_{\{\nu\}, \omega} \sum_{\{l\},\{\sigma\}} {\cal U}^{\nu\nu'\omega}_{l_1 l_2 l_3 l_4}
\overline{c}_{\nu \sigma l_1}^{\ast} \overline{c}_{\nu+\omega, \sigma l_2} 
\overline{c}_{\nu'+\omega, \sigma' l_4}^{\ast} \overline{c}_{\nu', \sigma' l_3},
\label{Seff-freq2}
\end{align}
where the static electron-electron interaction transforms to a three-frequency-dependent object:
\begin{align}
{\cal U}^{\nu\nu'\omega}_{l_1 l_2 l_3 l_4} = \sum_{\{j\}} U^{\phantom{*}}_{j_1 j_2 j_3 j_4} \big[\mathcal{R}^{\phantom{\dagger}}_{\nu}\big]_{l_1j_1} \big[\mathcal{R}^{\dagger}_{\nu+\omega}\big]_{j_2l_2}
\big[\mathcal{R}^{\phantom{\dagger}}_{\nu'+\omega}\big]_{l_4j_4} \big[\mathcal{R}^{\dagger}_{\nu'}\big]_{j_3l_3}.
\end{align}
Handling such a complex interaction is beyond the capabilities of existing impurity solvers.
Thus, neither option provides a viable solution: complete diagonalization of the hybridization function mitigates the sign problem but leads to an intractable form of the interaction, while keeping the interaction static is possible only for a non-diagonal form of the hybridization function, resulting in the sign problem.

To address this issue, in Refs.~\cite{PhysRevB.92.195126, PhysRevB.85.201101} the authors showed that for different types of clusters there exists a static (frequency-independent) transformation to an ``optimal'' single-particle basis for which the sign problem decreases significantly.
Yet, the sign problem cannot be completely  removed and still scales exponentially with the inverse of the temperature. 

In our work, we follow a similar idea and perform a basis transformation to diagonalize the local part of the single-particle Hamiltonian:
\begin{align}
\overline{\varepsilon}_{K} = \mathcal{R} \left[\sum_{K}\varepsilon_{K}\right] \mathcal{R}^{\dagger}.
\end{align}
This transformation minimizes the off-diagonal components of the hybridization function, as can be seen by applying the basis transformation directly to the hybridization function (see Appendix~\ref{justification-eqt-local-dispersion} for a detailed derivation):
\begin{align}
\overline{\Delta}_{\nu}  = \mathcal{R} \, \Delta_{\nu} \, \mathcal{R}^{\dagger} 
= \mathcal{R} \left[ \sum_{K} \varepsilon_{K} \right] \mathcal{R}^{\dagger} + \mathcal{O} \left( \sum_{K} \mathcal{R} \, \frac{ \left( \varepsilon_{K} + \Sigma^{\rm imp}_{\nu} \right)^2}{i \nu + \mu} \, \mathcal{R}^{\dagger} \right),
\label{rotated-hybr}
\end{align}
where $\Sigma^{\rm imp}_{\nu}$ is the self-energy of the cluster impurity problem.
However, the off-diagonal components of the hybridization function cannot be fully eliminated. 
In our approach, we take a further step by removing these off-diagonal terms through the flexibility in choosing the reference system for the dual diagrammatic expansion~\cite{Brener2020}. 
In dual approaches, the impurity problem~\eqref{Seff-freq} is separated from the original lattice action~\eqref{eq:actionlatt} by adding and subtracting the hybridization function $\Delta^{ll'}_{\nu}$, which may be chosen arbitrarily in the band space $\{l, l'\}$ (see, e.g., Ref.~\cite{10.21468/SciPostPhys.13.2.036}). 

After performing the basis transformation $\mathcal{R}$ that diagonalizes the local part of the single-particle Hamiltonian $\varepsilon_{\bf K}$, we construct the reference system with a strictly diagonal hybridization function $\Delta^{ll}_{\nu}$. 
The off-diagonal components of the hybridization matrix, which emerge naturally from the cluster construction, are deliberately suppressed in this formulation of the impurity problem. 
Consequently, the introduced reference impurity problem corresponds to a restricted form of the cluster dynamical mean-field theory (CDMFT), in which the hybridization between the bonding and antibonding ``orbitals'' is absent. 
For brevity, we will refer to these results as CDMFT results throughout this work. 

The off-diagonal components of the cluster self-energy, $\Sigma^{ll'}_{\nu}$ with ${l \neq l'}$, are subsequently generated through the \mbox{D-TRILEX} diagrammatic expansion. This approach coherently incorporates both momentum-independent (local) and momentum-dependent (non-local) correlation effects beyond those already captured by the diagonal impurity self-energy $\Sigma^{\rm imp}_{\nu,ll}$.
The proposed scheme is general and can be applied to arbitrary cluster problems.
The complexity of the cluster \mbox{D-TRILEX} approach can be estimated in the same way as for the multi-orbital formulation of the method. In Ref.~\cite{10.21468/SciPostPhys.13.2.036}, the complexity of the latter was estimated as
\begin{equation}
\mathcal{O}(N_\nu N_\omega)\,
\mathcal{O}\!\left(
   N_{\mathrm{imp}}^2 \times \sum_{i=1}^{N_{\mathrm{imp}}} N_{l_i}^8
\right)\,
\mathcal{O}(N_k \log N_k)\,,
\end{equation}
where $N_\nu$ ($N_\omega$) is the number of fermionic (bosonic) Matsubara frequencies, $N_{\mathrm{imp}}$ is the number of independent impurities in the reference problem, $N_{l_i}$ is the number of orbitals for the $i$-th impurity, and $N_k$ is the total number of $\mathbf{k}$-points in the Brillouin zone. 
The cluster calculation can then be viewed as a multi-band calculation with $N_{\mathrm{imp}}=1$ and $N_{l_i}$ being the total number of sites and orbitals in the cluster impurity problem.
Therefore, in practice, calculations for a up to a four-site cluster with one orbital per site, or two-site cluster with two orbitals per site, can be carried out within a reasonable time frame. 
Calculations for a six-site cluster with a single orbital per lattice site (or, equivalently, a three-site cluster with two orbitals per site) would already lie at the limit of feasibility with the current implementation.
In what follows, we demonstrate the applicability of the cluster \mbox{D-TRILEX} approach using the one-dimensional nano-ring Hubbard model with a dimer reference system as a representative example.

\subsection{Basis transformation for the 1D Hubbard model}

We start with the one-dimensional Hubbard Hamiltonian:
\begin{equation} 
H = - t \sum_{j, \tau, \sigma} c_{j, \sigma}^{\dagger} c^{\phantom{\dagger}}_{j + \tau,\sigma} + U \sum_{j} n_{j\uparrow}n_{j\downarrow}, 
\label{eq:Horiginal}
\end{equation}
where $j$ labels the atomic position, ${\tau=\pm1}$ denotes the nearest-neighbor site position difference, $t$ denotes the nearest-neighbor hopping integral, $c_{j, \sigma}^{(\dag)}$ is the annihilation (creation) operator for an electron at site $j$, with spin $\sigma$; $n_{j, \sigma} = c_{i, \sigma}^{\dagger}c^{\phantom{\dagger}}_{i, \sigma}$ is the density operator. 
For the remainder of this work, we set ${t = 1}$ and use it as the energy unit.
The chemical potential is set to ${\mu = U/2}$ so that the system is at half filling.

\begin{figure}[t!]
\includegraphics[width=0.90\linewidth]{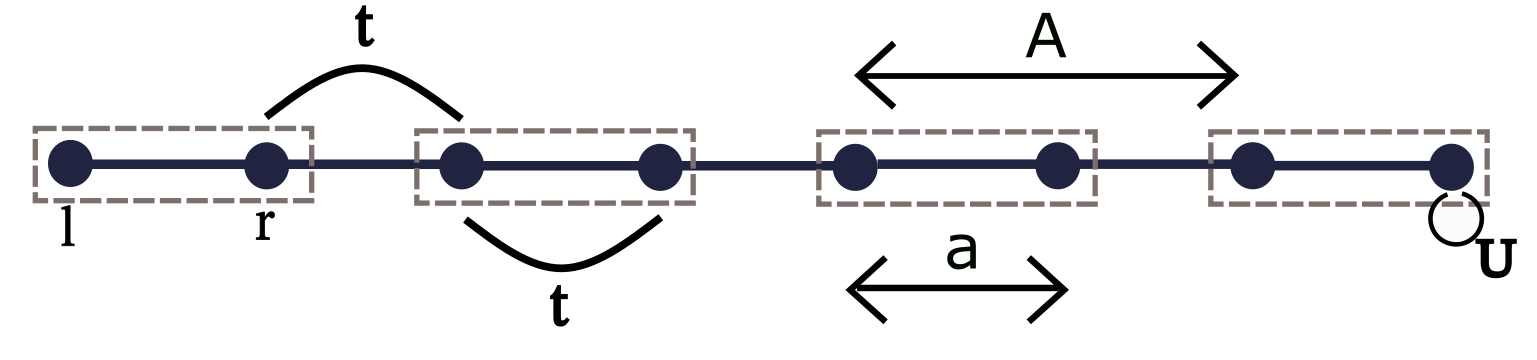}
\caption{Schematic representation of the one-dimensional Hubbard model. The lattice is tiled by identical two-site clusters, indicated by dashed boxes, consisting of the ``left'' (l) and ``right'' (r) sites. The nearest-neighbor hopping amplitude $t$ is considered the same within and between the clusters. The electrons interact via the on-site Coulomb repulsion \(U\). The distance between adjacent clusters is ${A = 2a}$, i.e., twice the lattice constant $a$.
\label{fig:1D-Hubbard_drawing}}
\end{figure}

In the case of a two-site cluster formulation of the problem one can introduce the spinors:
\begin{align}
C_{I} = 
\begin{pmatrix}
l_{I}\\
r_{I}
\end{pmatrix}
~~\text{and} ~~
C^{\dag}_{I} = 
\left( l_{I}^{\dag} ~ r_{I}^{\dag} \right),
\label{original-cluster-variable}
\end{align}
where $l^{(\dagger)}_{I}$ and $r^{(\dagger)}_{I}$ are the annihilation (creation) operators for the electron on the left and right sites in the cluster $I$, respectively (see Fig.~\ref{fig:1D-Hubbard_drawing}). 
The local part of the single-particle Hamiltonian becomes
\begin{align}
H_0^{\text{loc}} = -\sum_{I} C_{I}^{\dagger}
\begin{pmatrix}
0 & t\\
t & 0
\end{pmatrix} C_{I}.
\end{align}
It can be diagonalized by the following transformation to the bonding-antibonding basis: 
\begin{align}
\mathcal{R} =\frac{1}{\sqrt{2}}
\begin{pmatrix}
1 & 1\\
1 & -1
\end{pmatrix}.
\label{rotation-matrix}
\end{align} 
The full single-particle Hamiltonian has the following form:
\begin{align}
H_0 &=  -\sum_{I} C_{I}^{\dagger}
\Bigg\{ 
\begin{pmatrix}
0 & t\\
t & 0
\end{pmatrix}
C_{I} + 
\begin{pmatrix}
0 & t\\
0 & 0
\end{pmatrix}
C_{I-1} + 
\begin{pmatrix}
0 & 0\\
t & 0
\end{pmatrix} 
C_{I+1} \Bigg\} \notag \\
&= -\frac{t}{N_K} 
\sum_{K} C_{K}^{\dagger} 
\begin{pmatrix}
0 & 1 + e^{-iAK}\\
1 + e^{iAK} & 0
\end{pmatrix} 
C_{K},
\label{H0-1D}
\end{align}
where $N_K$ is the number of $K$-points in the reduced BZ, ${A=2a}$ is the distance in real space between the neighboring clusters (vector of translation), and $a$ is the lattice constant.
Upon the basis transformation, the single-particle Hamiltonian becomes
\begin{align}
\overline{H}_0  = \frac{1}{N_K} \sum_{K} \overline{C}^{\dagger}_{K} \,
\overline{\varepsilon}^{\phantom{\dagger}}_{K} \, \overline{C}^{\phantom{\dagger}}_{K},
\label{H0-stag}
\end{align}
where:
\begin{align}
\overline{\varepsilon}_{K} = -t 
\begin{pmatrix}
1 + \cos(A K) & i \sin(A K)\\
-i \sin(A K) & - 1 - \cos(A K)
\end{pmatrix}
\end{align}
and we also introduced a new spinor
\begin{align}
\overline{C} = \mathcal{R} \, C
= \frac{1}{\sqrt{2}}  
\left\{ 
\begin{pmatrix}
1\\
1
\end{pmatrix} 
l + 
\begin{pmatrix}
1\\
-1
\end{pmatrix}
r \right\}
= 
\begin{pmatrix}
a\\
b
\end{pmatrix}.
\label{new-fermionic-var}
\end{align}
The interaction part of the Hamiltonian:
\begin{align}
H_{U} = U \sum_{j} n_{j\uparrow} n_{j\downarrow}
\end{align}
transforms to the Kanamori-like form~\cite{10.1143/PTP.30.275} (${m,m'\in\{a,b\}}$)
\begin{gather}
\overline{H}_U = \frac{1}{2} \sum_{\substack{I,m\\ \sigma\sigma'}} {\cal U}\,n^{m}_{I, \sigma} n^{m}_{I, \sigma'}  
+ \frac{1}{2} \sum_{\substack{I,\sigma\sigma'\\m\neq m' }}\left({\cal U'} - {\cal J}\delta_{\sigma\sigma'}\right)n^{m}_{I, \sigma} n^{m'}_{I, \sigma'} \notag\\
-{\cal J}\sum_{I}\left(a^{\dag}_{I\uparrow} a^{\phantom{\dag}}_{I\downarrow} b^{\dag}_{I\downarrow} b^{\phantom{\dag}}_{I\uparrow} + b^{\dag}_{I\uparrow} b^{\phantom{\dag}}_{I\downarrow} a^{\dag}_{I\downarrow} a^{\phantom{\dag}}_{I\uparrow}\right) \notag\\
+{\cal J}\sum_{I}\left(a^{\dag}_{I\uparrow} a^{\dag}_{I\downarrow} b^{\phantom{\dag}}_{I\downarrow} b^{\phantom{\dag}}_{I\uparrow} + b^{\dag}_{I\uparrow} b^{\dag}_{I\downarrow} a^{\phantom{\dag}}_{I\downarrow} a^{\phantom{\dag}}_{I\uparrow} \right)
\label{interacting-part-H} 
\end{gather}
with an effective intra- (${{\cal U}}$) and inter-band (${{\cal U'}}$) Coulomb interactions and Hund's exchange coupling (${\cal J}$) equal to ${{\cal U} = {\cal U'} = {\cal J} = U/2}$.

\subsection{Computational scheme}

The lattice problem~\eqref{eq:Horiginal} in the bonding-antibonding basis
\begin{align}
\overline{H} = \overline{H}_0 + \overline{H}_U
\label{eq:modelH}
\end{align}
is solved using the multi-band \mbox{D-TRILEX} approach described in Ref.~\cite{10.21468/SciPostPhys.13.2.036}.
Below we provide the key steps of the computational workflow:\\[0.1cm]
{\it 1. DMFT calculation:}\\[0.1cm]
{\it a. Diagonalizing the local part of the single-particle Hamiltonian:}\\
Find the basis transformation ${\cal R}$ that diagonalizes the local part of the single-particle Hamiltonian.\\[0.1cm]
{\it b. Basis transformation:}\\
Perform the basis transformation ${\cal R}$ for the full Hamiltonian~\eqref{eq:Horiginal}.\\[0.1cm]
{\it c. DMFT solution:}\\
Solve the transformed Hamiltonian~\eqref{eq:modelH} using DMFT with the diagonal hybridization function $\Delta^{ll}_{\nu}$. In this work, the DMFT solution is obtained using the w2dynamics package~\cite{wallerberger2019}.\\[0.1cm]
{\it d. Evaluating impurity quantities:}\\
Once DMFT self-consistency is reached, the impurity Green's function $g_{\nu}$, the self-energy $\Sigma^{\rm imp}_{\nu}$, the fermionic hybridization function $\Delta_{\nu}$, the charge (${\varsigma={\rm ch}}$) and spin (${\varsigma={\rm sp}\in\{x,y,z\}}$) susceptibilities $\chi^{\varsigma}_{\omega}$, the polarization operator $\Pi_{{\rm imp}\,\omega}^{\varsigma}$ and the three-point vertex functions $\Lambda^{\varsigma}_{\nu\omega}$, needed for the diagrammatic part of the \mbox{D-TRILEX} scheme, are calculated from the converged DMFT solution.
Here, $\nu$ and $\omega$ refer respectively to fermionic and bosonic
Matsubara frequencies.
\\[0.1cm]
{\it 2. Self-consistent calculation of \mbox{D-TRILEX} diagrams:}\\
The self-consistent calculation of single- and two-
particle quantities in the dual space goes as follows:\\[0.1cm]
{\it a. Evaluating bare propagators:}\\
The output of the DMFT impurity problem is used to construct building blocks for the \mbox{D-TRILEX} diagrammatic expansion.

The bare dual Green's function ${\tilde{\cal G}^{ll'}_{K\nu} = \hat{G}^{ll'}_{K\nu} - \delta_{ll'}g^{ll}_{\nu}}$ corresponds to the difference between the DMFT $\hat{G}$ and impurity $g$ Green's functions, and thus is dressed only in the impurity self-energy $\Sigma^{\rm imp}$:
\begin{align}
\left[\hat{G}_{K\nu}\right]^{-1}_{ll'} = \left[G^{0}_{K\nu}\right]^{-1}_{ll'} - \delta_{ll'}\Sigma^{\rm imp}_{\nu,ll}.
\end{align}
In this expression, $G^{-1}_{0}$ is the inverse of the bare lattice Green's function defined as
\begin{align}
\left[G^{0}_{K\nu}\right]^{-1}_{ll'} = \delta_{ll'}\left(i\nu + \mu\right) - \overline{\varepsilon}^{ll'}_{K}.
\end{align}

The bare dual bosonic propagator (renormalized interaction) ${\tilde{\cal W}_{Q,\omega,\varsigma}^{l_1l_2;\,l_3l_4} = \hat{W}_{Q,\omega,\varsigma}^{l_1l_2;\,l_3l_4} - \frac12U^{\varsigma}_{l_1l_2l_3l_4}}$ is the bare interaction $U^{\varsigma}$ renormalized by the local polarization operator of the impurity problem $\Pi_{\rm imp}^{\varsigma}$ in the charge and spin channels:
\begin{align}
\left[\hat{W}_{Q,\omega,\varsigma}\right]^{-1}_{l_1l_2;\,l_3l_4} = \Big[U^{\varsigma}\Big]^{-1}_{l_1l_2;\,l_3l_4} - \Pi_{{\rm imp}\,\omega,\varsigma}^{l_1l_2;\,l_3l_4}.
\end{align}
Here, $\frac12U^{\varsigma}$ is subtracted from $\hat{W}^{\varsigma}$ in order to prevent double counting of the bare interaction between different channels. $Q$ denotes the bosonic momentum in the reduced BZ.\\[0.1cm]
{\it b. Computing the dual polarization operator:}\\
The dual polarization operator $\tilde{\Pi}^{\varsigma}$ in the charge and spin channel is expressed in terms of the dual Green's function $\tilde{G}$ and the vertex function $\Lambda^{\varsigma}$:
\begin{align}
\tilde{\Pi}_{Q,\omega,\varsigma}^{l_1 l_2;\, l_7 l_8}  
&= \,2\sum_{{\bf k},\nu,\{l\}} \Lambda^{\hspace{-0.05cm}l_4, l_3;\, l_2 l_1}_{\nu+\omega,-\omega, \varsigma} \, \tilde{G}_{K,\nu}^{l_3 l_5} \, \tilde{G}_{K+Q,\nu+\omega}^{l_6 l_4} \,\Lambda^{\hspace{-0.05cm}l_5, l_6;\, l_7 l_8}_{\nu,\omega,\varsigma}.
\label{eq:dual_pol}
\end{align}
\noindent
{\it c. Computing the dressed dual renormalized interaction:}\\
The dressed dual renormalized interaction $\tilde{W}^{\varsigma}$ in the charge and spin channel is calculated using the following expression:
\begin{align}
\left[\tilde{W}_{Q,\omega,\varsigma}\right]^{-1}_{l_1l_2;\,l_3l_4} = \left[\tilde{\cal W}_{Q,\omega,\varsigma}\right]^{-1}_{l_1l_2;\,l_3l_4} - \tilde{\Pi}_{Q,\omega,\varsigma}^{l_1l_2;\,l_3l_4}.
\end{align}
\noindent
{\it d. Computing the dual self-energy:}\\
The dual self-energy $\tilde{\Sigma}$ is expressed in terms of the dual Green's function $\tilde{G}$, the renormalized interaction $\tilde{W}^{\varsigma}$, and the vertex function $\Lambda^{\varsigma}$:
\begin{align}
\tilde{\Sigma}_{K\nu}^{l_1 l_7} 
=2&\sum_{K',\nu',\{l\}} 
\Lambda_{\nu,\omega=0,{\rm ch}}^{\hspace{-0.05cm}l_1, l_7;\, l_3 l_4} \,
\tilde{\cal W}_{Q=0,\omega=0,{\rm ch}}^{l_3 l_4;\,l_5 l_6} \, 
\Lambda_{\nu',\omega=0,{\rm ch}}^{\hspace{-0.05cm}l_8, l_2;\, l_6 l_5} \, \tilde{G}_{K'\nu'}^{l_2 l_8} \notag\\
-&\sum_{{\bf q},\omega,\varsigma,\{l\}}
\Lambda_{\nu,\omega,\varsigma}^{\hspace{-0.05cm}l_1, l_2;\, l_3 l_4} \, \tilde{G}_{K+Q,\nu+\omega}^{l_2 l_8} \, \tilde{W}_{Q,\omega,\varsigma}^{l_3 l_4;\,l_5 l_6} \, \Lambda_{\nu+\omega,-\omega,\varsigma}^{\hspace{-0.05cm}l_8, l_7;\, l_6 l_5}.
\label{eq:dual_sigma_ex}
\end{align}
\noindent
{\it e. Computing the dressed dual Green's function:}\\
The dressed dual Green's function  $\tilde{G}$ is calculated using the following expression:
\begin{align}
\left[\tilde{G}_{K\nu}\right]^{-1}_{ll'} = \left[\tilde{\cal G}_{K\nu}\right]^{-1}_{ll'} - \tilde{\Sigma}^{ll'}_{K\nu}.
\end{align}
\noindent
{\it f. Go back to step b. and iterate until convergence.}\\[0.1cm]
{\it 3. Evaluating lattice quantities:}\\[0.1cm]
{\it a. Evaluating the lattice self-energy and Green's function:}\\
The lattice self-energy $\overline{\Sigma}$ is obtained from the exact relation:
\begin{align}
\overline{\Sigma}^{ll'}_{K\nu} = \delta_{ll'}\Sigma^{\rm imp}_{\nu,ll} + \sum_{l_1}\tilde{\Sigma}^{ll_1}_{K\nu} \left[\mathbb{1} + g_{\nu}\cdot\tilde{\Sigma}_{K\nu}\right]^{-1}_{l_1l'}.
\end{align}
The lattice Green's function is then obtained as
\begin{align}
\Big[\overline{G}_{K\nu}\Big]^{-1}_{ll'} = \left[G^{0}_{K\nu}\right]^{-1}_{ll'} - \overline{\Sigma}^{ll'}_{K\nu}.
\end{align}
\noindent
{\it b. Transforming lattice quantities to the original basis:}\\[0.1cm]
In order to obtain the self-energy and the Green's function in the original basis, we perform the following transformation:
\begin{align}
O_{K\nu} = \mathcal{R}^{\dagger} \, \overline{O}_{K\nu} \,\mathcal{R}
\end{align}
with $O_{K\nu}$ being the self-energy $\Sigma_{K\nu}$ or Green's function $G_{K\nu}$.\\[0.1cm]
\noindent
{\it c. Periodization from the cluster to single-site form:}\\[0.1cm]
To obtain the lattice self-energy and Green's function corresponding to the single-site unit cell from the cluster quantities, we perform the following periodization step by imposing the translational invariance of the original lattice problem:
\begin{align}
\label{eq:L_def}
O^{\rm latt}_{k\nu} = \mathcal L_{k}[O^{ll'}_{k\nu}] = \frac{1}{N_{\rm c}}\sum_{ll'}e^{-ik(r_{l}-r_{l'})}O^{ll'}_{k\nu},
\end{align}
where $r_{l}$ is the position of the $l$-th atom in the unit cell and $k$ corresponds to the original (extended) BZ.
In the dimer case, this relation reduces to
\begin{align}
O^{\rm latt}_{k\nu} = 
\tfrac12\left(O^{11}_{k\nu} + O^{22}_{k\nu}\right) + {\rm Re}\,O^{12}_{k\nu}\,\cos(ka) + {\rm Im}\,O^{12}_{k\nu}\,\sin(ka).
\label{eq:Dimer_periodization}
\end{align}
The quantity $O^{ll'}_{k\nu}$ in the extended BZ can be obtained from the cluster quantity $O^{ll'}_{K\nu}$ in the reduced BZ using the periodicity in momentum space. 

\section{Results} 
\label{sec:results}

To demonstrate the performance of the \mbox{D-TRILEX} method, we apply it to the one-dimensional nano-ring Hubbard model. 
We focus on periodic chains with $N_{\text{c}} = 4$, 6, and 8 lattice sites, for which results can be directly compared to the Hirsch-Fye QMC solution and to the more involved D$\Gamma$A method in the ladder and parquet implementations, for which the data are available from Ref.~\cite{PhysRevB.91.115115}. 
Although Hirsch–Fye QMC entails a systematic Trotter time–discretization error, we can regard those QMC results as reliable benchmarks for the present study. As documented in the Appendix ``Computational Details'' of Ref.~\cite{PhysRevB.91.115115}, the authors assessed and controlled discretization effects (by varying $\Delta\tau$ and checking convergence of the observables relevant here).
For consistency with that work, all calculations are performed for ${t = 1}$, ${U = 2}$, and inverse temperature ${\beta = 10}$.

\begin{figure}[t!]
\includegraphics[width=\linewidth]{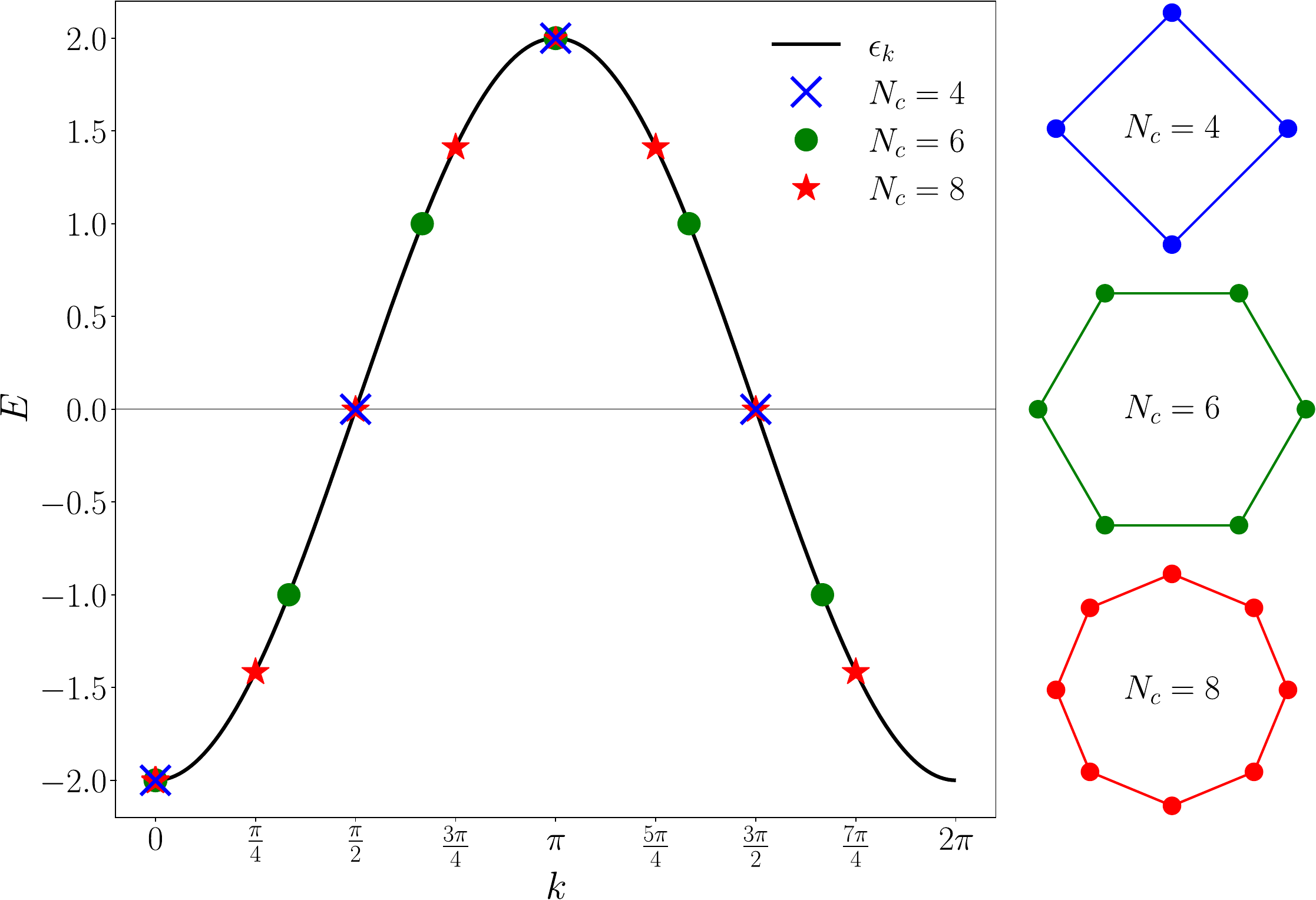}
\caption{The left panel shows the discretized dispersion $\varepsilon_k$ along the first Brillouin zone (lattice constant ${a = 1}$). The full line represents the dispersion of an infinite one-dimensional chain, whereas the discrete symbols correspond to the finite number of lattice sites $N_{\rm c}$. 
The right column depicts the ring geometries: a four-site (blue, top), a six-site (green, middle), and an eight-site (red, bottom) rings.  
In the dispersion plot their spectra are indicated by blue crosses, green circles, and red stars, respectively.
\label{fig:nanoring}}
\end{figure}

\begin{figure}[b!]
\includegraphics[width=1\linewidth]{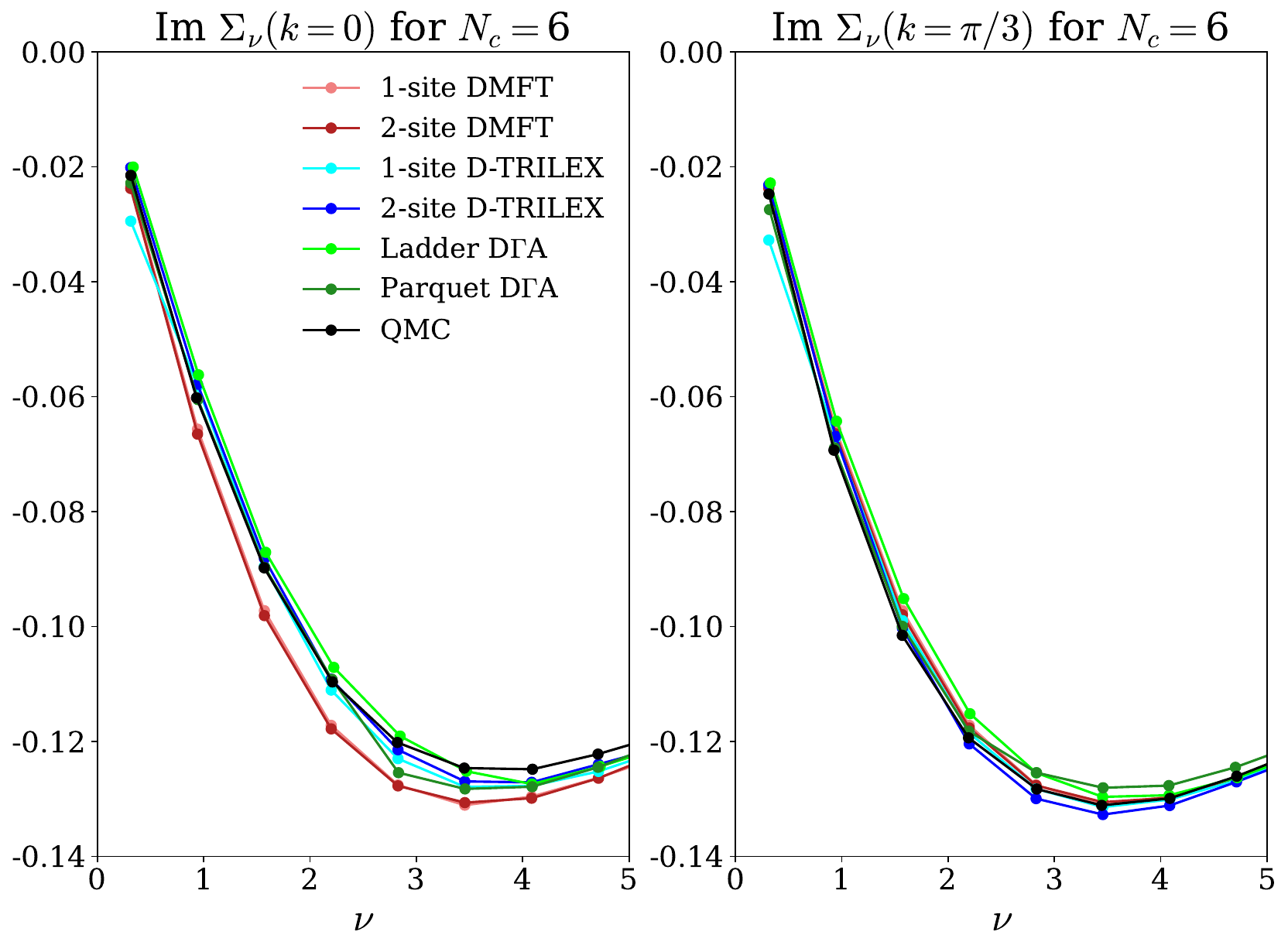}
\caption{The imaginary part of the self-energy calculated as a function of Matsubara frequency $\nu$ at two momenta ${k=0}$ (left) and ${\pi/3}$ (right). The results are obtained for the case of ${N_{\text{c}}=6}$ at ${U = 2}$ and ${\beta = 10}$ using different methods indicated in the legend. The ladder D$\Gamma$A, parquet D$\Gamma$A, and QMC results are taken from Ref.~\cite{PhysRevB.91.115115}.
\label{figs_N6_Im}}
\end{figure}

\subsection{Insulating system, $\boldsymbol{N=6}$}

We refer to the configuration with ${N_{\rm c}=6}$ sites as an insulating system, since its non-interacting spectral function exhibits an energy gap separating the occupied states at \mbox{$k = 0,~\pi/3,~5\pi/3$} from the unoccupied states at \mbox{$k = 2\pi/3,~\pi,~4\pi/3$}. This behavior is illustrated in Fig.~\ref{fig:nanoring}, where the ${N_{\rm c}=6}$ case is shown in green.

In Fig.~\ref{figs_N6_Im} we plot the imaginary part of the lattice self-energy calculated at ${k=0}$ (a) and ${k=\pi/3}$ (b) momenta as a function of Matsubara frequency $\nu$.
The results are obtained using the single-site DMFT (light red), two-site cluster DMFT (dark red), single-site \mbox{D-TRILEX} (cyan), two-site cluster \mbox{D-TRILEX} (blue) and are compared with the ladder D$\Gamma$A (light green), parquet D$\Gamma$A (dark green) and QMC (black) data of Ref.~\cite{PhysRevB.91.115115}.
The two-site cluster DMFT and \mbox{D-TRILEX} self-energies are periodized from the cluster space corresponding to the reduced BZ to the single-site form corresponding to the extended BZ using Eq.~\eqref{eq:L_def}.  
This will be discussed in more detailed later, see sec.~\ref{sec:periodization}. The small momentum-dependence of the imaginary part of the self-energy, that can be seen in Fig.~\ref{figs_N6_Im}, allows for the DMFT results to be close to the benchmark QMC result. Both D$\Gamma$A and \mbox{D-TRILEX} methods improve upon DMFT and reproduce a slight change of the self-energy between the ${k=0}$ and ${k=\pi/3}$ points. 
We observe that the single-site \mbox{D-TRILEX} calculations slightly overestimate the self-energy at low frequencies, but the cluster extension of the method cures this problem and is in a very good agreement with the QMC result. 
Overall, we find that the imaginary part of the self-energy is reproduced with good accuracy by all methods in the insulating case of ${N_{\rm c}=6}$.

\begin{figure}[t!]
\includegraphics[width=1\linewidth]{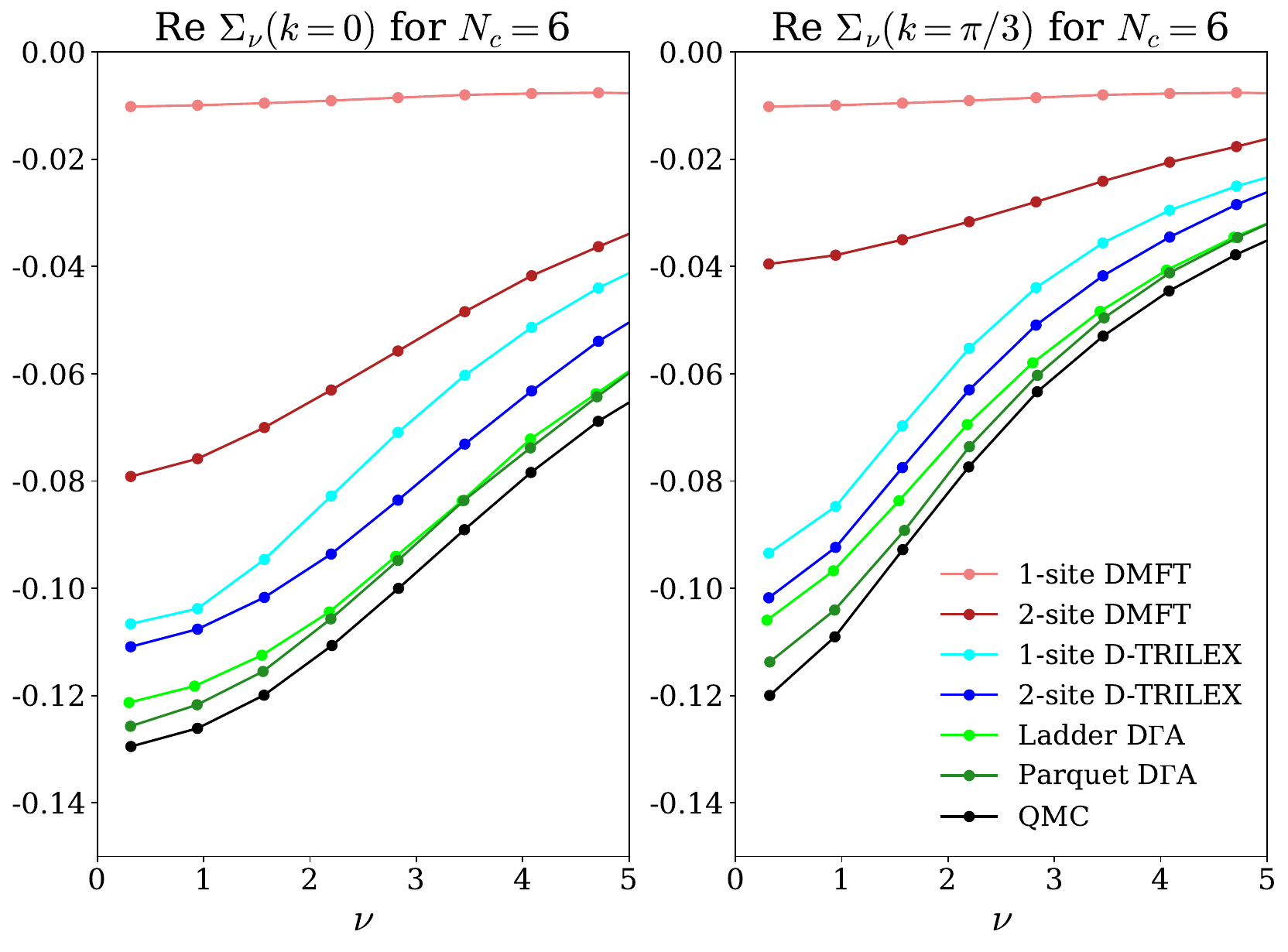}
\caption{The real part of the self-energy calculated as a function of Matsubara frequency $\nu$ at two momenta ${k=0}$ (left) and ${k=\pi/3}$ (right). The results are obtained for the case of ${N_{\text{c}}=6}$ at ${U = 2}$ and ${\beta = 10}$ using different methods indicated in the legend. The ladder D$\Gamma$A, parquet D$\Gamma$A, and benchmark QMC results are taken from Ref.~\cite{PhysRevB.91.115115}.
\label{figs-N_6_Re}}
\end{figure}

In contrast, the real part of the self-energy, shown in Fig.~\ref{figs-N_6_Re}, exhibits more pronounced differences among the considered methods. The single-site DMFT result fails dramatically at both ${k = 0}$ and ${k = \pi/3}$ points.
The CDMFT approach improves upon single-site DMFT, but the results remain far from the benchmark QMC data at both momenta. 
The single-site \mbox{D-TRILEX} provides a substantial improvement over CDMFT. 
This result suggests, that in the insulating ${N_{c}=6}$ case the long-range collective fluctuations, captured diagrammatically, play a more important role than the non-perturbative short-range correlations treated within the cluster reference problem.
The cluster \mbox{D-TRILEX} approach further enhances the accuracy compared to its single-site version. Nevertheless, both \mbox{D-TRILEX} versions exhibit an approximately constant offset with respect to the QMC reference, suggesting that the discrepancy is largely momentum-independent. 
At both ${k = 0}$ and ${k = \pi/3}$, the ladder D$\Gamma$A yields more accurate results than cluster \mbox{D-TRILEX}, while the parquet D$\Gamma$A achieves the best agreement with the QMC benchmark among all methods considered. 
These observations hold consistently for both momentum points, indicating that the hierarchy of approximations remains robust across different regions of the Brillouin zone in this insulating case.
We also note that, although the \mbox{D-TRILEX} results for the real part of the self-energy in the ${N_{\rm c}=6}$ case are less accurate than those of D$\Gamma$A methods, the self-energy itself remains very small compared to the electronic dispersion, ${|\text{Re}\,\Sigma|\ll|\varepsilon_{k}|}$.
Therefore, the observed discrepancy is not expected to have a significant impact.

\subsection{Metallic systems, $\boldsymbol{N=4}$ and $\boldsymbol{8}$} 

\begin{figure}[t!]
\includegraphics[width=1\linewidth]{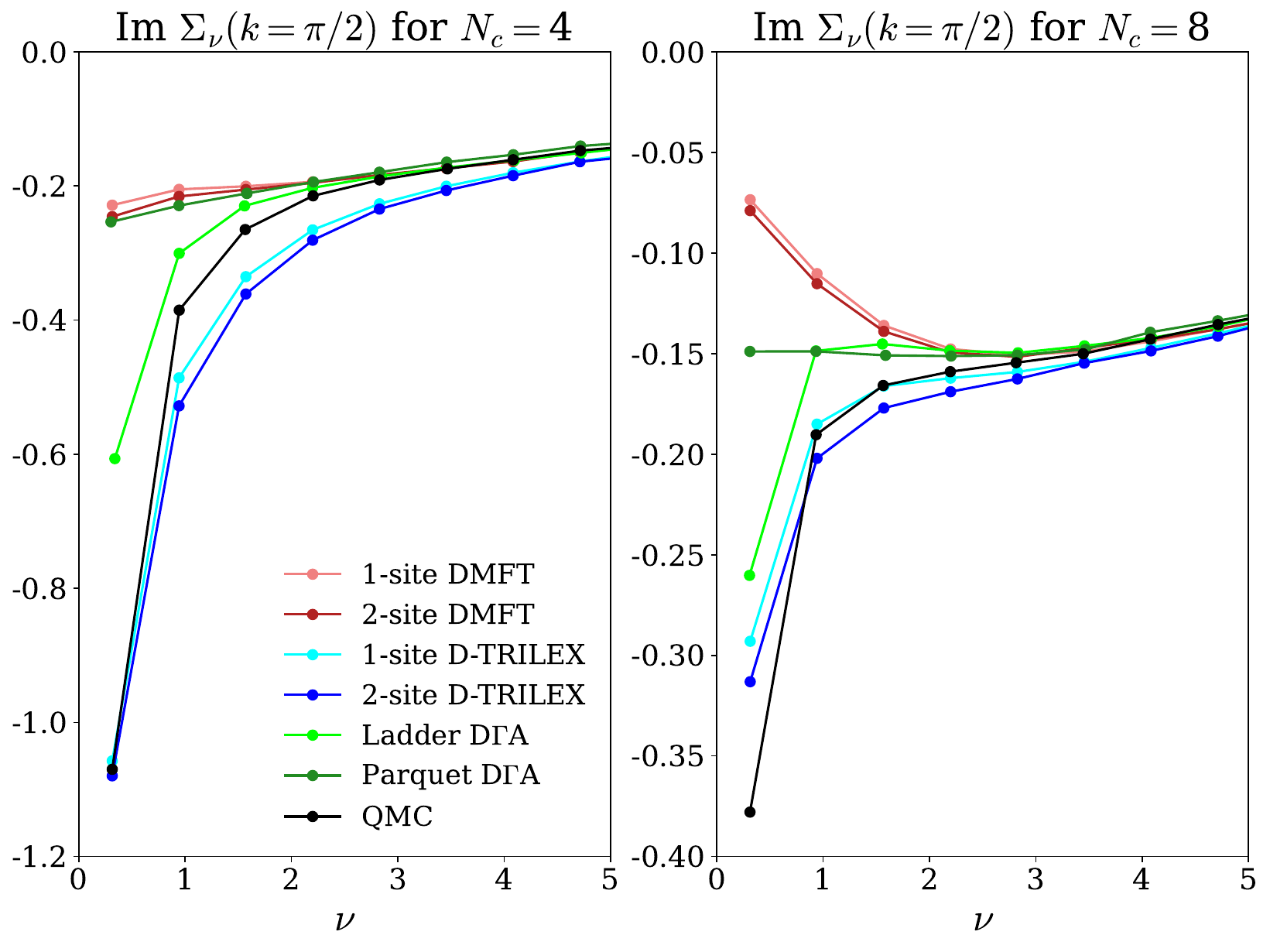}
\caption{The imaginary part of the self-energy as a function of Matsubara frequency $\nu$ at momentum $k=\pi/2$ for the systems with $N_{\text{c}}=4$ (left) and $N_{\text{c}}=8$ (right).  
All results are computed at $U = 2$ and $\beta = 10$ using the methods indicated in the legend.
The ladder D$\Gamma$A, parquet D$\Gamma$A, and benchmark QMC results are taken from Ref.~\cite{PhysRevB.91.115115}.
\label{figs-Im-k_pi2}}
\end{figure}

In the cases of ${N_{\rm c}=4}$ and ${8}$, which we refer to as metallic systems, the non-interacting spectral function exhibits a doubly degenerate state at the Fermi level, corresponding to the momenta ${k=\pi/2}$ and ${3\pi/2}$.
This behavior is illustrated by the blue and red colors in Fig.~\ref{fig:nanoring}, respectively.
This suggests that the self-energies at non-equivalent $k$-points may exhibit significant momentum-dependent variation, particularly when comparing values at the Fermi energy to those further away from it.
Single-site DMFT is not able to reproduce the momentum dependence of the self-energy, and more elaborate methods are required to capture these effects.

\begin{figure}[t!]
\includegraphics[width=1\linewidth]{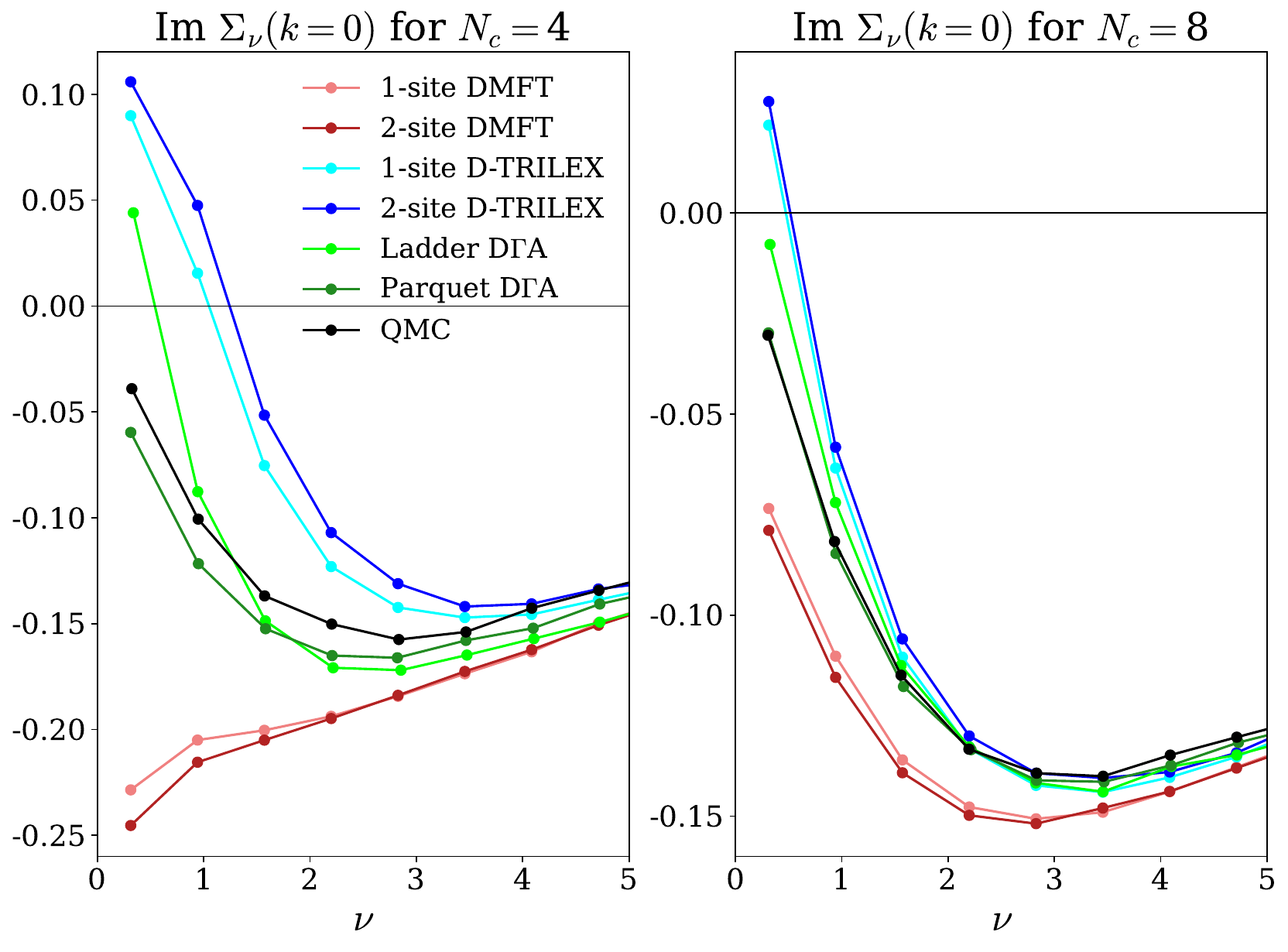}
\caption{The imaginary part of the self-energy as a function of Matsubara frequency $\nu$ at momentum $k = 0$ for lattices with $N_{\text{c}} = 4$ (left) and $N_{\text{c}} = 8$ (right). All curves are calculated at $U = 2$ and $\beta = 10$ using the methods specified in the legend.
The ladder D$\Gamma$A, parquet D$\Gamma$A, and benchmark QMC results are taken from Ref.~\cite{PhysRevB.91.115115}
\label{figs-Im-k_0}}
\end{figure}

In Fig.~\ref{figs-Im-k_pi2} we show the imaginary part of the self-energy calculated at ${k=\pi/2}$ as a function of frequency for the case of ${N_{c}=4}$ (left) and ${N_{c}=8}$ (right). 
Note, that ${{\rm Re}\,\Sigma(k=\pi/2)=0}$ in our case.
This $k$-point corresponds to the Fermi energy, and the self-energy at ${k=\pi/2}$ is much larger than the one obtained at ${k=0}$ (see Fig.~\ref{figs-Im-k_0}).   
First, the ${{\rm Im}\,\Sigma(k=\pi/2)}$ provided by QMC for both ${N_{c}=4}$ and ${N_{c}=8}$ cases show an insulating (divergent at ${\nu\to0}$) behavior.
Interestingly, the cluster DMFT result fails to reproduce this behavior and nearly coincides with the single-site DMFT result. This indicates that long-range collective electronic fluctuations play a crucial role in the metallic case, particularly at the Fermi energy.
Among all considered approaches, \mbox{D-TRILEX} provides the most accurate result for the self-energy at ${k=\pi/2}$. 
At ${N_{\rm c}=4}$, the single-site and cluster \mbox{D-TRILEX} results are similar. 
This trend persists for ${N_{\rm c}=8}$, although the difference between the single-site and cluster \mbox{D-TRILEX} results becomes more visible, with the cluster version being closer to the exact result at the lowest Matsubara frequency, while the single-site approximation is more accurate at higher frequencies.
The ladder D$\Gamma$A captures the insulating behavior of the self-energy but is substantially less accurate than both versions of \mbox{D-TRILEX}. 
Surprisingly, we find that although the parquet D$\Gamma$A is a diagrammatic extension of the ladder version, it performs significantly worse: it fails to capture the insulating behavior of the self-energy and even remains at the level of DMFT for ${N_{\rm c}=4}$.
Ref.~\cite{PhysRevB.91.115115} attributes the better performance of ladder D$\Gamma$A at \mbox{$k=\pi/2$} to the so-called $\lambda$-correction, arguing that it effectively emulates an outer self-consistency by modifying the impurity problem. 
Our comparison challenges this interpretation, since the single-site \mbox{D-TRILEX} approach, which features a similar (ladder-like) diagrammatic structure, uses the same DMFT impurity problem as a reference for the diagrammatic expansion, and also does not perform an outer self-consistency loop, yet achieves better accuracy without employing the $\lambda$-correction.

\begin{figure}[t!]
\includegraphics[width=1\linewidth]{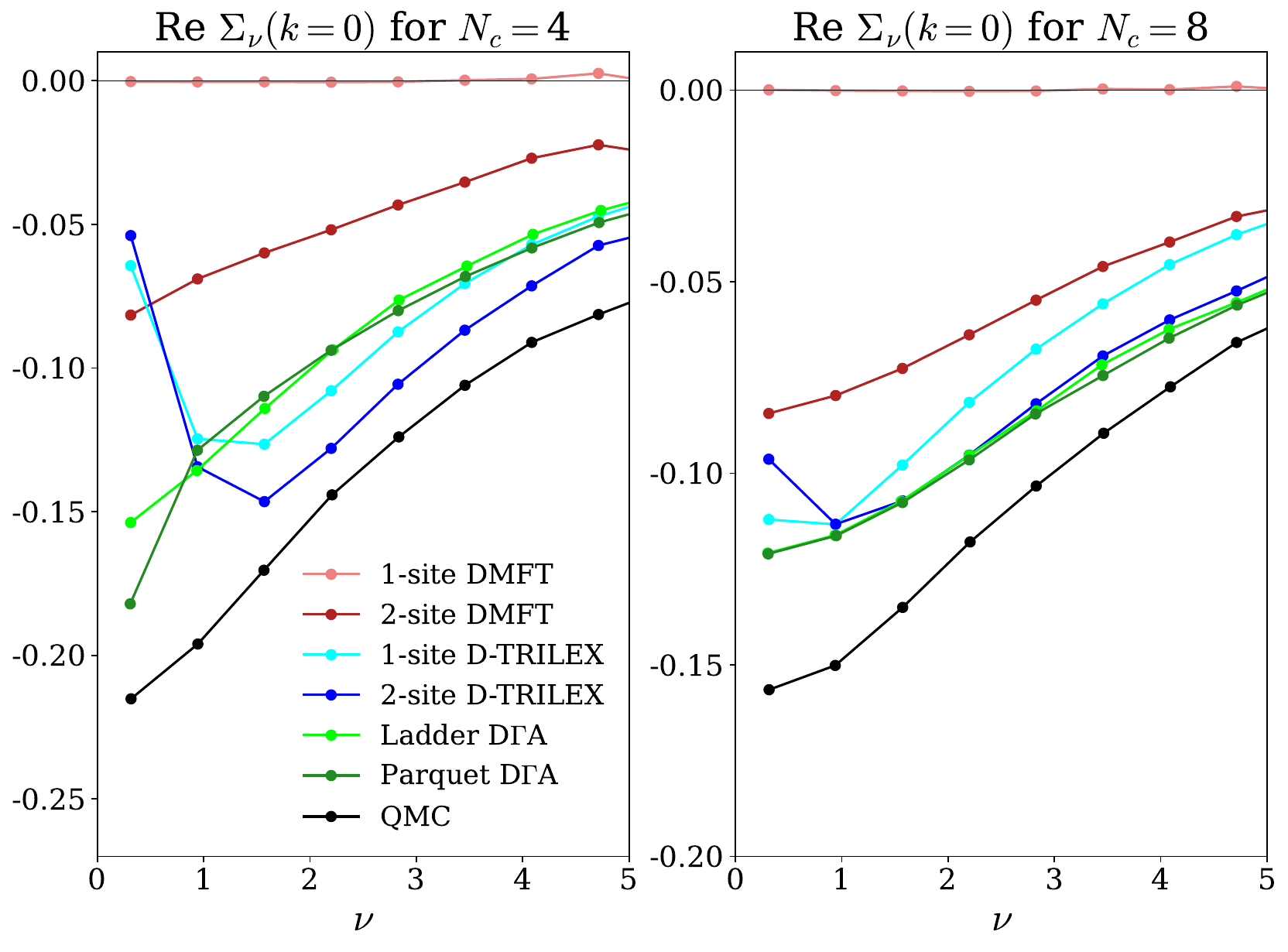} 
\caption{The real part of the self-energy as a function of Matsubara frequency $\nu$ at momentum ${k = 0}$ for lattices with ${N_{\text{c}} = 4}$ (left) and ${N_{\text{c}} = 8}$ (right).  
All results are computed at ${U = 2}$ and ${\beta = 10}$ using the methods indicated in the legend.
The ladder D$\Gamma$A, parquet D$\Gamma$A, and benchmark QMC results are taken from Ref.~\cite{PhysRevB.91.115115}\label{figs-Re-k_0}}
\end{figure}

At the Brillouin zone center, ${k=0}$, the parquet D$\Gamma$A provides the most accurate description of both the imaginary (Fig.~\ref{figs-Im-k_0}) and real (Fig.~\ref{figs-Re-k_0}) parts of the self-energy, while both DMFT versions are the least accurate.
We note that ${{\rm Im}\,\Sigma(k=0)}$ for ${N_{\rm c}=4}$ predicted by the single-site and cluster \mbox{D-TRILEX} approaches shows a non-causal behavior for the two lowest Matsubara frequencies. 
The anomaly carries over to the real part of the self-energy, where the first two Matsubara points deviate strongly from the exact trend. 
This behavior improves with increasing the number of lattice sites to ${N_{\rm c}=8}$, but the imaginary part of the \mbox{D-TRILEX} self-energy still remains positive at the lowest Matsubara frequency. 
A similar behavior for ${{\rm Im}\,\Sigma(k=0)}$ is also found in Ref.~\cite{PhysRevB.91.115115} for the ladder D$\Gamma$A approach in the case of ${N_{\rm c}=4}$ and is attributed to the neglect of particle–particle diagrams inherent in the particle–hole ladder approximation.   
The current \mbox{D-TRILEX} implementation also omits particle–particle correlations.
However, the fact that the cluster extension of the single-site \mbox{D-TRILEX} does not lead to significant improvements suggests that the missing diagrammatic contributions are likely long-ranged and go beyond a perturbative ladder-like approximation.

The momentum-dependent \mbox{D-TRILEX} correction to the local DMFT self-energy can be related to the leading eigenvalue (LE) of collective electronic fluctuations in the charge and spin channels. In the absence of these fluctuations, the LE is zero and DMFT and \mbox{D-TRILEX} results become identical. 
On the contrary, the ${\text{LE}=1}$ leads to a divergence in the considered channel and signals the phase transition to the ordered state.
In our cease, the charge LE is very small (${\text{LE}_{\rm ch} \simeq 0.06}$) indicating that the charge fluctuations are nearly absent in the system. 
The spin LE is instead rather large (${\text{LE}_{\rm sp} \simeq 0.65}$) and therefore the magnetic fluctuations represent the leading collective electronic fluctuations.

\begin{figure}[t!]
\includegraphics[width=0.85\linewidth]{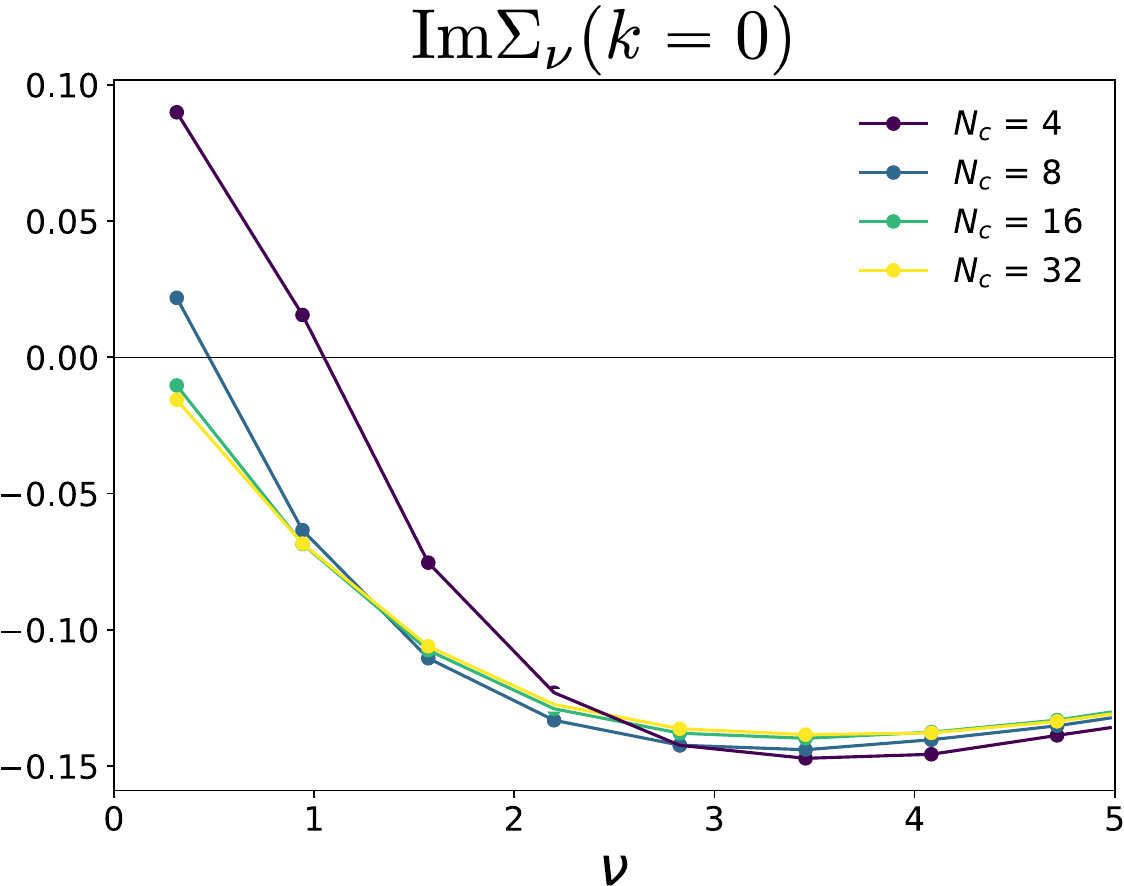}
\caption{Imaginary part of the self-energy at the $\Gamma$ point ($k=0$) as a function of Matsubara frequency $\nu$ for different cluster sizes $N_c$. The non-causality problem (positive values of $\mathrm{Im}\,\Sigma_\nu$) is clearly visible for small clusters ($N_c=4$ and $8$), but progressively diminishes and essentially disappears for $N_c \geq 16$.
\label{fig:non-causality-evol-with-N}}
\end{figure}

The non-causal behavior observed in the imaginary part of the self-energy at small system sizes is directly connected to these strong magnetic fluctuations. 
Figure~\ref{fig:non-causality-evol-with-N} shows that the non-causal behavior decreases as the system size increases and disappears entirely for ${N_c \geq 16}$. 
This suggests that the numerical instability originates from the treatment of spin fluctuations in the diagrammatic expansion, whose magnitude is highly sensitive to the discretization of momentum space.
In particular, both the density of states (DOS) and the spin susceptibility can display sharp, quasi-$\delta$-function features that are numerically difficult to resolve. 
This issue is especially pronounced in 1D, where the DOS exhibits strong singularities at the band edges corresponding to the boundary $k$-points of the BZ.
In small-scale systems with coarse momentum grids, these peaked structures are represented by single $k$-points, leading to an overestimation of their amplitude and consequently to stronger numerical instabilities in the self-energy. As the system size increases and the momentum grid becomes finer, the same physical peaks are distributed over multiple neighboring $k$-points. 
It contributes in smoothing the discretized susceptibility and DOS, and leads to competition for magnetic fluctuations between modes characterized by the $k$-points physical peaks are distributed over.
This improved momentum resolution reduces the numerical artifacts and explains the suppression of non-causal behavior observed for ${N \geq 16}$. 
However, this convergence behavior reflects a competition between two effects: while finer grids better resolve the intrinsic width of the spin susceptibility peaks, they also capture more accurately their sharp structure. The observed systematic improvement with system size indicates that the smoothing effect dominates, and that the quasi-$\delta$-function features possess a finite intrinsic width that becomes properly resolved only at sufficiently fine momentum discretization.

The non-causality issue in small-scale systems, where the number of $k$-points is fixed by the number of lattice sites, calls for a more elaborate treatment of correlation effects.
For example, in Ref.~\cite{PhysRevB.105.245115} the authors proposed a modified self-consistency scheme by reformulating the Dyson equation to enforce causality of the total self-energy by construction, given the nonlocal self-energy correction. In the \mbox{D-TRILEX} framework, the non-local self-energy correction is not an external input but rather the primary quantity to be determined. Nevertheless, implementing an outer self-consistency scheme between the cluster reference system and the \mbox{D-TRILEX} diagrammatic expansion could, in principle, address the non-causality issue by iteratively adjusting the cluster impurity problem until the total self-energy becomes causal. However, this would require recalculating the expensive two-particle impurity quantities at each iteration of the outer loop, which would increase the numerical cost tremendously since the computation of these two-particle vertex functions represents the most expensive step of the calculation. 

Focusing on Fig.~\ref{figs-Re-k_0}, we observe that, with the exception of the two lowest Matsubara frequencies, the real part of the self-energy obtained from the \mbox{D-TRILEX} approaches exhibits comparable accuracy to the parquet D$\Gamma$A method, and in some cases even approaches the QMC benchmark more closely. The \mbox{D-TRILEX} diagrammatic contributions lead to an overcorrection of the (C)DMFT results exclusively at ${k=0}$ and only for the lowest Matsubara frequencies. We note that this discrepancy is localized both in momentum space (appearing only at the Brillouin zone center) and in frequency space (affecting only the low-frequency regime). 
We anticipate that implementing an outer self-consistency loop, which would update the reference impurity problem would mitigate this overcorrection and further improve the overall agreement with the benchmark results.

In general, as in the insulating ${N_{\rm c}=6}$ case, the real part of the self-energy away from the Fermi energy remains relatively small compared to the electronic dispersion, ${|\text{Re}\,\Sigma(k=0)|\ll|\varepsilon_{k}|}$, so discrepancies with the benchmark QMC results are not expected to be significant for describing the electronic properties of the system. 

\subsection{Periodization and its impact on the lattice self‑energy}
\label{sec:periodization}

The cluster formulation inherently produces self-energies and Green's functions with a matrix structure in the space of cluster sites, corresponding to the tiling of the dimerized reference system. To obtain physical observables for the original lattice problem with a single-site periodicity, a periodization procedure is required to map these cluster quantities onto scalar momentum-dependent functions defined in the extended/original Brillouin zone.
Let us compare the self-energies obtained using two distinct periodization schemes: 
(i) applying the periodization operation $\mathcal{L}_k$ to the cluster self-energy~\eqref{eq:L_def}, which is proportional to the inverse of the Green's function, i.e., to $G^{-1}$, and (ii) applying it to the cluster Green's function $G$ and then computing the self-energy from the periodized Green's function by inverting the Dyson equation.
Since $\mathcal{L}_k$ is a linear operation, whereas matrix inversion is not, the two procedures (although identical for the exact solution of the problem) may yield different results within approximate methods:
\begin{align}
\text{(i)}~\Sigma_{k\nu}^{\text{latt}} 
& = \mathcal{L}_k\left[\Sigma_{k\nu}\right] \notag\\
&=\mathcal{L}_k\left[\mathbb{1}(i\nu+\mu) - \varepsilon^{ll'}_{k} - G^{-1}_{k\nu}\right] \notag\\
& = i\nu+\mu - \varepsilon_{k} - \mathcal{L}_k \left[G^{-1}_{k\nu}\right],
\label{eq:Sigma_L_inverse} \\
\text{(ii)}~\Sigma_{k\nu}^{\text{latt}}
& = i\nu+\mu - \varepsilon_{k} - \left(\mathcal{L}_k\left[G_{k\nu}\right]\right)^{-1},
\label{eq:Sigma_L_green}
\end{align}
where $\varepsilon^{ll'}_{k}$ is the cluster version of the ``original'' dispersion $\varepsilon_{k}$.
In the following, we refer to Eqs.~\eqref{eq:Sigma_L_inverse} and~\eqref{eq:Sigma_L_green} as the $\Sigma$- and $G$-periodization schemes, respectively.

In the case of the two-site cluster, the difference between the $\Sigma$- and $G$-periodization schemes can be obtained analytically.
In the particle–hole–symmetric case, the two-site cluster Green function reduces to
\begin{align}
G_{K\nu} =
\begin{pmatrix}
G^{11}_{K\nu} & G^{12}_{K\nu} \\[0.1cm]
G^{12}_{K\nu} & G^{22}_{K\nu} \\
\end{pmatrix},
\end{align}
where ${G^{11}_{K\nu}\in \mathbb C}$, and ${G^{12}_{K\nu}\in\mathbb R}$.
The difference between the two periodization schemes for the self-energy becomes
\begin{align}
\label{eq:mismatch_formula}
\Sigma_{k\nu}^{\text{(ii)}} - \Sigma_{k\nu}^{\text{(i)}} 
&= \big( \mathcal L_k[G_{k\nu}] \big)^{-1} - \mathcal L_k\bigl[G^{-1}_{k\nu}\bigr] \\
&= \frac{ \left(G^{12}_{k\nu}\right)^{2} \sin^{2}(ka) }
  { \left[ \left(G^{12}_{k\nu}\right)^{2} - \left(G^{11}_{k\nu}\right)^{2} \right] \left[ G^{11}_{k\nu} + G^{12}_{k\nu}\cos(ka) \right] }.
\end{align}
Therefore, the two periodization schemes give identical results at ${k=0}$, and the discrepancy is the largest at ${ka=\pi/2}$, i.e. at the Fermi energy. 

\begin{figure}[b!]
\includegraphics[width=1\linewidth]{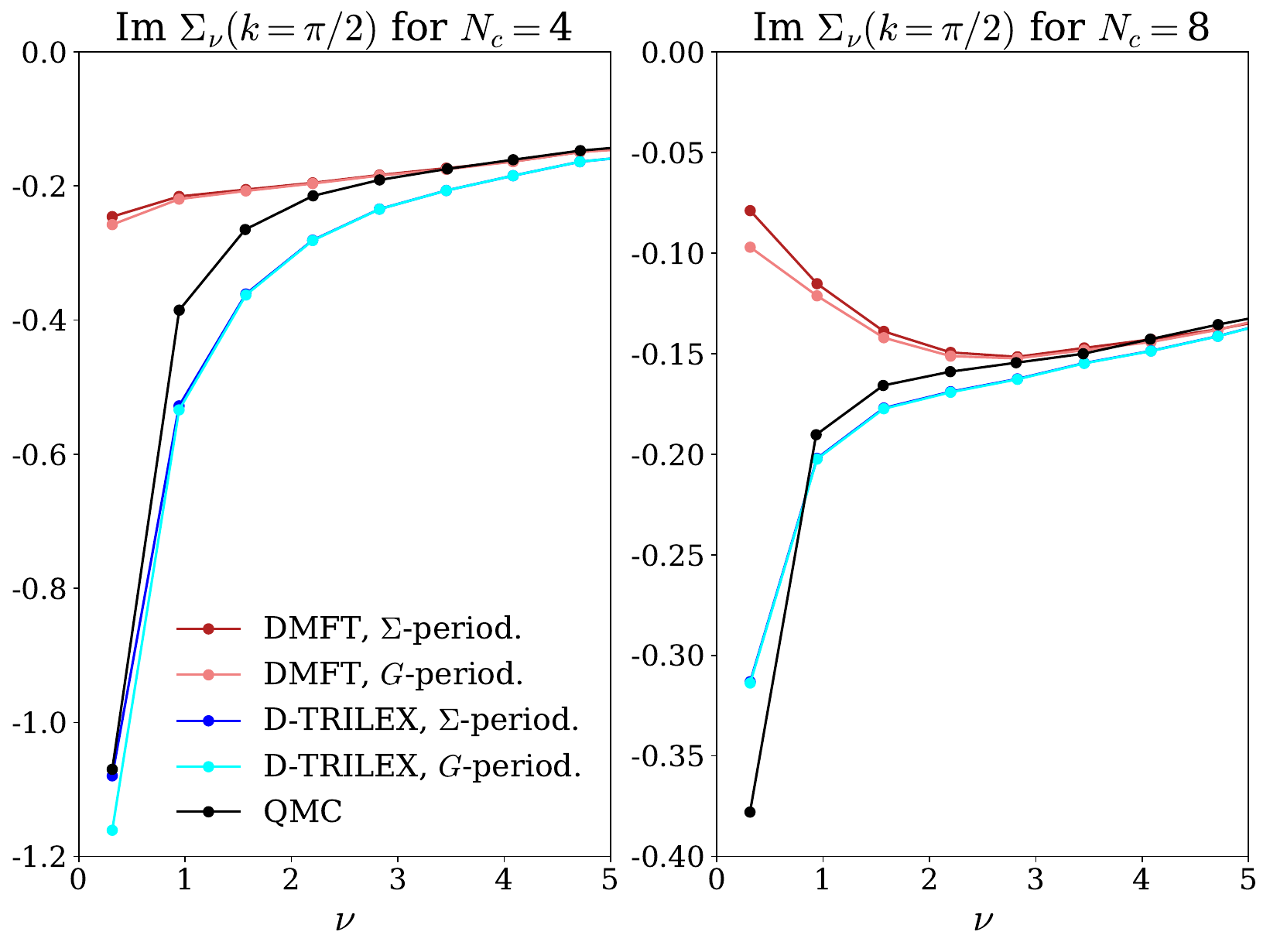}
\caption{The imaginary part of the lattice self-energy obtained at ${k=\pi/2}$ from the cluster DMFT (red colors) and cluster \mbox{D-TRILEX} (blue colors) methods using $\Sigma$- (light colors) and $G$- (dark colors) periodization schemes. The black curve corresponds to the benchmark QMC result taken from Ref.~\cite{PhysRevB.91.115115}.
\label{fig:periodization_comparison}}
\end{figure}

Figure~\ref{fig:periodization_comparison} compares the imaginary part of the self-energy at $k=\pi/2$ (${a=1}$) obtained using the $\Sigma$- and $G$-periodization schemes for both CDMFT and cluster \mbox{D-TRILEX} methods, applied to nanorings composed of ${N_{\text{c}} = 4}$ (left figure) and ${N_{\text{c}} = 8}$ (right figure) sites.
For a smaller cluster of ${N_{\text{c}} = 4}$ sites, both periodization schemes yield very similar results within each method, with discrepancies appearing only at the lowest Matsubara frequency. Quantitatively, the absolute difference between the two periodization schemes is larger for \mbox{D-TRILEX} (${|\Delta \text{Im}\Sigma| \approx 0.081}$) than for DMFT (${|\Delta \text{Im}\Sigma| \approx 0.011}$). However, when normalized by the magnitude of the self-energy itself, the relative discrepancies are comparable (7.0\% for \mbox{D-TRILEX} {\it vs} 4.6\% for DMFT). For a larger cluster of ${N_{\rm c}=8}$ sites, a qualitatively different behavior emerges: the periodization ambiguity becomes more pronounced in CDMFT, whereas it is substantially reduced in cluster \mbox{D-TRILEX}, with the two periodization schemes yielding nearly indistinguishable results across all Matsubara frequencies. 
This trend demonstrates that the \mbox{D-TRILEX} diagrammatic correction, by incorporating inter-cluster correlations, effectively mitigates the periodization ambiguity inherent in cluster approaches.
The convergence of the two periodization schemes as the cluster size increases suggests that \mbox{D-TRILEX} provides a more consistent framework for mapping cluster quantities back to the original lattice, thereby addressing one of the fundamental limitations of conventional cluster methods.

\subsection{Removing off-diagonal terms from the hybridization and partial restoration of the translation invariance}

Let us first justify the neglect of the off-diagonal contributions to the hybridization function. 
We perform CDMFT calculations for the two-site cluster impurity problem using the w2dynamics package~\cite{wallerberger2019}, which allows inclusion of a (small) off-diagonal hybridization in the calculation of single-particle quantities. 
By carrying out CDMFT with the full hybridization function, we find that transforming the two-site cluster to the bonding–antibonding basis not only eliminates the fermionic sign problem but is also well justified numerically, as the off-diagonal hybridization components remain below the numerical noise across the entire explored parameter range.

\begin{figure}[t!]
\includegraphics[width=0.65\linewidth]{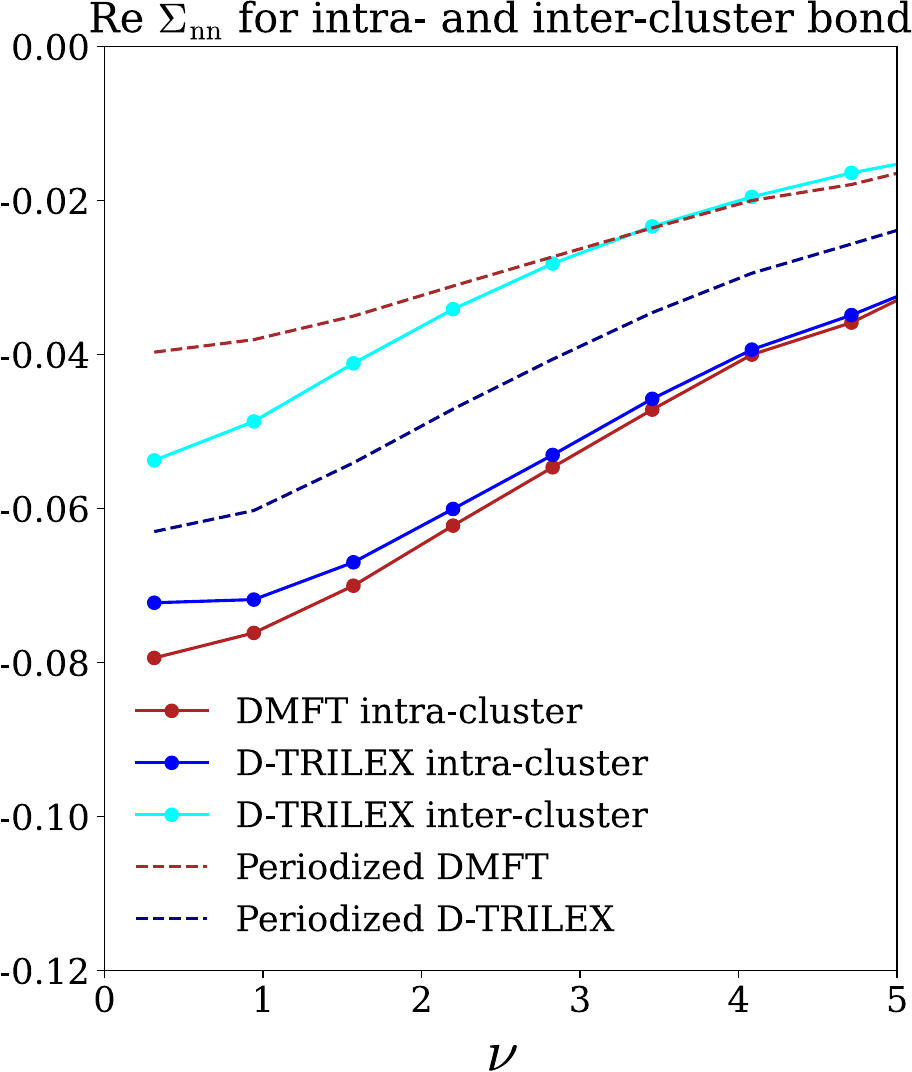}
\caption{Real part of the nearest-neighbor self-energy in real space as a function of Matsubara frequency $\nu$ for the ${N_{\rm c}=8}$ lattice. The results distinguish between intra-cluster bonds (connections within a two-site cluster) and inter-cluster bonds (connections between adjacent clusters). For CDMFT: intra-cluster contribution (dark red solid line), inter-cluster contribution (light red solid line, left panel only; identically zero for self-energy), and periodized result obtained via Eq.~\eqref{eq:Sigma_L_inverse} (red dashed line). For cluster \mbox{D-TRILEX}: intra-cluster contribution (dark blue solid line), inter-cluster contribution generated diagrammatically (cyan solid line), and periodized result (blue dashed line). In a translationally invariant system, intra- and inter-cluster contributions should coincide and equal the periodized value. The deviation between these quantities quantifies the degree of translational symmetry breaking inherent to the cluster approach. All calculations performed at ${U=2}$, ${\beta=10}$, $t=1$, and half-filling.
\label{fig:Sigma_comparison}}
\end{figure}

Let us now analyze whether the diagrammatic contributions introduced beyond CDMFT within the cluster \mbox{D-TRILEX} scheme are able to, at least partially, restore the translational symmetry that is broken by introducing the cluster reference system at the CDMFT step. To this aim, in Fig.~\ref{fig:Sigma_comparison} we compare the inter- and intra-cluster quantities corresponding to nearest-neighbor lattice sites in real space for the case of ${N_{\rm c}=8}$ lattice sites. In a perfectly translationally invariant system, these quantities should be identical for all equivalent bonds, i.e., ${\Sigma^{\text{intra}}_{\nu} = \Sigma^{\text{inter}}_{\nu}}$.
The differences between intra-cluster, inter-cluster, and periodized quantities therefore provide convenient metrics for quantifying the degree to which translational symmetry is broken.

Figure~\ref{fig:Sigma_comparison} displays the real part of the nearest-neighbor self-energy in real space, distinguishing between intra-cluster (within a two-site cluster) and inter-cluster (between adjacent clusters) contributions, for all non-vanishing components.
In CDMFT, the inter-cluster self-energy vanishes identically by construction, as a direct consequence of the DMFT assumption that the self-energy is purely local in the cluster site basis.
Consequently, the intra-cluster self-energy (solid red line) is far from the periodized value (dashed red line), with the discrepancy serving as a direct measure of translational symmetry breaking. 
In the cluster \mbox{D-TRILEX} approach, the inter-cluster contribution to the self-energy (light cyan) is generated diagrammatically through the inclusion of long-range collective fluctuations. However, this inter-cluster self-energy remains different from both the intra-cluster (dark blue) and periodized (dashed blue) values, indicating that translational symmetry is partially restored. Nevertheless, the overall discrepancy between intra-cluster and periodized self-energies is substantially reduced in \mbox{D-TRILEX} compared to DMFT, representing a clear improvement in the direction of restoring translational invariance.

A striking feature in Fig.~\ref{fig:Sigma_comparison} is the proximity between intra-cluster self-energies obtained from DMFT (dark red) and \mbox{D-TRILEX} (dark blue). This close agreement can be attributed to the strong, non-perturbative nature of the correlations between neighboring sites in this small-scale one-dimensional system. The additional perturbative contributions included in the \mbox{D-TRILEX} diagrammatic expansion are insufficient to significantly modify the intra-cluster self-energy, precisely because the dominant physics at this length scale is of non-perturbative character and is already encoded in the reference system. The primary contribution of the \mbox{D-TRILEX} diagrams is therefore to generate the inter-cluster self-energy (light cyan), which, when combined with the intra-cluster component through periodization, yields a result (dashed blue) that is in good agreement with the benchmark QMC data, as demonstrated in the preceding sections.

These observations lead us to conclude that the CDMFT reference system constitutes the limiting factor preventing complete restoration of translational invariance within the cluster \mbox{D-TRILEX} framework. The diagrammatic corrections, while successfully generating inter-cluster correlations, operate within the constraint imposed by the fixed CDMFT solution, which inherently treats intra- and inter-cluster bonds asymmetrically. A fully self-consistent algorithm incorporating an outer self-consistency loop would be required to achieve complete restoration of translational invariance. In such a scheme, the reference impurity problem would be iteratively updated to ensure self-consistency between the impurity-level and lattice-level descriptions: specifically, a chosen local observable (e.g., the local Green's function or self-energy) computed by applying the full \mbox{D-TRILEX} diagrammatic expansion should reproduce the corresponding quantity obtained directly from the cluster impurity problem.
Nevertheless, the present results demonstrate that even without this outer loop, cluster \mbox{D-TRILEX} achieves a meaningful reduction in translational symmetry breaking compared to CDMFT.

\section{Summary and outlook}
\label{sec:conclusion}

In this work, we have developed and benchmarked a cluster extension of the \mbox{D-TRILEX} approach for treating non-local electronic correlations in strongly interacting systems. We demonstrate that the method performs remarkably well for small-scale low-dimensional systems, exemplified here by the one-dimensional nano-ring Hubbard model with ${N_c = 4, 6, 8}$ lattice sites at intermediate coupling strength ${U=2t}$ and inverse temperature ${\beta=10}$.
Our results establish a clear hierarchy of approximations. The cluster \mbox{D-TRILEX} approach systematically improves upon both single-site DMFT and its cluster extension (CDMFT) by incorporating long-range collective charge and spin fluctuations through its diagrammatic expansion. Furthermore, the cluster \mbox{D-TRILEX} yields more accurate results than its single-site counterpart, demonstrating the benefit of combining exact short-range correlations within the cluster with the efficient treatment of long-wavelength fluctuations. When compared to the computationally more demanding dynamical vertex approximation (D$\Gamma$A), we find that while the parquet D$\Gamma$A achieves better accuracy in certain regions of momentum space far from the Fermi level, single-site and cluster \mbox{D-TRILEX} deliver superior performance at the physically most critical momenta associated with Fermi energy. This is particularly significant because these momentum points, where the self-energy becomes comparable in magnitude to the bare electronic dispersion, govern the low-energy physics 

We have identified a limitation of the current implementation of the method. 
At ${k=0}$ momentum at the Brillouin zone boundary, the \mbox{D-TRILEX} approach exhibits non-causal behavior in the imaginary part of the self-energy, manifesting as positive values at the lowest Matsubara frequencies for small system sizes. 
This pathology, which may arise intrinsically in some DMFT extensions~\cite{PhysRevB.105.245115, PhysRevB.97.125141}, such as diagrammatic \mbox{$GW$+EDMFT}~\cite{PhysRevB.95.155104, PhysRevB.104.195146, PhysRevMaterials.1.043803, PhysRevB.105.085102} and in ladder D$\Gamma$A~\cite{PhysRevB.91.115115, Titvinidze_2025} methods, diminishes systematically with increasing system size.
In the cases examined to elucidate the origin of causality violations, we find that this issue is suppressed for system sizes ${N_c \geq 16}$. Additionally, implementing an outer self-consistency loop between the reference system and the diagrammatic expansion could, in principle, is expected to remedy this problem, albeit at significantly increased computational cost.

Another focus of our analysis concerns the periodization ambiguity and the translational-symmetry breaking inherent to cluster methods. We have demonstrated that the \mbox{D-TRILEX} diagrammatic corrections substantially reduce the periodization ambiguity compared to CDMFT: for ${N_c=8}$, the ${\Sigma}$- and ${G}$-periodization schemes yield nearly identical results within \mbox{D-TRILEX} approach, whereas significant discrepancies can be observed in CDMFT. By examining the real-space structure of intra-cluster and inter-cluster contributions to the self-energy, we have shown that \mbox{D-TRILEX} partially restores translational invariance by generating non-zero inter-cluster correlations through its diagrammatic expansion.

The modest computational cost combined with the ability to capture non-local correlations and reduce periodization ambiguity, positions cluster \mbox{D-TRILEX} as a promising tool for exploring symmetry-broken states in realistic materials. In particular, the method is well-suited for investigating phase transitions associated with both local order parameters (such as charge density waves or magnetic ordering) and non-local order parameters (such as $d$-wave superconductivity), where the interplay between short-range non-perturbative correlations and long-range collective fluctuations plays a decisive role. Future work will focus on applications to multi-orbital systems, exploration of different cluster geometries, and investigation of symmetry-broken phases in realistic materials.

\section*{Acknowledgements}
The authors acknowledge support from IDRIS/GENCI under grant number 091393, and are thankful to the CPHT computer support team for their help.

\appendix

\section{Rotation of the hybridization function} 
\label{justification-eqt-local-dispersion}

In this appendix we present the detained derivation of Eq.~\eqref{rotated-hybr}.
Throughout the derivation we work with rotated dispersion $\overline{\varepsilon}_{k}$ and impurity self-energy $\overline{\Sigma}^{\rm imp}_{\nu}$ defined as
\begin{align}
\overline{\varepsilon}_{K} = 
\mathcal R\,\varepsilon_{K}\,\mathcal R^{\dagger} ~~ \text{and} ~~
\overline{\Sigma}_{\nu}^{\text{imp}} = 
\mathcal R\,\Sigma^{\text{imp}}_{\nu}\,\mathcal R^{\dagger}.   
\end{align}
For sake of simplicity, in what follows we will omit the superscript ``imp'' in the impurity self-energy. 
The expression for the rotated hybridization function can be found via the Dyson equation for the impurity Green's function:
\begin{align}
\overline{\Delta}_{\nu}
= \mathbb{1}(i\nu+\mu) - \overline{\Sigma}_{\nu} - \left\{\sum_{K}
\left[\mathbb{1}(i\nu+\mu) - \overline{\varepsilon}_{K} - \overline{\Sigma}_{\nu}\right]^{-1}\right\}^{-1}.
    \label{tilde_delta}
\end{align}
In this expression, the term inside the curly brackets corresponds to the local part of the CDMFT Green's function $\hat{G}_{K\nu}$, which, according to the DMFT self-consistency condition is equal to the impurity Green's function: $\sum_{K}\hat{G}_{K\nu} = g^{\rm imp}_{\nu}$.
Let us perform the high-frequency expansion for the last term on the right‐hand side:
\begin{equation}
\sum_{K}\left[\mathbb{1}(i\nu+\mu) - (\overline{\varepsilon}_{K}+\overline{\Sigma}_{\nu})\right]^{-1}
\simeq \frac{1}{i\nu+\mu} \left[\mathbb{1} + \sum_{n\ge 1} \sum_{K} 
\left(\tfrac{\overline{\varepsilon}_{K}+\overline{\Sigma}_{\nu}}{i\nu+\mu}\right)^{n} \right].
\end{equation}
The inverse of this expression reads
\begin{equation}
\Biggl[ \mathbb{1} + \sum_{n\ge 1} \sum_{K} 
\Bigl(\tfrac{\overline{\varepsilon}_{K}+\overline{\Sigma}_{\nu}}{i\nu+\mu}\Bigr)^{n}
    \Biggr]^{-1}
  = \sum_{m\ge 0}
     \Biggl( -
        \sum_{n\ge 1}\sum_{K}
        \Bigl(\tfrac{\overline{\varepsilon}_{K}+\overline{\Sigma}_{\nu}}{i\nu+\mu}\Bigr)^{\!n}
     \Biggr)^{m}\;.
\end{equation}
We now disentangle the lowest–order pieces of the double series.
Specifically, we single out the ${m=0}$ slice, which produces the identity matrix ${\mathbb{1}(i\nu+\mu)}$, and the unique ${m=1}$, ${n=1}$ contribution, which yields the linear term ${(i\nu+\mu)\sum_{K} \bigl(\overline{\varepsilon}_{K}+\overline{\Sigma}_{\nu} \bigr) / (i\nu+\mu) = \langle\overline{\varepsilon}_{K}\rangle + \overline{\Sigma}_{\nu}}$.
All other configurations are collected into the two residual sums:
(i) ${m=1}$ with ${n\ge 2}$, and (ii) ${m\ge 2}$ with any ${n\ge 1}$.  
Both start at the second order in the small parameter
$(\overline{\varepsilon}_{K}+\overline{\Sigma}_{\nu})/(i\nu+\mu)$ and therefore
constitute genuinely higher‐order corrections:
\begin{gather} 
\left\{\sum_{K}\left[\mathbb{1}(i\nu+\mu) - (\overline{\varepsilon}_{K}+\overline{\Sigma}_{\nu})\right]^{-1}
\right\}^{-1} =  \notag\\
= \mathbb{1}(i\nu+\mu) - \langle\overline{\varepsilon}_{K}\rangle - \overline{\Sigma}_{\nu} - (i\nu+\mu) \sum_{n\ge 2}\sum_{K} \Bigl(\tfrac{\overline{\varepsilon}_{K}+\overline{\Sigma}_{\nu}}{i\nu+\mu}\Bigr)^{\!n} \notag\\
+(i\nu+\mu) \sum_{m\ge 2} \Biggl( -\sum_{n\ge 1}\sum_{K}\Bigl(\tfrac{\overline{\varepsilon}_{K}+\overline{\Sigma}_{\nu}}{i\nu+\mu}\Bigr)^{\!n}\Biggr)^{m}.
\end{gather}
We find, that the terms \mbox{$\mathbb{1}(i\nu+\mu)$} and $\overline{\Sigma}_{\nu}$ cancel out with the corresponding terms in the definition of $\overline{\Delta}_{\nu}$~\eqref{tilde_delta}. The final expression for the rotated hybridization becomes
\begin{align}
\overline{\Delta}(\nu) = \langle\overline{\varepsilon}_{K}\rangle &- (i\nu+\mu) \sum_{n\ge 2}\sum_{K}\Bigl(\tfrac{\overline{\varepsilon}_{K}+\overline{\Sigma}_{\nu}} {i\nu+\mu}\Bigr)^{\!n} \notag \\
&+ (i\nu+\mu) \sum_{m\ge 2}\Biggl( -\sum_{n\ge 1}\sum_{K} \Bigl(\tfrac{\overline{\varepsilon}_{K}+\overline{\Sigma}_{\nu}} {i\nu+\mu}\Bigr)^{\!n}\Biggr)^{m}.
\end{align}
This relation implies that the most optimal basis that reduces off-diagonal components of the hybridization is the one that diagonalizes the local part of the electronic dispersion $\langle\overline{\varepsilon}_{K}\rangle$. 

\bibliography{paper-1D-Hubbard.bib}

\begin{thebibliography}{134}%
\makeatletter
\providecommand \@ifxundefined [1]{%
 \@ifx{#1\undefined}
}%
\providecommand \@ifnum [1]{%
 \ifnum #1\expandafter \@firstoftwo
 \else \expandafter \@secondoftwo
 \fi
}%
\providecommand \@ifx [1]{%
 \ifx #1\expandafter \@firstoftwo
 \else \expandafter \@secondoftwo
 \fi
}%
\providecommand \natexlab [1]{#1}%
\providecommand \enquote  [1]{``#1''}%
\providecommand \bibnamefont  [1]{#1}%
\providecommand \bibfnamefont [1]{#1}%
\providecommand \citenamefont [1]{#1}%
\providecommand \href@noop [0]{\@secondoftwo}%
\providecommand \href [0]{\begingroup \@sanitize@url \@href}%
\providecommand \@href[1]{\@@startlink{#1}\@@href}%
\providecommand \@@href[1]{\endgroup#1\@@endlink}%
\providecommand \@sanitize@url [0]{\catcode `\\12\catcode `\$12\catcode `\&12\catcode `\#12\catcode `\^12\catcode `\_12\catcode `\%12\relax}%
\providecommand \@@startlink[1]{}%
\providecommand \@@endlink[0]{}%
\providecommand \url  [0]{\begingroup\@sanitize@url \@url }%
\providecommand \@url [1]{\endgroup\@href {#1}{\urlprefix }}%
\providecommand \urlprefix  [0]{URL }%
\providecommand \Eprint [0]{\href }%
\providecommand \doibase [0]{https://doi.org/}%
\providecommand \selectlanguage [0]{\@gobble}%
\providecommand \bibinfo  [0]{\@secondoftwo}%
\providecommand \bibfield  [0]{\@secondoftwo}%
\providecommand \translation [1]{[#1]}%
\providecommand \BibitemOpen [0]{}%
\providecommand \bibitemStop [0]{}%
\providecommand \bibitemNoStop [0]{.\EOS\space}%
\providecommand \EOS [0]{\spacefactor3000\relax}%
\providecommand \BibitemShut  [1]{\csname bibitem#1\endcsname}%
\let\auto@bib@innerbib\@empty
\bibitem [{\citenamefont {Abrikosov}\ and\ \citenamefont {Khalatnikov}(1959)}]{abrikosov1959}%
  \BibitemOpen
  \bibfield  {author} {\bibinfo {author} {\bibfnamefont {A.~A.}\ \bibnamefont {Abrikosov}}\ and\ \bibinfo {author} {\bibfnamefont {I.~M.}\ \bibnamefont {Khalatnikov}},\ }\bibfield  {title} {\bibinfo {title} {{The theory of a Fermi liquid (the properties of liquid 3He at low temperatures)}},\ }\href {https://doi.org/10.1088/0034-4885/22/1/310} {\bibfield  {journal} {\bibinfo  {journal} {Rep. Prog. Phys.}\ }\textbf {\bibinfo {volume} {22}},\ \bibinfo {pages} {329} (\bibinfo {year} {1959})}\BibitemShut {NoStop}%
\bibitem [{\citenamefont {Nozieres}(2018)}]{nozieres2018}%
  \BibitemOpen
  \bibfield  {author} {\bibinfo {author} {\bibfnamefont {P.}~\bibnamefont {Nozieres}},\ }\href@noop {} {\emph {\bibinfo {title} {{Theory {{Of Quantum Liquids}}}}}}\ (\bibinfo  {publisher} {{CRC Press}},\ \bibinfo {year} {2018})\BibitemShut {NoStop}%
\bibitem [{\citenamefont {Hedin}(1965)}]{GW1}%
  \BibitemOpen
  \bibfield  {author} {\bibinfo {author} {\bibfnamefont {L.}~\bibnamefont {Hedin}},\ }\bibfield  {title} {\bibinfo {title} {{New Method for Calculating the One-Particle Green's Function with Application to the Electron-Gas Problem}},\ }\href {https://doi.org/10.1103/PhysRev.139.A796} {\bibfield  {journal} {\bibinfo  {journal} {Phys. Rev.}\ }\textbf {\bibinfo {volume} {139}},\ \bibinfo {pages} {A796} (\bibinfo {year} {1965})}\BibitemShut {NoStop}%
\bibitem [{\citenamefont {Aryasetiawan}\ and\ \citenamefont {Gunnarsson}(1998)}]{GW2}%
  \BibitemOpen
  \bibfield  {author} {\bibinfo {author} {\bibfnamefont {F.}~\bibnamefont {Aryasetiawan}}\ and\ \bibinfo {author} {\bibfnamefont {O.}~\bibnamefont {Gunnarsson}},\ }\bibfield  {title} {\bibinfo {title} {{The $GW$ method}},\ }\href {http://stacks.iop.org/0034-4885/61/i=3/a=002} {\bibfield  {journal} {\bibinfo  {journal} {Rep. Prog. Phys.}\ }\textbf {\bibinfo {volume} {61}},\ \bibinfo {pages} {237} (\bibinfo {year} {1998})}\BibitemShut {NoStop}%
\bibitem [{\citenamefont {Hedin}(1999)}]{GW3}%
  \BibitemOpen
  \bibfield  {author} {\bibinfo {author} {\bibfnamefont {L.}~\bibnamefont {Hedin}},\ }\bibfield  {title} {\bibinfo {title} {{On correlation effects in electron spectroscopies and the $GW$ approximation}},\ }\href {http://stacks.iop.org/0953-8984/11/i=42/a=201} {\bibfield  {journal} {\bibinfo  {journal} {J. Phys. Condens. Matter}\ }\textbf {\bibinfo {volume} {11}},\ \bibinfo {pages} {R489} (\bibinfo {year} {1999})}\BibitemShut {NoStop}%
\bibitem [{\citenamefont {Bickers}\ and\ \citenamefont {Scalapino}(1989)}]{bickers1989}%
  \BibitemOpen
  \bibfield  {author} {\bibinfo {author} {\bibfnamefont {N.~E.}\ \bibnamefont {Bickers}}\ and\ \bibinfo {author} {\bibfnamefont {D.~J.}\ \bibnamefont {Scalapino}},\ }\bibfield  {title} {\bibinfo {title} {{Conserving approximations for strongly fluctuating electron systems. I. Formalism and calculational approach}},\ }\href {https://doi.org/https://doi.org/10.1016/0003-4916(89)90359-X} {\bibfield  {journal} {\bibinfo  {journal} {Ann. Phys.}\ }\textbf {\bibinfo {volume} {193}},\ \bibinfo {pages} {206} (\bibinfo {year} {1989})}\BibitemShut {NoStop}%
\bibitem [{\citenamefont {Bickers}\ \emph {et~al.}(1989)\citenamefont {Bickers}, \citenamefont {Scalapino},\ and\ \citenamefont {White}}]{PhysRevLett.62.961}%
  \BibitemOpen
  \bibfield  {author} {\bibinfo {author} {\bibfnamefont {N.~E.}\ \bibnamefont {Bickers}}, \bibinfo {author} {\bibfnamefont {D.~J.}\ \bibnamefont {Scalapino}},\ and\ \bibinfo {author} {\bibfnamefont {S.~R.}\ \bibnamefont {White}},\ }\bibfield  {title} {\bibinfo {title} {{Conserving Approximations for Strongly Correlated Electron Systems: Bethe-Salpeter Equation and Dynamics for the Two-Dimensional Hubbard Model}},\ }\href {https://doi.org/10.1103/PhysRevLett.62.961} {\bibfield  {journal} {\bibinfo  {journal} {Phys. Rev. Lett.}\ }\textbf {\bibinfo {volume} {62}},\ \bibinfo {pages} {961} (\bibinfo {year} {1989})}\BibitemShut {NoStop}%
\bibitem [{\citenamefont {Lichtenstein}\ and\ \citenamefont {Katsnelson}(1998)}]{PhysRevB.57.6884}%
  \BibitemOpen
  \bibfield  {author} {\bibinfo {author} {\bibfnamefont {A.~I.}\ \bibnamefont {Lichtenstein}}\ and\ \bibinfo {author} {\bibfnamefont {M.~I.}\ \bibnamefont {Katsnelson}},\ }\bibfield  {title} {\bibinfo {title} {{Ab initio calculations of quasiparticle band structure in correlated systems: LDA++ approach}},\ }\href {https://doi.org/10.1103/PhysRevB.57.6884} {\bibfield  {journal} {\bibinfo  {journal} {Phys. Rev. B}\ }\textbf {\bibinfo {volume} {57}},\ \bibinfo {pages} {6884} (\bibinfo {year} {1998})}\BibitemShut {NoStop}%
\bibitem [{\citenamefont {Georges}\ \emph {et~al.}(1996)\citenamefont {Georges}, \citenamefont {Kotliar}, \citenamefont {Krauth},\ and\ \citenamefont {Rozenberg}}]{RevModPhys.68.13}%
  \BibitemOpen
  \bibfield  {author} {\bibinfo {author} {\bibfnamefont {A.}~\bibnamefont {Georges}}, \bibinfo {author} {\bibfnamefont {G.}~\bibnamefont {Kotliar}}, \bibinfo {author} {\bibfnamefont {W.}~\bibnamefont {Krauth}},\ and\ \bibinfo {author} {\bibfnamefont {M.~J.}\ \bibnamefont {Rozenberg}},\ }\bibfield  {title} {\bibinfo {title} {{Dynamical mean-field theory of strongly correlated fermion systems and the limit of infinite dimensions}},\ }\href {https://doi.org/10.1103/RevModPhys.68.13} {\bibfield  {journal} {\bibinfo  {journal} {Rev. Mod. Phys.}\ }\textbf {\bibinfo {volume} {68}},\ \bibinfo {pages} {13} (\bibinfo {year} {1996})}\BibitemShut {NoStop}%
\bibitem [{\citenamefont {Mott}(1974)}]{Mott}%
  \BibitemOpen
  \bibfield  {author} {\bibinfo {author} {\bibfnamefont {N.~F.}\ \bibnamefont {Mott}},\ }\href@noop {} {\emph {\bibinfo {title} {{Metal-insulator transitions}}}}\ (\bibinfo  {publisher} {London: Taylor \& Francis},\ \bibinfo {year} {1974})\BibitemShut {NoStop}%
\bibitem [{\citenamefont {Imada}\ \emph {et~al.}(1998)\citenamefont {Imada}, \citenamefont {Fujimori},\ and\ \citenamefont {Tokura}}]{RevModPhys.70.1039}%
  \BibitemOpen
  \bibfield  {author} {\bibinfo {author} {\bibfnamefont {M.}~\bibnamefont {Imada}}, \bibinfo {author} {\bibfnamefont {A.}~\bibnamefont {Fujimori}},\ and\ \bibinfo {author} {\bibfnamefont {Y.}~\bibnamefont {Tokura}},\ }\bibfield  {title} {\bibinfo {title} {{Metal-insulator transitions}},\ }\href {https://doi.org/10.1103/RevModPhys.70.1039} {\bibfield  {journal} {\bibinfo  {journal} {Rev. Mod. Phys.}\ }\textbf {\bibinfo {volume} {70}},\ \bibinfo {pages} {1039} (\bibinfo {year} {1998})}\BibitemShut {NoStop}%
\bibitem [{\citenamefont {Georges}\ \emph {et~al.}(2013)\citenamefont {Georges}, \citenamefont {de~Medici},\ and\ \citenamefont {Mravlje}}]{Hunds_metals1}%
  \BibitemOpen
  \bibfield  {author} {\bibinfo {author} {\bibfnamefont {A.}~\bibnamefont {Georges}}, \bibinfo {author} {\bibfnamefont {L.}~\bibnamefont {de~Medici}},\ and\ \bibinfo {author} {\bibfnamefont {J.}~\bibnamefont {Mravlje}},\ }\bibfield  {title} {\bibinfo {title} {{Strong Correlations from Hund’s Coupling}},\ }\href {https://doi.org/https://doi.org/10.1146/annurev-conmatphys-020911-125045} {\bibfield  {journal} {\bibinfo  {journal} {Annu. Rev. Condens. Matter Phys.}\ }\textbf {\bibinfo {volume} {4}},\ \bibinfo {pages} {137} (\bibinfo {year} {2013})}\BibitemShut {NoStop}%
\bibitem [{\citenamefont {Georges}\ and\ \citenamefont {Kotliar}(2024)}]{Hunds_metals2}%
  \BibitemOpen
  \bibfield  {author} {\bibinfo {author} {\bibfnamefont {A.}~\bibnamefont {Georges}}\ and\ \bibinfo {author} {\bibfnamefont {G.}~\bibnamefont {Kotliar}},\ }\bibfield  {title} {\bibinfo {title} {{The Hund-metal path to strong electronic correlations}},\ }\href {https://doi.org/10.1063/pt.wqrz.qpjx} {\bibfield  {journal} {\bibinfo  {journal} {Physics Today}\ }\textbf {\bibinfo {volume} {77}},\ \bibinfo {pages} {46} (\bibinfo {year} {2024})}\BibitemShut {NoStop}%
\bibitem [{\citenamefont {Metzner}\ and\ \citenamefont {Vollhardt}(1989)}]{PhysRevLett.62.324}%
  \BibitemOpen
  \bibfield  {author} {\bibinfo {author} {\bibfnamefont {W.}~\bibnamefont {Metzner}}\ and\ \bibinfo {author} {\bibfnamefont {D.}~\bibnamefont {Vollhardt}},\ }\bibfield  {title} {\bibinfo {title} {{Correlated Lattice Fermions in $d=\ensuremath{\infty}$ Dimensions}},\ }\href {https://doi.org/10.1103/PhysRevLett.62.324} {\bibfield  {journal} {\bibinfo  {journal} {Phys. Rev. Lett.}\ }\textbf {\bibinfo {volume} {62}},\ \bibinfo {pages} {324} (\bibinfo {year} {1989})}\BibitemShut {NoStop}%
\bibitem [{\citenamefont {Müller-Hartmann}(1989)}]{muller-hartmann1989}%
  \BibitemOpen
  \bibfield  {author} {\bibinfo {author} {\bibfnamefont {E.}~\bibnamefont {Müller-Hartmann}},\ }\bibfield  {title} {\bibinfo {title} {{Correlated Fermions on a Lattice in High Dimensions}},\ }\href {https://doi.org/10.1007/BF01311397} {\bibfield  {journal} {\bibinfo  {journal} {Z. Phys.}\ }\textbf {\bibinfo {volume} {74}},\ \bibinfo {pages} {507} (\bibinfo {year} {1989})}\BibitemShut {NoStop}%
\bibitem [{\citenamefont {Chatzieleftheriou}\ \emph {et~al.}(2024)\citenamefont {Chatzieleftheriou}, \citenamefont {Biermann},\ and\ \citenamefont {Stepanov}}]{PhysRevLett.132.236504}%
  \BibitemOpen
  \bibfield  {author} {\bibinfo {author} {\bibfnamefont {M.}~\bibnamefont {Chatzieleftheriou}}, \bibinfo {author} {\bibfnamefont {S.}~\bibnamefont {Biermann}},\ and\ \bibinfo {author} {\bibfnamefont {E.~A.}\ \bibnamefont {Stepanov}},\ }\bibfield  {title} {\bibinfo {title} {{Local and Nonlocal Electronic Correlations at the Metal-Insulator Transition in the Two-Dimensional Hubbard Model}},\ }\href {https://doi.org/10.1103/PhysRevLett.132.236504} {\bibfield  {journal} {\bibinfo  {journal} {Phys. Rev. Lett.}\ }\textbf {\bibinfo {volume} {132}},\ \bibinfo {pages} {236504} (\bibinfo {year} {2024})}\BibitemShut {NoStop}%
\bibitem [{\citenamefont {Vandelli}\ \emph {et~al.}(2022)\citenamefont {Vandelli}, \citenamefont {Kaufmann}, \citenamefont {El-Nabulsi}, \citenamefont {Harkov}, \citenamefont {Lichtenstein},\ and\ \citenamefont {Stepanov}}]{10.21468/SciPostPhys.13.2.036}%
  \BibitemOpen
  \bibfield  {author} {\bibinfo {author} {\bibfnamefont {M.}~\bibnamefont {Vandelli}}, \bibinfo {author} {\bibfnamefont {J.}~\bibnamefont {Kaufmann}}, \bibinfo {author} {\bibfnamefont {M.}~\bibnamefont {El-Nabulsi}}, \bibinfo {author} {\bibfnamefont {V.}~\bibnamefont {Harkov}}, \bibinfo {author} {\bibfnamefont {A.~I.}\ \bibnamefont {Lichtenstein}},\ and\ \bibinfo {author} {\bibfnamefont {E.~A.}\ \bibnamefont {Stepanov}},\ }\bibfield  {title} {\bibinfo {title} {{{Multi-band D-TRILEX approach to materials with strong electronic correlations}}},\ }\href {https://doi.org/10.21468/SciPostPhys.13.2.036} {\bibfield  {journal} {\bibinfo  {journal} {SciPost Phys.}\ }\textbf {\bibinfo {volume} {13}},\ \bibinfo {pages} {036} (\bibinfo {year} {2022})}\BibitemShut {NoStop}%
\bibitem [{\citenamefont {Valli}\ \emph {et~al.}(2015)\citenamefont {Valli}, \citenamefont {Sch\"afer}, \citenamefont {Thunstr\"om}, \citenamefont {Rohringer}, \citenamefont {Andergassen}, \citenamefont {Sangiovanni}, \citenamefont {Held},\ and\ \citenamefont {Toschi}}]{PhysRevB.91.115115}%
  \BibitemOpen
  \bibfield  {author} {\bibinfo {author} {\bibfnamefont {A.}~\bibnamefont {Valli}}, \bibinfo {author} {\bibfnamefont {T.}~\bibnamefont {Sch\"afer}}, \bibinfo {author} {\bibfnamefont {P.}~\bibnamefont {Thunstr\"om}}, \bibinfo {author} {\bibfnamefont {G.}~\bibnamefont {Rohringer}}, \bibinfo {author} {\bibfnamefont {S.}~\bibnamefont {Andergassen}}, \bibinfo {author} {\bibfnamefont {G.}~\bibnamefont {Sangiovanni}}, \bibinfo {author} {\bibfnamefont {K.}~\bibnamefont {Held}},\ and\ \bibinfo {author} {\bibfnamefont {A.}~\bibnamefont {Toschi}},\ }\bibfield  {title} {\bibinfo {title} {{Dynamical vertex approximation in its parquet implementation: Application to Hubbard nanorings}},\ }\href {https://doi.org/10.1103/PhysRevB.91.115115} {\bibfield  {journal} {\bibinfo  {journal} {Phys. Rev. B}\ }\textbf {\bibinfo {volume} {91}},\ \bibinfo {pages} {115115} (\bibinfo {year} {2015})}\BibitemShut {NoStop}%
\bibitem [{\citenamefont {Hettler}\ \emph {et~al.}(1998)\citenamefont {Hettler}, \citenamefont {Tahvildar-Zadeh}, \citenamefont {Jarrell}, \citenamefont {Pruschke},\ and\ \citenamefont {Krishnamurthy}}]{PhysRevB.58.R7475}%
  \BibitemOpen
  \bibfield  {author} {\bibinfo {author} {\bibfnamefont {M.~H.}\ \bibnamefont {Hettler}}, \bibinfo {author} {\bibfnamefont {A.~N.}\ \bibnamefont {Tahvildar-Zadeh}}, \bibinfo {author} {\bibfnamefont {M.}~\bibnamefont {Jarrell}}, \bibinfo {author} {\bibfnamefont {T.}~\bibnamefont {Pruschke}},\ and\ \bibinfo {author} {\bibfnamefont {H.~R.}\ \bibnamefont {Krishnamurthy}},\ }\bibfield  {title} {\bibinfo {title} {{Nonlocal dynamical correlations of strongly interacting electron systems}},\ }\href {https://doi.org/10.1103/PhysRevB.58.R7475} {\bibfield  {journal} {\bibinfo  {journal} {Phys. Rev. B}\ }\textbf {\bibinfo {volume} {58}},\ \bibinfo {pages} {R7475} (\bibinfo {year} {1998})}\BibitemShut {NoStop}%
\bibitem [{\citenamefont {Lichtenstein}\ and\ \citenamefont {Katsnelson}(2000)}]{PhysRevB.62.R9283}%
  \BibitemOpen
  \bibfield  {author} {\bibinfo {author} {\bibfnamefont {A.~I.}\ \bibnamefont {Lichtenstein}}\ and\ \bibinfo {author} {\bibfnamefont {M.~I.}\ \bibnamefont {Katsnelson}},\ }\bibfield  {title} {\bibinfo {title} {{Antiferromagnetism and d-wave superconductivity in cuprates: A cluster dynamical mean-field theory}},\ }\href {https://doi.org/10.1103/PhysRevB.62.R9283} {\bibfield  {journal} {\bibinfo  {journal} {Phys. Rev. B}\ }\textbf {\bibinfo {volume} {62}},\ \bibinfo {pages} {R9283} (\bibinfo {year} {2000})}\BibitemShut {NoStop}%
\bibitem [{\citenamefont {Kotliar}\ \emph {et~al.}(2001)\citenamefont {Kotliar}, \citenamefont {Savrasov}, \citenamefont {P\'alsson},\ and\ \citenamefont {Biroli}}]{PhysRevLett.87.186401}%
  \BibitemOpen
  \bibfield  {author} {\bibinfo {author} {\bibfnamefont {G.}~\bibnamefont {Kotliar}}, \bibinfo {author} {\bibfnamefont {S.~Y.}\ \bibnamefont {Savrasov}}, \bibinfo {author} {\bibfnamefont {G.}~\bibnamefont {P\'alsson}},\ and\ \bibinfo {author} {\bibfnamefont {G.}~\bibnamefont {Biroli}},\ }\bibfield  {title} {\bibinfo {title} {{Cellular Dynamical Mean Field Approach to Strongly Correlated Systems}},\ }\href {https://doi.org/10.1103/PhysRevLett.87.186401} {\bibfield  {journal} {\bibinfo  {journal} {Phys. Rev. Lett.}\ }\textbf {\bibinfo {volume} {87}},\ \bibinfo {pages} {186401} (\bibinfo {year} {2001})}\BibitemShut {NoStop}%
\bibitem [{\citenamefont {Maier}\ \emph {et~al.}(2005)\citenamefont {Maier}, \citenamefont {Jarrell}, \citenamefont {Pruschke},\ and\ \citenamefont {Hettler}}]{RevModPhys.77.1027}%
  \BibitemOpen
  \bibfield  {author} {\bibinfo {author} {\bibfnamefont {T.}~\bibnamefont {Maier}}, \bibinfo {author} {\bibfnamefont {M.}~\bibnamefont {Jarrell}}, \bibinfo {author} {\bibfnamefont {T.}~\bibnamefont {Pruschke}},\ and\ \bibinfo {author} {\bibfnamefont {M.~H.}\ \bibnamefont {Hettler}},\ }\bibfield  {title} {\bibinfo {title} {{Quantum cluster theories}},\ }\href {https://doi.org/10.1103/RevModPhys.77.1027} {\bibfield  {journal} {\bibinfo  {journal} {Rev. Mod. Phys.}\ }\textbf {\bibinfo {volume} {77}},\ \bibinfo {pages} {1027} (\bibinfo {year} {2005})}\BibitemShut {NoStop}%
\bibitem [{\citenamefont {Tremblay}\ \emph {et~al.}(2006)\citenamefont {Tremblay}, \citenamefont {Kyung},\ and\ \citenamefont {S\'en\'echal}}]{doi:10.1063/1.2199446}%
  \BibitemOpen
  \bibfield  {author} {\bibinfo {author} {\bibfnamefont {A.-M.~S.}\ \bibnamefont {Tremblay}}, \bibinfo {author} {\bibfnamefont {B.}~\bibnamefont {Kyung}},\ and\ \bibinfo {author} {\bibfnamefont {D.}~\bibnamefont {S\'en\'echal}},\ }\bibfield  {title} {\bibinfo {title} {{Pseudogap and high-temperature superconductivity from weak to strong coupling. Towards a quantitative theory (Review Article)}},\ }\href {https://doi.org/10.1063/1.2199446} {\bibfield  {journal} {\bibinfo  {journal} {Low Temp. Phys.}\ }\textbf {\bibinfo {volume} {32}},\ \bibinfo {pages} {424} (\bibinfo {year} {2006})}\BibitemShut {NoStop}%
\bibitem [{\citenamefont {Kotliar}\ \emph {et~al.}(2006)\citenamefont {Kotliar}, \citenamefont {Savrasov}, \citenamefont {Haule}, \citenamefont {Oudovenko}, \citenamefont {Parcollet},\ and\ \citenamefont {Marianetti}}]{RevModPhys.78.865}%
  \BibitemOpen
  \bibfield  {author} {\bibinfo {author} {\bibfnamefont {G.}~\bibnamefont {Kotliar}}, \bibinfo {author} {\bibfnamefont {S.~Y.}\ \bibnamefont {Savrasov}}, \bibinfo {author} {\bibfnamefont {K.}~\bibnamefont {Haule}}, \bibinfo {author} {\bibfnamefont {V.~S.}\ \bibnamefont {Oudovenko}}, \bibinfo {author} {\bibfnamefont {O.}~\bibnamefont {Parcollet}},\ and\ \bibinfo {author} {\bibfnamefont {C.~A.}\ \bibnamefont {Marianetti}},\ }\bibfield  {title} {\bibinfo {title} {{Electronic structure calculations with dynamical mean-field theory}},\ }\href {https://doi.org/10.1103/RevModPhys.78.865} {\bibfield  {journal} {\bibinfo  {journal} {Rev. Mod. Phys.}\ }\textbf {\bibinfo {volume} {78}},\ \bibinfo {pages} {865} (\bibinfo {year} {2006})}\BibitemShut {NoStop}%
\bibitem [{\citenamefont {Park}\ \emph {et~al.}(2008)\citenamefont {Park}, \citenamefont {Haule},\ and\ \citenamefont {Kotliar}}]{PhysRevLett.101.186403}%
  \BibitemOpen
  \bibfield  {author} {\bibinfo {author} {\bibfnamefont {H.}~\bibnamefont {Park}}, \bibinfo {author} {\bibfnamefont {K.}~\bibnamefont {Haule}},\ and\ \bibinfo {author} {\bibfnamefont {G.}~\bibnamefont {Kotliar}},\ }\bibfield  {title} {\bibinfo {title} {{Cluster Dynamical Mean Field Theory of the Mott Transition}},\ }\href {https://doi.org/10.1103/PhysRevLett.101.186403} {\bibfield  {journal} {\bibinfo  {journal} {Phys. Rev. Lett.}\ }\textbf {\bibinfo {volume} {101}},\ \bibinfo {pages} {186403} (\bibinfo {year} {2008})}\BibitemShut {NoStop}%
\bibitem [{\citenamefont {Haule}\ and\ \citenamefont {Kotliar}(2007)}]{PhysRevB.76.104509}%
  \BibitemOpen
  \bibfield  {author} {\bibinfo {author} {\bibfnamefont {K.}~\bibnamefont {Haule}}\ and\ \bibinfo {author} {\bibfnamefont {G.}~\bibnamefont {Kotliar}},\ }\bibfield  {title} {\bibinfo {title} {{Strongly correlated superconductivity: A plaquette dynamical mean-field theory study}},\ }\href {https://doi.org/10.1103/PhysRevB.76.104509} {\bibfield  {journal} {\bibinfo  {journal} {Phys. Rev. B}\ }\textbf {\bibinfo {volume} {76}},\ \bibinfo {pages} {104509} (\bibinfo {year} {2007})}\BibitemShut {NoStop}%
\bibitem [{\citenamefont {Civelli}\ \emph {et~al.}(2008)\citenamefont {Civelli}, \citenamefont {Capone}, \citenamefont {Georges}, \citenamefont {Haule}, \citenamefont {Parcollet}, \citenamefont {Stanescu},\ and\ \citenamefont {Kotliar}}]{PhysRevLett.100.046402}%
  \BibitemOpen
  \bibfield  {author} {\bibinfo {author} {\bibfnamefont {M.}~\bibnamefont {Civelli}}, \bibinfo {author} {\bibfnamefont {M.}~\bibnamefont {Capone}}, \bibinfo {author} {\bibfnamefont {A.}~\bibnamefont {Georges}}, \bibinfo {author} {\bibfnamefont {K.}~\bibnamefont {Haule}}, \bibinfo {author} {\bibfnamefont {O.}~\bibnamefont {Parcollet}}, \bibinfo {author} {\bibfnamefont {T.~D.}\ \bibnamefont {Stanescu}},\ and\ \bibinfo {author} {\bibfnamefont {G.}~\bibnamefont {Kotliar}},\ }\bibfield  {title} {\bibinfo {title} {{Nodal-Antinodal Dichotomy and the Two Gaps of a Superconducting Doped Mott Insulator}},\ }\href {https://doi.org/10.1103/PhysRevLett.100.046402} {\bibfield  {journal} {\bibinfo  {journal} {Phys. Rev. Lett.}\ }\textbf {\bibinfo {volume} {100}},\ \bibinfo {pages} {046402} (\bibinfo {year} {2008})}\BibitemShut {NoStop}%
\bibitem [{\citenamefont {Harland}\ \emph {et~al.}(2016)\citenamefont {Harland}, \citenamefont {Katsnelson},\ and\ \citenamefont {Lichtenstein}}]{PhysRevB.94.125133}%
  \BibitemOpen
  \bibfield  {author} {\bibinfo {author} {\bibfnamefont {M.}~\bibnamefont {Harland}}, \bibinfo {author} {\bibfnamefont {M.~I.}\ \bibnamefont {Katsnelson}},\ and\ \bibinfo {author} {\bibfnamefont {A.~I.}\ \bibnamefont {Lichtenstein}},\ }\bibfield  {title} {\bibinfo {title} {{Plaquette valence bond theory of high-temperature superconductivity}},\ }\href {https://doi.org/10.1103/PhysRevB.94.125133} {\bibfield  {journal} {\bibinfo  {journal} {Phys. Rev. B}\ }\textbf {\bibinfo {volume} {94}},\ \bibinfo {pages} {125133} (\bibinfo {year} {2016})}\BibitemShut {NoStop}%
\bibitem [{\citenamefont {Harland}\ \emph {et~al.}(2020)\citenamefont {Harland}, \citenamefont {Brener}, \citenamefont {Katsnelson},\ and\ \citenamefont {Lichtenstein}}]{PhysRevB.101.045119}%
  \BibitemOpen
  \bibfield  {author} {\bibinfo {author} {\bibfnamefont {M.}~\bibnamefont {Harland}}, \bibinfo {author} {\bibfnamefont {S.}~\bibnamefont {Brener}}, \bibinfo {author} {\bibfnamefont {M.~I.}\ \bibnamefont {Katsnelson}},\ and\ \bibinfo {author} {\bibfnamefont {A.~I.}\ \bibnamefont {Lichtenstein}},\ }\bibfield  {title} {\bibinfo {title} {{Exactly solvable model of strongly correlated $d$-wave superconductivity}},\ }\href {https://doi.org/10.1103/PhysRevB.101.045119} {\bibfield  {journal} {\bibinfo  {journal} {Phys. Rev. B}\ }\textbf {\bibinfo {volume} {101}},\ \bibinfo {pages} {045119} (\bibinfo {year} {2020})}\BibitemShut {NoStop}%
\bibitem [{\citenamefont {Danilov}\ \emph {et~al.}(2022)\citenamefont {Danilov}, \citenamefont {van Loon}, \citenamefont {Brener}, \citenamefont {Iskakov}, \citenamefont {Katsnelson},\ and\ \citenamefont {Lichtenstein}}]{danilov2022}%
  \BibitemOpen
  \bibfield  {author} {\bibinfo {author} {\bibfnamefont {M.}~\bibnamefont {Danilov}}, \bibinfo {author} {\bibfnamefont {E.~G. C.~P.}\ \bibnamefont {van Loon}}, \bibinfo {author} {\bibfnamefont {S.}~\bibnamefont {Brener}}, \bibinfo {author} {\bibfnamefont {S.}~\bibnamefont {Iskakov}}, \bibinfo {author} {\bibfnamefont {M.~I.}\ \bibnamefont {Katsnelson}},\ and\ \bibinfo {author} {\bibfnamefont {A.~I.}\ \bibnamefont {Lichtenstein}},\ }\bibfield  {title} {\bibinfo {title} {{Degenerate plaquette physics as key ingredient of high-temperature superconductivity in cuprates}},\ }\href {https://doi.org/10.1038/s41535-022-00454-6} {\bibfield  {journal} {\bibinfo  {journal} {npj Quantum Mater.}\ }\textbf {\bibinfo {volume} {7}},\ \bibinfo {pages} {50} (\bibinfo {year} {2022})}\BibitemShut {NoStop}%
\bibitem [{\citenamefont {Dong}\ \emph {et~al.}(2022)\citenamefont {Dong}, \citenamefont {Gull},\ and\ \citenamefont {Millis}}]{dong2022}%
  \BibitemOpen
  \bibfield  {author} {\bibinfo {author} {\bibfnamefont {X.}~\bibnamefont {Dong}}, \bibinfo {author} {\bibfnamefont {E.}~\bibnamefont {Gull}},\ and\ \bibinfo {author} {\bibfnamefont {A.~J.}\ \bibnamefont {Millis}},\ }\bibfield  {title} {\bibinfo {title} {{Quantifying the role of antiferromagnetic fluctuations in the superconductivity of the doped Hubbard model}},\ }\href {https://doi.org/10.1038/s41567-022-01710-z} {\bibfield  {journal} {\bibinfo  {journal} {Nature Physics}\ }\textbf {\bibinfo {volume} {18}},\ \bibinfo {pages} {1293} (\bibinfo {year} {2022})}\BibitemShut {NoStop}%
\bibitem [{\citenamefont {Wu}\ \emph {et~al.}(2018)\citenamefont {Wu}, \citenamefont {Scheurer}, \citenamefont {Chatterjee}, \citenamefont {Sachdev}, \citenamefont {Georges},\ and\ \citenamefont {Ferrero}}]{PhysRevX.8.021048}%
  \BibitemOpen
  \bibfield  {author} {\bibinfo {author} {\bibfnamefont {W.}~\bibnamefont {Wu}}, \bibinfo {author} {\bibfnamefont {M.~S.}\ \bibnamefont {Scheurer}}, \bibinfo {author} {\bibfnamefont {S.}~\bibnamefont {Chatterjee}}, \bibinfo {author} {\bibfnamefont {S.}~\bibnamefont {Sachdev}}, \bibinfo {author} {\bibfnamefont {A.}~\bibnamefont {Georges}},\ and\ \bibinfo {author} {\bibfnamefont {M.}~\bibnamefont {Ferrero}},\ }\bibfield  {title} {\bibinfo {title} {{Pseudogap and Fermi-Surface Topology in the Two-Dimensional Hubbard Model}},\ }\href {https://doi.org/10.1103/PhysRevX.8.021048} {\bibfield  {journal} {\bibinfo  {journal} {Phys. Rev. X}\ }\textbf {\bibinfo {volume} {8}},\ \bibinfo {pages} {021048} (\bibinfo {year} {2018})}\BibitemShut {NoStop}%
\bibitem [{\citenamefont {Stepanov}\ \emph {et~al.}(2025)\citenamefont {Stepanov}, \citenamefont {Iskakov}, \citenamefont {Katsnelson},\ and\ \citenamefont {Lichtenstein}}]{Cuprates}%
  \BibitemOpen
  \bibfield  {author} {\bibinfo {author} {\bibfnamefont {E.~A.}\ \bibnamefont {Stepanov}}, \bibinfo {author} {\bibfnamefont {S.}~\bibnamefont {Iskakov}}, \bibinfo {author} {\bibfnamefont {M.~I.}\ \bibnamefont {Katsnelson}},\ and\ \bibinfo {author} {\bibfnamefont {A.~I.}\ \bibnamefont {Lichtenstein}},\ }\href {https://doi.org/10.48550/arXiv.2502.08635} {\bibinfo {title} {{Superconductivity of Bad Fermions: Origin of Two Gaps in HTSC Cuprates}}},\ \bibinfo {howpublished} {Preprint arXiv:2502.08635} (\bibinfo {year} {2025})\BibitemShut {NoStop}%
\bibitem [{\citenamefont {Hirsch}\ and\ \citenamefont {Fye}(1986)}]{PhysRevLett.56.2521}%
  \BibitemOpen
  \bibfield  {author} {\bibinfo {author} {\bibfnamefont {J.~E.}\ \bibnamefont {Hirsch}}\ and\ \bibinfo {author} {\bibfnamefont {R.~M.}\ \bibnamefont {Fye}},\ }\bibfield  {title} {\bibinfo {title} {{Monte Carlo Method for Magnetic Impurities in Metals}},\ }\href {https://doi.org/10.1103/PhysRevLett.56.2521} {\bibfield  {journal} {\bibinfo  {journal} {Phys. Rev. Lett.}\ }\textbf {\bibinfo {volume} {56}},\ \bibinfo {pages} {2521} (\bibinfo {year} {1986})}\BibitemShut {NoStop}%
\bibitem [{\citenamefont {Beard}\ and\ \citenamefont {Wiese}(1996)}]{PhysRevLett.77.5130}%
  \BibitemOpen
  \bibfield  {author} {\bibinfo {author} {\bibfnamefont {B.~B.}\ \bibnamefont {Beard}}\ and\ \bibinfo {author} {\bibfnamefont {U.-J.}\ \bibnamefont {Wiese}},\ }\bibfield  {title} {\bibinfo {title} {{Simulations of Discrete Quantum Systems in Continuous Euclidean Time}},\ }\href {https://doi.org/10.1103/PhysRevLett.77.5130} {\bibfield  {journal} {\bibinfo  {journal} {Phys. Rev. Lett.}\ }\textbf {\bibinfo {volume} {77}},\ \bibinfo {pages} {5130} (\bibinfo {year} {1996})}\BibitemShut {NoStop}%
\bibitem [{\citenamefont {Prokof'ev}\ \emph {et~al.}(1996)\citenamefont {Prokof'ev}, \citenamefont {Svistunov},\ and\ \citenamefont {Tupitsyn}}]{prokofev1996}%
  \BibitemOpen
  \bibfield  {author} {\bibinfo {author} {\bibfnamefont {N.~V.}\ \bibnamefont {Prokof'ev}}, \bibinfo {author} {\bibfnamefont {B.~V.}\ \bibnamefont {Svistunov}},\ and\ \bibinfo {author} {\bibfnamefont {I.~S.}\ \bibnamefont {Tupitsyn}},\ }\bibfield  {title} {\bibinfo {title} {{Exact Quantum {{Monte Carlo}} Process for the Statistics of Discrete Systems}},\ }\href {https://doi.org/10.1134/1.567243} {\bibfield  {journal} {\bibinfo  {journal} {JETP Letters}\ }\textbf {\bibinfo {volume} {64}},\ \bibinfo {pages} {911} (\bibinfo {year} {1996})}\BibitemShut {NoStop}%
\bibitem [{\citenamefont {Rubtsov}\ and\ \citenamefont {Lichtenstein}(2004)}]{rubtsov2004}%
  \BibitemOpen
  \bibfield  {author} {\bibinfo {author} {\bibfnamefont {A.~N.}\ \bibnamefont {Rubtsov}}\ and\ \bibinfo {author} {\bibfnamefont {A.~I.}\ \bibnamefont {Lichtenstein}},\ }\bibfield  {title} {\bibinfo {title} {{Continuous-Time Quantum {{Monte Carlo}} Method for Fermions: {{Beyond}} Auxiliary Field Framework}},\ }\href {https://doi.org/10.1134/1.1800216} {\bibfield  {journal} {\bibinfo  {journal} {JETP Letters}\ }\textbf {\bibinfo {volume} {80}},\ \bibinfo {pages} {61} (\bibinfo {year} {2004})}\BibitemShut {NoStop}%
\bibitem [{\citenamefont {Rubtsov}\ \emph {et~al.}(2005)\citenamefont {Rubtsov}, \citenamefont {Savkin},\ and\ \citenamefont {Lichtenstein}}]{PhysRevB.72.035122}%
  \BibitemOpen
  \bibfield  {author} {\bibinfo {author} {\bibfnamefont {A.~N.}\ \bibnamefont {Rubtsov}}, \bibinfo {author} {\bibfnamefont {V.~V.}\ \bibnamefont {Savkin}},\ and\ \bibinfo {author} {\bibfnamefont {A.~I.}\ \bibnamefont {Lichtenstein}},\ }\bibfield  {title} {\bibinfo {title} {{Continuous-time quantum Monte Carlo method for fermions}},\ }\href {https://doi.org/10.1103/PhysRevB.72.035122} {\bibfield  {journal} {\bibinfo  {journal} {Phys. Rev. B}\ }\textbf {\bibinfo {volume} {72}},\ \bibinfo {pages} {035122} (\bibinfo {year} {2005})}\BibitemShut {NoStop}%
\bibitem [{\citenamefont {Werner}\ \emph {et~al.}(2006)\citenamefont {Werner}, \citenamefont {Comanac}, \citenamefont {de' Medici}, \citenamefont {Troyer},\ and\ \citenamefont {Millis}}]{PhysRevLett.97.076405}%
  \BibitemOpen
  \bibfield  {author} {\bibinfo {author} {\bibfnamefont {P.}~\bibnamefont {Werner}}, \bibinfo {author} {\bibfnamefont {A.}~\bibnamefont {Comanac}}, \bibinfo {author} {\bibfnamefont {L.}~\bibnamefont {de' Medici}}, \bibinfo {author} {\bibfnamefont {M.}~\bibnamefont {Troyer}},\ and\ \bibinfo {author} {\bibfnamefont {A.~J.}\ \bibnamefont {Millis}},\ }\bibfield  {title} {\bibinfo {title} {{Continuous-Time Solver for Quantum Impurity Models}},\ }\href {https://doi.org/10.1103/PhysRevLett.97.076405} {\bibfield  {journal} {\bibinfo  {journal} {Phys. Rev. Lett.}\ }\textbf {\bibinfo {volume} {97}},\ \bibinfo {pages} {076405} (\bibinfo {year} {2006})}\BibitemShut {NoStop}%
\bibitem [{\citenamefont {Gull}\ \emph {et~al.}(2008)\citenamefont {Gull}, \citenamefont {Werner}, \citenamefont {Wang}, \citenamefont {Troyer},\ and\ \citenamefont {Millis}}]{Gull2008}%
  \BibitemOpen
  \bibfield  {author} {\bibinfo {author} {\bibfnamefont {E.}~\bibnamefont {Gull}}, \bibinfo {author} {\bibfnamefont {P.}~\bibnamefont {Werner}}, \bibinfo {author} {\bibfnamefont {X.}~\bibnamefont {Wang}}, \bibinfo {author} {\bibfnamefont {M.}~\bibnamefont {Troyer}},\ and\ \bibinfo {author} {\bibfnamefont {A.~J.}\ \bibnamefont {Millis}},\ }\bibfield  {title} {\bibinfo {title} {{Local Order and the Gapped Phase of the {{Hubbard}} Model: {{A}} Plaquette Dynamical Mean-Field Investigation}},\ }\href {https://doi.org/10.1209/0295-5075/84/37009} {\bibfield  {journal} {\bibinfo  {journal} {EPL}\ }\textbf {\bibinfo {volume} {84}},\ \bibinfo {pages} {37009} (\bibinfo {year} {2008})}\BibitemShut {NoStop}%
\bibitem [{\citenamefont {Gull}\ \emph {et~al.}(2011)\citenamefont {Gull}, \citenamefont {Millis}, \citenamefont {Lichtenstein}, \citenamefont {Rubtsov}, \citenamefont {Troyer},\ and\ \citenamefont {Werner}}]{RevModPhys.83.349}%
  \BibitemOpen
  \bibfield  {author} {\bibinfo {author} {\bibfnamefont {E.}~\bibnamefont {Gull}}, \bibinfo {author} {\bibfnamefont {A.~J.}\ \bibnamefont {Millis}}, \bibinfo {author} {\bibfnamefont {A.~I.}\ \bibnamefont {Lichtenstein}}, \bibinfo {author} {\bibfnamefont {A.~N.}\ \bibnamefont {Rubtsov}}, \bibinfo {author} {\bibfnamefont {M.}~\bibnamefont {Troyer}},\ and\ \bibinfo {author} {\bibfnamefont {P.}~\bibnamefont {Werner}},\ }\bibfield  {title} {\bibinfo {title} {{Continuous-time Monte Carlo methods for quantum impurity models}},\ }\href {https://doi.org/10.1103/RevModPhys.83.349} {\bibfield  {journal} {\bibinfo  {journal} {Rev. Mod. Phys.}\ }\textbf {\bibinfo {volume} {83}},\ \bibinfo {pages} {349} (\bibinfo {year} {2011})}\BibitemShut {NoStop}%
\bibitem [{\citenamefont {Loh}\ \emph {et~al.}(1990)\citenamefont {Loh}, \citenamefont {Gubernatis}, \citenamefont {Scalettar}, \citenamefont {White}, \citenamefont {Scalapino},\ and\ \citenamefont {Sugar}}]{PhysRevB.41.9301}%
  \BibitemOpen
  \bibfield  {author} {\bibinfo {author} {\bibfnamefont {E.~Y.}\ \bibnamefont {Loh}}, \bibinfo {author} {\bibfnamefont {J.~E.}\ \bibnamefont {Gubernatis}}, \bibinfo {author} {\bibfnamefont {R.~T.}\ \bibnamefont {Scalettar}}, \bibinfo {author} {\bibfnamefont {S.~R.}\ \bibnamefont {White}}, \bibinfo {author} {\bibfnamefont {D.~J.}\ \bibnamefont {Scalapino}},\ and\ \bibinfo {author} {\bibfnamefont {R.~L.}\ \bibnamefont {Sugar}},\ }\bibfield  {title} {\bibinfo {title} {{Sign problem in the numerical simulation of many-electron systems}},\ }\href {https://doi.org/10.1103/PhysRevB.41.9301} {\bibfield  {journal} {\bibinfo  {journal} {Phys. Rev. B}\ }\textbf {\bibinfo {volume} {41}},\ \bibinfo {pages} {9301} (\bibinfo {year} {1990})}\BibitemShut {NoStop}%
\bibitem [{\citenamefont {Gorelov}\ \emph {et~al.}(2009)\citenamefont {Gorelov}, \citenamefont {Wehling}, \citenamefont {Rubtsov}, \citenamefont {Katsnelson},\ and\ \citenamefont {Lichtenstein}}]{PhysRevB.80.155132}%
  \BibitemOpen
  \bibfield  {author} {\bibinfo {author} {\bibfnamefont {E.}~\bibnamefont {Gorelov}}, \bibinfo {author} {\bibfnamefont {T.~O.}\ \bibnamefont {Wehling}}, \bibinfo {author} {\bibfnamefont {A.~N.}\ \bibnamefont {Rubtsov}}, \bibinfo {author} {\bibfnamefont {M.~I.}\ \bibnamefont {Katsnelson}},\ and\ \bibinfo {author} {\bibfnamefont {A.~I.}\ \bibnamefont {Lichtenstein}},\ }\bibfield  {title} {\bibinfo {title} {{Relevance of the complete Coulomb interaction matrix for the Kondo problem: Co impurities in Cu hosts}},\ }\href {https://doi.org/10.1103/PhysRevB.80.155132} {\bibfield  {journal} {\bibinfo  {journal} {Phys. Rev. B}\ }\textbf {\bibinfo {volume} {80}},\ \bibinfo {pages} {155132} (\bibinfo {year} {2009})}\BibitemShut {NoStop}%
\bibitem [{\citenamefont {Wang}\ and\ \citenamefont {Millis}(2010)}]{PhysRevB.81.045106}%
  \BibitemOpen
  \bibfield  {author} {\bibinfo {author} {\bibfnamefont {X.}~\bibnamefont {Wang}}\ and\ \bibinfo {author} {\bibfnamefont {A.~J.}\ \bibnamefont {Millis}},\ }\bibfield  {title} {\bibinfo {title} {{Quantum criticality and non-Fermi-liquid behavior in a two-level two-lead quantum dot}},\ }\href {https://doi.org/10.1103/PhysRevB.81.045106} {\bibfield  {journal} {\bibinfo  {journal} {Phys. Rev. B}\ }\textbf {\bibinfo {volume} {81}},\ \bibinfo {pages} {045106} (\bibinfo {year} {2010})}\BibitemShut {NoStop}%
\bibitem [{\citenamefont {Werner}\ and\ \citenamefont {Millis}(2006)}]{PhysRevB.74.155107}%
  \BibitemOpen
  \bibfield  {author} {\bibinfo {author} {\bibfnamefont {P.}~\bibnamefont {Werner}}\ and\ \bibinfo {author} {\bibfnamefont {A.~J.}\ \bibnamefont {Millis}},\ }\bibfield  {title} {\bibinfo {title} {{Hybridization expansion impurity solver: General formulation and application to Kondo lattice and two-orbital models}},\ }\href {https://doi.org/10.1103/PhysRevB.74.155107} {\bibfield  {journal} {\bibinfo  {journal} {Phys. Rev. B}\ }\textbf {\bibinfo {volume} {74}},\ \bibinfo {pages} {155107} (\bibinfo {year} {2006})}\BibitemShut {NoStop}%
\bibitem [{\citenamefont {Civelli}\ \emph {et~al.}(2005)\citenamefont {Civelli}, \citenamefont {Capone}, \citenamefont {Kancharla}, \citenamefont {Parcollet},\ and\ \citenamefont {Kotliar}}]{PhysRevLett.95.106402}%
  \BibitemOpen
  \bibfield  {author} {\bibinfo {author} {\bibfnamefont {M.}~\bibnamefont {Civelli}}, \bibinfo {author} {\bibfnamefont {M.}~\bibnamefont {Capone}}, \bibinfo {author} {\bibfnamefont {S.~S.}\ \bibnamefont {Kancharla}}, \bibinfo {author} {\bibfnamefont {O.}~\bibnamefont {Parcollet}},\ and\ \bibinfo {author} {\bibfnamefont {G.}~\bibnamefont {Kotliar}},\ }\bibfield  {title} {\bibinfo {title} {{Dynamical Breakup of the Fermi Surface in a Doped Mott Insulator}},\ }\href {https://doi.org/10.1103/PhysRevLett.95.106402} {\bibfield  {journal} {\bibinfo  {journal} {Phys. Rev. Lett.}\ }\textbf {\bibinfo {volume} {95}},\ \bibinfo {pages} {106402} (\bibinfo {year} {2005})}\BibitemShut {NoStop}%
\bibitem [{\citenamefont {Stanescu}\ and\ \citenamefont {Kotliar}(2006)}]{PhysRevB.74.125110}%
  \BibitemOpen
  \bibfield  {author} {\bibinfo {author} {\bibfnamefont {T.~D.}\ \bibnamefont {Stanescu}}\ and\ \bibinfo {author} {\bibfnamefont {G.}~\bibnamefont {Kotliar}},\ }\bibfield  {title} {\bibinfo {title} {{Fermi arcs and hidden zeros of the Green function in the pseudogap state}},\ }\href {https://doi.org/10.1103/PhysRevB.74.125110} {\bibfield  {journal} {\bibinfo  {journal} {Phys. Rev. B}\ }\textbf {\bibinfo {volume} {74}},\ \bibinfo {pages} {125110} (\bibinfo {year} {2006})}\BibitemShut {NoStop}%
\bibitem [{\citenamefont {Sakai}\ \emph {et~al.}(2012)\citenamefont {Sakai}, \citenamefont {Sangiovanni}, \citenamefont {Civelli}, \citenamefont {Motome}, \citenamefont {Held},\ and\ \citenamefont {Imada}}]{PhysRevB.85.035102}%
  \BibitemOpen
  \bibfield  {author} {\bibinfo {author} {\bibfnamefont {S.}~\bibnamefont {Sakai}}, \bibinfo {author} {\bibfnamefont {G.}~\bibnamefont {Sangiovanni}}, \bibinfo {author} {\bibfnamefont {M.}~\bibnamefont {Civelli}}, \bibinfo {author} {\bibfnamefont {Y.}~\bibnamefont {Motome}}, \bibinfo {author} {\bibfnamefont {K.}~\bibnamefont {Held}},\ and\ \bibinfo {author} {\bibfnamefont {M.}~\bibnamefont {Imada}},\ }\bibfield  {title} {\bibinfo {title} {{Cluster-size dependence in cellular dynamical mean-field theory}},\ }\href {https://doi.org/10.1103/PhysRevB.85.035102} {\bibfield  {journal} {\bibinfo  {journal} {Phys. Rev. B}\ }\textbf {\bibinfo {volume} {85}},\ \bibinfo {pages} {035102} (\bibinfo {year} {2012})}\BibitemShut {NoStop}%
\bibitem [{\citenamefont {Katanin}\ \emph {et~al.}(2009)\citenamefont {Katanin}, \citenamefont {Toschi},\ and\ \citenamefont {Held}}]{PhysRevB.80.075104}%
  \BibitemOpen
  \bibfield  {author} {\bibinfo {author} {\bibfnamefont {A.~A.}\ \bibnamefont {Katanin}}, \bibinfo {author} {\bibfnamefont {A.}~\bibnamefont {Toschi}},\ and\ \bibinfo {author} {\bibfnamefont {K.}~\bibnamefont {Held}},\ }\bibfield  {title} {\bibinfo {title} {{Comparing pertinent effects of antiferromagnetic fluctuations in the two- and three-dimensional Hubbard model}},\ }\href {https://doi.org/10.1103/PhysRevB.80.075104} {\bibfield  {journal} {\bibinfo  {journal} {Phys. Rev. B}\ }\textbf {\bibinfo {volume} {80}},\ \bibinfo {pages} {075104} (\bibinfo {year} {2009})}\BibitemShut {NoStop}%
\bibitem [{\citenamefont {Rohringer}\ \emph {et~al.}(2018)\citenamefont {Rohringer}, \citenamefont {Hafermann}, \citenamefont {Toschi}, \citenamefont {Katanin}, \citenamefont {Antipov}, \citenamefont {Katsnelson}, \citenamefont {Lichtenstein}, \citenamefont {Rubtsov},\ and\ \citenamefont {Held}}]{RevModPhys.90.025003}%
  \BibitemOpen
  \bibfield  {author} {\bibinfo {author} {\bibfnamefont {G.}~\bibnamefont {Rohringer}}, \bibinfo {author} {\bibfnamefont {H.}~\bibnamefont {Hafermann}}, \bibinfo {author} {\bibfnamefont {A.}~\bibnamefont {Toschi}}, \bibinfo {author} {\bibfnamefont {A.~A.}\ \bibnamefont {Katanin}}, \bibinfo {author} {\bibfnamefont {A.~E.}\ \bibnamefont {Antipov}}, \bibinfo {author} {\bibfnamefont {M.~I.}\ \bibnamefont {Katsnelson}}, \bibinfo {author} {\bibfnamefont {A.~I.}\ \bibnamefont {Lichtenstein}}, \bibinfo {author} {\bibfnamefont {A.~N.}\ \bibnamefont {Rubtsov}},\ and\ \bibinfo {author} {\bibfnamefont {K.}~\bibnamefont {Held}},\ }\bibfield  {title} {\bibinfo {title} {{Diagrammatic routes to nonlocal correlations beyond dynamical mean field theory}},\ }\href {https://doi.org/10.1103/RevModPhys.90.025003} {\bibfield  {journal} {\bibinfo  {journal} {Rev. Mod. Phys.}\ }\textbf {\bibinfo {volume} {90}},\ \bibinfo {pages} {025003} (\bibinfo {year} {2018})}\BibitemShut {NoStop}%
\bibitem [{\citenamefont {Lyakhova}\ \emph {et~al.}(2023)\citenamefont {Lyakhova}, \citenamefont {Astretsov},\ and\ \citenamefont {Rubtsov}}]{Lyakhova_review}%
  \BibitemOpen
  \bibfield  {author} {\bibinfo {author} {\bibfnamefont {Y.~S.}\ \bibnamefont {Lyakhova}}, \bibinfo {author} {\bibfnamefont {G.~V.}\ \bibnamefont {Astretsov}},\ and\ \bibinfo {author} {\bibfnamefont {A.~N.}\ \bibnamefont {Rubtsov}},\ }\bibfield  {title} {\bibinfo {title} {{The mean-field concept and post-DMFT methods in the contemporary theory of correlated systems}},\ }\href {https://doi.org/10.3367/UFNe.2022.09.039231} {\bibfield  {journal} {\bibinfo  {journal} {PHYS-USP+}\ }\textbf {\bibinfo {volume} {193}},\ \bibinfo {pages} {825–844} (\bibinfo {year} {2023})},\ \bibinfo {note} {[Phys. Usp. 66 775–793 (2023)]}\BibitemShut {NoStop}%
\bibitem [{\citenamefont {Iskakov}\ \emph {et~al.}(2016)\citenamefont {Iskakov}, \citenamefont {Antipov},\ and\ \citenamefont {Gull}}]{PhysRevB.94.035102}%
  \BibitemOpen
  \bibfield  {author} {\bibinfo {author} {\bibfnamefont {S.}~\bibnamefont {Iskakov}}, \bibinfo {author} {\bibfnamefont {A.~E.}\ \bibnamefont {Antipov}},\ and\ \bibinfo {author} {\bibfnamefont {E.}~\bibnamefont {Gull}},\ }\bibfield  {title} {\bibinfo {title} {{Diagrammatic Monte Carlo for dual fermions}},\ }\href {https://doi.org/10.1103/PhysRevB.94.035102} {\bibfield  {journal} {\bibinfo  {journal} {Phys. Rev. B}\ }\textbf {\bibinfo {volume} {94}},\ \bibinfo {pages} {035102} (\bibinfo {year} {2016})}\BibitemShut {NoStop}%
\bibitem [{\citenamefont {Gukelberger}\ \emph {et~al.}(2017)\citenamefont {Gukelberger}, \citenamefont {Kozik},\ and\ \citenamefont {Hafermann}}]{PhysRevB.96.035152}%
  \BibitemOpen
  \bibfield  {author} {\bibinfo {author} {\bibfnamefont {J.}~\bibnamefont {Gukelberger}}, \bibinfo {author} {\bibfnamefont {E.}~\bibnamefont {Kozik}},\ and\ \bibinfo {author} {\bibfnamefont {H.}~\bibnamefont {Hafermann}},\ }\bibfield  {title} {\bibinfo {title} {{Diagrammatic Monte Carlo approach for diagrammatic extensions of dynamical mean-field theory: Convergence analysis of the dual fermion technique}},\ }\href {https://doi.org/10.1103/PhysRevB.96.035152} {\bibfield  {journal} {\bibinfo  {journal} {Phys. Rev. B}\ }\textbf {\bibinfo {volume} {96}},\ \bibinfo {pages} {035152} (\bibinfo {year} {2017})}\BibitemShut {NoStop}%
\bibitem [{\citenamefont {Vandelli}\ \emph {et~al.}(2020)\citenamefont {Vandelli}, \citenamefont {Harkov}, \citenamefont {Stepanov}, \citenamefont {Gukelberger}, \citenamefont {Kozik}, \citenamefont {Rubio},\ and\ \citenamefont {Lichtenstein}}]{PhysRevB.102.195109}%
  \BibitemOpen
  \bibfield  {author} {\bibinfo {author} {\bibfnamefont {M.}~\bibnamefont {Vandelli}}, \bibinfo {author} {\bibfnamefont {V.}~\bibnamefont {Harkov}}, \bibinfo {author} {\bibfnamefont {E.~A.}\ \bibnamefont {Stepanov}}, \bibinfo {author} {\bibfnamefont {J.}~\bibnamefont {Gukelberger}}, \bibinfo {author} {\bibfnamefont {E.}~\bibnamefont {Kozik}}, \bibinfo {author} {\bibfnamefont {A.}~\bibnamefont {Rubio}},\ and\ \bibinfo {author} {\bibfnamefont {A.~I.}\ \bibnamefont {Lichtenstein}},\ }\bibfield  {title} {\bibinfo {title} {{Dual boson diagrammatic Monte Carlo approach applied to the extended Hubbard model}},\ }\href {https://doi.org/10.1103/PhysRevB.102.195109} {\bibfield  {journal} {\bibinfo  {journal} {Phys. Rev. B}\ }\textbf {\bibinfo {volume} {102}},\ \bibinfo {pages} {195109} (\bibinfo {year} {2020})}\BibitemShut {NoStop}%
\bibitem [{\citenamefont {Prokof'ev}\ and\ \citenamefont {Svistunov}(1998)}]{PhysRevLett.81.2514}%
  \BibitemOpen
  \bibfield  {author} {\bibinfo {author} {\bibfnamefont {N.~V.}\ \bibnamefont {Prokof'ev}}\ and\ \bibinfo {author} {\bibfnamefont {B.~V.}\ \bibnamefont {Svistunov}},\ }\bibfield  {title} {\bibinfo {title} {{Polaron Problem by Diagrammatic Quantum Monte Carlo}},\ }\href {https://doi.org/10.1103/PhysRevLett.81.2514} {\bibfield  {journal} {\bibinfo  {journal} {Phys. Rev. Lett.}\ }\textbf {\bibinfo {volume} {81}},\ \bibinfo {pages} {2514} (\bibinfo {year} {1998})}\BibitemShut {NoStop}%
\bibitem [{\citenamefont {Kozik}\ \emph {et~al.}(2010)\citenamefont {Kozik}, \citenamefont {Van~Houcke}, \citenamefont {Gull}, \citenamefont {Pollet}, \citenamefont {Prokof'ev}, \citenamefont {Svistunov},\ and\ \citenamefont {Troyer}}]{Kozik_2010}%
  \BibitemOpen
  \bibfield  {author} {\bibinfo {author} {\bibfnamefont {E.}~\bibnamefont {Kozik}}, \bibinfo {author} {\bibfnamefont {K.}~\bibnamefont {Van~Houcke}}, \bibinfo {author} {\bibfnamefont {E.}~\bibnamefont {Gull}}, \bibinfo {author} {\bibfnamefont {L.}~\bibnamefont {Pollet}}, \bibinfo {author} {\bibfnamefont {N.}~\bibnamefont {Prokof'ev}}, \bibinfo {author} {\bibfnamefont {B.}~\bibnamefont {Svistunov}},\ and\ \bibinfo {author} {\bibfnamefont {M.}~\bibnamefont {Troyer}},\ }\bibfield  {title} {\bibinfo {title} {{Diagrammatic Monte Carlo for correlated fermions}},\ }\href {https://doi.org/10.1209/0295-5075/90/10004} {\bibfield  {journal} {\bibinfo  {journal} {EPL}\ }\textbf {\bibinfo {volume} {90}},\ \bibinfo {pages} {10004} (\bibinfo {year} {2010})}\BibitemShut {NoStop}%
\bibitem [{\citenamefont {Biermann}\ \emph {et~al.}(2003)\citenamefont {Biermann}, \citenamefont {Aryasetiawan},\ and\ \citenamefont {Georges}}]{PhysRevLett.90.086402}%
  \BibitemOpen
  \bibfield  {author} {\bibinfo {author} {\bibfnamefont {S.}~\bibnamefont {Biermann}}, \bibinfo {author} {\bibfnamefont {F.}~\bibnamefont {Aryasetiawan}},\ and\ \bibinfo {author} {\bibfnamefont {A.}~\bibnamefont {Georges}},\ }\bibfield  {title} {\bibinfo {title} {{First-Principles Approach to the Electronic Structure of Strongly Correlated Systems: Combining the $GW$ Approximation and Dynamical Mean-Field Theory}},\ }\href {https://doi.org/10.1103/PhysRevLett.90.086402} {\bibfield  {journal} {\bibinfo  {journal} {Phys. Rev. Lett.}\ }\textbf {\bibinfo {volume} {90}},\ \bibinfo {pages} {086402} (\bibinfo {year} {2003})}\BibitemShut {NoStop}%
\bibitem [{\citenamefont {Sun}\ and\ \citenamefont {Kotliar}(2004)}]{PhysRevLett.92.196402}%
  \BibitemOpen
  \bibfield  {author} {\bibinfo {author} {\bibfnamefont {P.}~\bibnamefont {Sun}}\ and\ \bibinfo {author} {\bibfnamefont {G.}~\bibnamefont {Kotliar}},\ }\bibfield  {title} {\bibinfo {title} {{Many-Body Approximation Scheme beyond GW}},\ }\href {https://doi.org/10.1103/PhysRevLett.92.196402} {\bibfield  {journal} {\bibinfo  {journal} {Phys. Rev. Lett.}\ }\textbf {\bibinfo {volume} {92}},\ \bibinfo {pages} {196402} (\bibinfo {year} {2004})}\BibitemShut {NoStop}%
\bibitem [{\citenamefont {Ayral}\ \emph {et~al.}(2012)\citenamefont {Ayral}, \citenamefont {Werner},\ and\ \citenamefont {Biermann}}]{PhysRevLett.109.226401}%
  \BibitemOpen
  \bibfield  {author} {\bibinfo {author} {\bibfnamefont {T.}~\bibnamefont {Ayral}}, \bibinfo {author} {\bibfnamefont {P.}~\bibnamefont {Werner}},\ and\ \bibinfo {author} {\bibfnamefont {S.}~\bibnamefont {Biermann}},\ }\bibfield  {title} {\bibinfo {title} {{Spectral Properties of Correlated Materials: Local Vertex and Nonlocal Two-Particle Correlations from Combined $GW$ and Dynamical Mean Field Theory}},\ }\href {https://doi.org/10.1103/PhysRevLett.109.226401} {\bibfield  {journal} {\bibinfo  {journal} {Phys. Rev. Lett.}\ }\textbf {\bibinfo {volume} {109}},\ \bibinfo {pages} {226401} (\bibinfo {year} {2012})}\BibitemShut {NoStop}%
\bibitem [{\citenamefont {Ayral}\ \emph {et~al.}(2013)\citenamefont {Ayral}, \citenamefont {Biermann},\ and\ \citenamefont {Werner}}]{PhysRevB.87.125149}%
  \BibitemOpen
  \bibfield  {author} {\bibinfo {author} {\bibfnamefont {T.}~\bibnamefont {Ayral}}, \bibinfo {author} {\bibfnamefont {S.}~\bibnamefont {Biermann}},\ and\ \bibinfo {author} {\bibfnamefont {P.}~\bibnamefont {Werner}},\ }\bibfield  {title} {\bibinfo {title} {{Screening and nonlocal correlations in the extended Hubbard model from self-consistent combined $GW$ and dynamical mean field theory}},\ }\href {https://doi.org/10.1103/PhysRevB.87.125149} {\bibfield  {journal} {\bibinfo  {journal} {Phys. Rev. B}\ }\textbf {\bibinfo {volume} {87}},\ \bibinfo {pages} {125149} (\bibinfo {year} {2013})}\BibitemShut {NoStop}%
\bibitem [{\citenamefont {Huang}\ \emph {et~al.}(2014)\citenamefont {Huang}, \citenamefont {Ayral}, \citenamefont {Biermann},\ and\ \citenamefont {Werner}}]{PhysRevB.90.195114}%
  \BibitemOpen
  \bibfield  {author} {\bibinfo {author} {\bibfnamefont {L.}~\bibnamefont {Huang}}, \bibinfo {author} {\bibfnamefont {T.}~\bibnamefont {Ayral}}, \bibinfo {author} {\bibfnamefont {S.}~\bibnamefont {Biermann}},\ and\ \bibinfo {author} {\bibfnamefont {P.}~\bibnamefont {Werner}},\ }\bibfield  {title} {\bibinfo {title} {{Extended dynamical mean-field study of the Hubbard model with long-range interactions}},\ }\href {https://doi.org/10.1103/PhysRevB.90.195114} {\bibfield  {journal} {\bibinfo  {journal} {Phys. Rev. B}\ }\textbf {\bibinfo {volume} {90}},\ \bibinfo {pages} {195114} (\bibinfo {year} {2014})}\BibitemShut {NoStop}%
\bibitem [{\citenamefont {Boehnke}\ \emph {et~al.}(2016)\citenamefont {Boehnke}, \citenamefont {Nilsson}, \citenamefont {Aryasetiawan},\ and\ \citenamefont {Werner}}]{PhysRevB.94.201106}%
  \BibitemOpen
  \bibfield  {author} {\bibinfo {author} {\bibfnamefont {L.}~\bibnamefont {Boehnke}}, \bibinfo {author} {\bibfnamefont {F.}~\bibnamefont {Nilsson}}, \bibinfo {author} {\bibfnamefont {F.}~\bibnamefont {Aryasetiawan}},\ and\ \bibinfo {author} {\bibfnamefont {P.}~\bibnamefont {Werner}},\ }\bibfield  {title} {\bibinfo {title} {{When strong correlations become weak: Consistent merging of $GW$ and DMFT}},\ }\href {https://doi.org/10.1103/PhysRevB.94.201106} {\bibfield  {journal} {\bibinfo  {journal} {Phys. Rev. B}\ }\textbf {\bibinfo {volume} {94}},\ \bibinfo {pages} {201106} (\bibinfo {year} {2016})}\BibitemShut {NoStop}%
\bibitem [{\citenamefont {Ayral}\ \emph {et~al.}(2017{\natexlab{a}})\citenamefont {Ayral}, \citenamefont {Biermann}, \citenamefont {Werner},\ and\ \citenamefont {Boehnke}}]{PhysRevB.95.245130}%
  \BibitemOpen
  \bibfield  {author} {\bibinfo {author} {\bibfnamefont {T.}~\bibnamefont {Ayral}}, \bibinfo {author} {\bibfnamefont {S.}~\bibnamefont {Biermann}}, \bibinfo {author} {\bibfnamefont {P.}~\bibnamefont {Werner}},\ and\ \bibinfo {author} {\bibfnamefont {L.}~\bibnamefont {Boehnke}},\ }\bibfield  {title} {\bibinfo {title} {{Influence of Fock exchange in combined many-body perturbation and dynamical mean field theory}},\ }\href {https://doi.org/10.1103/PhysRevB.95.245130} {\bibfield  {journal} {\bibinfo  {journal} {Phys. Rev. B}\ }\textbf {\bibinfo {volume} {95}},\ \bibinfo {pages} {245130} (\bibinfo {year} {2017}{\natexlab{a}})}\BibitemShut {NoStop}%
\bibitem [{\citenamefont {Martin}\ \emph {et~al.}(2023)\citenamefont {Martin}, \citenamefont {Gauvin-Ndiaye},\ and\ \citenamefont {Tremblay}}]{PhysRevB.107.075158}%
  \BibitemOpen
  \bibfield  {author} {\bibinfo {author} {\bibfnamefont {N.}~\bibnamefont {Martin}}, \bibinfo {author} {\bibfnamefont {C.}~\bibnamefont {Gauvin-Ndiaye}},\ and\ \bibinfo {author} {\bibfnamefont {A.-M.~S.}\ \bibnamefont {Tremblay}},\ }\bibfield  {title} {\bibinfo {title} {{Nonlocal corrections to dynamical mean-field theory from the two-particle self-consistent method}},\ }\href {https://doi.org/10.1103/PhysRevB.107.075158} {\bibfield  {journal} {\bibinfo  {journal} {Phys. Rev. B}\ }\textbf {\bibinfo {volume} {107}},\ \bibinfo {pages} {075158} (\bibinfo {year} {2023})}\BibitemShut {NoStop}%
\bibitem [{\citenamefont {Zantout}\ \emph {et~al.}(2023)\citenamefont {Zantout}, \citenamefont {Backes}, \citenamefont {Razpopov}, \citenamefont {Lessnich},\ and\ \citenamefont {Valent\'{\i}}}]{PhysRevB.107.235101}%
  \BibitemOpen
  \bibfield  {author} {\bibinfo {author} {\bibfnamefont {K.}~\bibnamefont {Zantout}}, \bibinfo {author} {\bibfnamefont {S.}~\bibnamefont {Backes}}, \bibinfo {author} {\bibfnamefont {A.}~\bibnamefont {Razpopov}}, \bibinfo {author} {\bibfnamefont {D.}~\bibnamefont {Lessnich}},\ and\ \bibinfo {author} {\bibfnamefont {R.}~\bibnamefont {Valent\'{\i}}},\ }\bibfield  {title} {\bibinfo {title} {{Improved effective vertices in the multiorbital two-particle self-consistent method from dynamical mean-field theory}},\ }\href {https://doi.org/10.1103/PhysRevB.107.235101} {\bibfield  {journal} {\bibinfo  {journal} {Phys. Rev. B}\ }\textbf {\bibinfo {volume} {107}},\ \bibinfo {pages} {235101} (\bibinfo {year} {2023})}\BibitemShut {NoStop}%
\bibitem [{\citenamefont {Simard}\ and\ \citenamefont {Werner}(2023)}]{PhysRevB.107.245137}%
  \BibitemOpen
  \bibfield  {author} {\bibinfo {author} {\bibfnamefont {O.}~\bibnamefont {Simard}}\ and\ \bibinfo {author} {\bibfnamefont {P.}~\bibnamefont {Werner}},\ }\bibfield  {title} {\bibinfo {title} {{Dynamical mean field theory extension to the nonequilibrium two-particle self-consistent approach}},\ }\href {https://doi.org/10.1103/PhysRevB.107.245137} {\bibfield  {journal} {\bibinfo  {journal} {Phys. Rev. B}\ }\textbf {\bibinfo {volume} {107}},\ \bibinfo {pages} {245137} (\bibinfo {year} {2023})}\BibitemShut {NoStop}%
\bibitem [{\citenamefont {Geng}\ \emph {et~al.}(2025)\citenamefont {Geng}, \citenamefont {Yan},\ and\ \citenamefont {Werner}}]{PhysRevB.111.115143}%
  \BibitemOpen
  \bibfield  {author} {\bibinfo {author} {\bibfnamefont {L.}~\bibnamefont {Geng}}, \bibinfo {author} {\bibfnamefont {J.}~\bibnamefont {Yan}},\ and\ \bibinfo {author} {\bibfnamefont {P.}~\bibnamefont {Werner}},\ }\bibfield  {title} {\bibinfo {title} {{Two-particle self-consistent approach combined with dynamical mean field theory: A real-frequency study of the square-lattice Hubbard model}},\ }\href {https://doi.org/10.1103/PhysRevB.111.115143} {\bibfield  {journal} {\bibinfo  {journal} {Phys. Rev. B}\ }\textbf {\bibinfo {volume} {111}},\ \bibinfo {pages} {115143} (\bibinfo {year} {2025})}\BibitemShut {NoStop}%
\bibitem [{\citenamefont {Profe}\ \emph {et~al.}(2025)\citenamefont {Profe}, \citenamefont {Yan}, \citenamefont {Zantout}, \citenamefont {Werner},\ and\ \citenamefont {Valentí}}]{tpscdmft}%
  \BibitemOpen
  \bibfield  {author} {\bibinfo {author} {\bibfnamefont {J.~B.}\ \bibnamefont {Profe}}, \bibinfo {author} {\bibfnamefont {J.}~\bibnamefont {Yan}}, \bibinfo {author} {\bibfnamefont {K.}~\bibnamefont {Zantout}}, \bibinfo {author} {\bibfnamefont {P.}~\bibnamefont {Werner}},\ and\ \bibinfo {author} {\bibfnamefont {R.}~\bibnamefont {Valentí}},\ }\bibfield  {title} {\bibinfo {title} {{Multi-orbital two-particle self-consistent approach – Strengths and limitations}},\ }\href {https://doi.org/10.21468/SciPostPhys.19.1.026} {\bibfield  {journal} {\bibinfo  {journal} {SciPost Phys.}\ }\textbf {\bibinfo {volume} {19}},\ \bibinfo {pages} {026} (\bibinfo {year} {2025})}\BibitemShut {NoStop}%
\bibitem [{\citenamefont {Rubtsov}\ \emph {et~al.}(2008)\citenamefont {Rubtsov}, \citenamefont {Katsnelson},\ and\ \citenamefont {Lichtenstein}}]{PhysRevB.77.033101}%
  \BibitemOpen
  \bibfield  {author} {\bibinfo {author} {\bibfnamefont {A.~N.}\ \bibnamefont {Rubtsov}}, \bibinfo {author} {\bibfnamefont {M.~I.}\ \bibnamefont {Katsnelson}},\ and\ \bibinfo {author} {\bibfnamefont {A.~I.}\ \bibnamefont {Lichtenstein}},\ }\bibfield  {title} {\bibinfo {title} {{Dual fermion approach to nonlocal correlations in the Hubbard model}},\ }\href {https://doi.org/10.1103/PhysRevB.77.033101} {\bibfield  {journal} {\bibinfo  {journal} {Phys. Rev. B}\ }\textbf {\bibinfo {volume} {77}},\ \bibinfo {pages} {033101} (\bibinfo {year} {2008})}\BibitemShut {NoStop}%
\bibitem [{\citenamefont {Rubtsov}\ \emph {et~al.}(2009)\citenamefont {Rubtsov}, \citenamefont {Katsnelson}, \citenamefont {Lichtenstein},\ and\ \citenamefont {Georges}}]{PhysRevB.79.045133}%
  \BibitemOpen
  \bibfield  {author} {\bibinfo {author} {\bibfnamefont {A.~N.}\ \bibnamefont {Rubtsov}}, \bibinfo {author} {\bibfnamefont {M.~I.}\ \bibnamefont {Katsnelson}}, \bibinfo {author} {\bibfnamefont {A.~I.}\ \bibnamefont {Lichtenstein}},\ and\ \bibinfo {author} {\bibfnamefont {A.}~\bibnamefont {Georges}},\ }\bibfield  {title} {\bibinfo {title} {{Dual fermion approach to the two-dimensional Hubbard model: Antiferromagnetic fluctuations and Fermi arcs}},\ }\href {https://doi.org/10.1103/PhysRevB.79.045133} {\bibfield  {journal} {\bibinfo  {journal} {Phys. Rev. B}\ }\textbf {\bibinfo {volume} {79}},\ \bibinfo {pages} {045133} (\bibinfo {year} {2009})}\BibitemShut {NoStop}%
\bibitem [{\citenamefont {Hafermann}\ \emph {et~al.}(2009)\citenamefont {Hafermann}, \citenamefont {Li}, \citenamefont {Rubtsov}, \citenamefont {Katsnelson}, \citenamefont {Lichtenstein},\ and\ \citenamefont {Monien}}]{PhysRevLett.102.206401}%
  \BibitemOpen
  \bibfield  {author} {\bibinfo {author} {\bibfnamefont {H.}~\bibnamefont {Hafermann}}, \bibinfo {author} {\bibfnamefont {G.}~\bibnamefont {Li}}, \bibinfo {author} {\bibfnamefont {A.~N.}\ \bibnamefont {Rubtsov}}, \bibinfo {author} {\bibfnamefont {M.~I.}\ \bibnamefont {Katsnelson}}, \bibinfo {author} {\bibfnamefont {A.~I.}\ \bibnamefont {Lichtenstein}},\ and\ \bibinfo {author} {\bibfnamefont {H.}~\bibnamefont {Monien}},\ }\bibfield  {title} {\bibinfo {title} {{Efficient Perturbation Theory for Quantum Lattice Models}},\ }\href {https://doi.org/10.1103/PhysRevLett.102.206401} {\bibfield  {journal} {\bibinfo  {journal} {Phys. Rev. Lett.}\ }\textbf {\bibinfo {volume} {102}},\ \bibinfo {pages} {206401} (\bibinfo {year} {2009})}\BibitemShut {NoStop}%
\bibitem [{\citenamefont {{Brener, S. and Stepanov, Evgeny A. and Rubtsov, Alexey N. and Katsnelson, Mikhail I. and Lichtenstein, Alexander I.}}(2020)}]{Brener2020}%
  \BibitemOpen
  \bibfield  {author} {\bibinfo {author} {\bibnamefont {{Brener, S. and Stepanov, Evgeny A. and Rubtsov, Alexey N. and Katsnelson, Mikhail I. and Lichtenstein, Alexander I.}}},\ }\bibfield  {title} {\bibinfo {title} {{Dual fermion method as a prototype of generic reference-system approach for correlated fermions}},\ }\href {https://doi.org/{10.1016/j.aop.2020.168310}} {\bibfield  {journal} {\bibinfo  {journal} {{Ann. Phys.}}\ }\textbf {\bibinfo {volume} {{422}}},\ \bibinfo {pages} {{168310}} (\bibinfo {year} {{2020}})}\BibitemShut {NoStop}%
\bibitem [{\citenamefont {Rubtsov}\ \emph {et~al.}(2012)\citenamefont {Rubtsov}, \citenamefont {Katsnelson},\ and\ \citenamefont {Lichtenstein}}]{rubtsov2012}%
  \BibitemOpen
  \bibfield  {author} {\bibinfo {author} {\bibfnamefont {A.~N.}\ \bibnamefont {Rubtsov}}, \bibinfo {author} {\bibfnamefont {M.~I.}\ \bibnamefont {Katsnelson}},\ and\ \bibinfo {author} {\bibfnamefont {A.~I.}\ \bibnamefont {Lichtenstein}},\ }\bibfield  {title} {\bibinfo {title} {{Dual boson approach to collective excitations in correlated fermionic systems}},\ }\href {https://doi.org/https://doi.org/10.1016/j.aop.2012.01.002} {\bibfield  {journal} {\bibinfo  {journal} {Ann. Phys.}\ }\textbf {\bibinfo {volume} {327}},\ \bibinfo {pages} {1320} (\bibinfo {year} {2012})}\BibitemShut {NoStop}%
\bibitem [{\citenamefont {van Loon}\ \emph {et~al.}(2014{\natexlab{a}})\citenamefont {van Loon}, \citenamefont {Lichtenstein}, \citenamefont {Katsnelson}, \citenamefont {Parcollet},\ and\ \citenamefont {Hafermann}}]{PhysRevB.90.235135}%
  \BibitemOpen
  \bibfield  {author} {\bibinfo {author} {\bibfnamefont {E.~G. C.~P.}\ \bibnamefont {van Loon}}, \bibinfo {author} {\bibfnamefont {A.~I.}\ \bibnamefont {Lichtenstein}}, \bibinfo {author} {\bibfnamefont {M.~I.}\ \bibnamefont {Katsnelson}}, \bibinfo {author} {\bibfnamefont {O.}~\bibnamefont {Parcollet}},\ and\ \bibinfo {author} {\bibfnamefont {H.}~\bibnamefont {Hafermann}},\ }\bibfield  {title} {\bibinfo {title} {{Beyond extended dynamical mean-field theory: Dual boson approach to the two-dimensional extended Hubbard model}},\ }\href {https://doi.org/10.1103/PhysRevB.90.235135} {\bibfield  {journal} {\bibinfo  {journal} {Phys. Rev. B}\ }\textbf {\bibinfo {volume} {90}},\ \bibinfo {pages} {235135} (\bibinfo {year} {2014}{\natexlab{a}})}\BibitemShut {NoStop}%
\bibitem [{\citenamefont {Stepanov}\ \emph {et~al.}(2016{\natexlab{a}})\citenamefont {Stepanov}, \citenamefont {van Loon}, \citenamefont {Katanin}, \citenamefont {Lichtenstein}, \citenamefont {Katsnelson},\ and\ \citenamefont {Rubtsov}}]{PhysRevB.93.045107}%
  \BibitemOpen
  \bibfield  {author} {\bibinfo {author} {\bibfnamefont {E.~A.}\ \bibnamefont {Stepanov}}, \bibinfo {author} {\bibfnamefont {E.~G. C.~P.}\ \bibnamefont {van Loon}}, \bibinfo {author} {\bibfnamefont {A.~A.}\ \bibnamefont {Katanin}}, \bibinfo {author} {\bibfnamefont {A.~I.}\ \bibnamefont {Lichtenstein}}, \bibinfo {author} {\bibfnamefont {M.~I.}\ \bibnamefont {Katsnelson}},\ and\ \bibinfo {author} {\bibfnamefont {A.~N.}\ \bibnamefont {Rubtsov}},\ }\bibfield  {title} {\bibinfo {title} {{Self-consistent dual boson approach to single-particle and collective excitations in correlated systems}},\ }\href {https://doi.org/10.1103/PhysRevB.93.045107} {\bibfield  {journal} {\bibinfo  {journal} {Phys. Rev. B}\ }\textbf {\bibinfo {volume} {93}},\ \bibinfo {pages} {045107} (\bibinfo {year} {2016}{\natexlab{a}})}\BibitemShut {NoStop}%
\bibitem [{\citenamefont {Stepanov}\ \emph {et~al.}(2016{\natexlab{b}})\citenamefont {Stepanov}, \citenamefont {Huber}, \citenamefont {van Loon}, \citenamefont {Lichtenstein},\ and\ \citenamefont {Katsnelson}}]{PhysRevB.94.205110}%
  \BibitemOpen
  \bibfield  {author} {\bibinfo {author} {\bibfnamefont {E.~A.}\ \bibnamefont {Stepanov}}, \bibinfo {author} {\bibfnamefont {A.}~\bibnamefont {Huber}}, \bibinfo {author} {\bibfnamefont {E.~G. C.~P.}\ \bibnamefont {van Loon}}, \bibinfo {author} {\bibfnamefont {A.~I.}\ \bibnamefont {Lichtenstein}},\ and\ \bibinfo {author} {\bibfnamefont {M.~I.}\ \bibnamefont {Katsnelson}},\ }\bibfield  {title} {\bibinfo {title} {{From local to nonlocal correlations: The Dual Boson perspective}},\ }\href {https://doi.org/10.1103/PhysRevB.94.205110} {\bibfield  {journal} {\bibinfo  {journal} {Phys. Rev. B}\ }\textbf {\bibinfo {volume} {94}},\ \bibinfo {pages} {205110} (\bibinfo {year} {2016}{\natexlab{b}})}\BibitemShut {NoStop}%
\bibitem [{\citenamefont {Stepanov}\ \emph {et~al.}(2018)\citenamefont {Stepanov}, \citenamefont {Peters}, \citenamefont {Krivenko}, \citenamefont {Lichtenstein}, \citenamefont {Katsnelson},\ and\ \citenamefont {Rubtsov}}]{stepanov2018}%
  \BibitemOpen
  \bibfield  {author} {\bibinfo {author} {\bibfnamefont {E.~A.}\ \bibnamefont {Stepanov}}, \bibinfo {author} {\bibfnamefont {L.}~\bibnamefont {Peters}}, \bibinfo {author} {\bibfnamefont {I.~S.}\ \bibnamefont {Krivenko}}, \bibinfo {author} {\bibfnamefont {A.~I.}\ \bibnamefont {Lichtenstein}}, \bibinfo {author} {\bibfnamefont {M.~I.}\ \bibnamefont {Katsnelson}},\ and\ \bibinfo {author} {\bibfnamefont {A.~N.}\ \bibnamefont {Rubtsov}},\ }\bibfield  {title} {\bibinfo {title} {{Quantum spin fluctuations and evolution of electronic structure in cuprates}},\ }\href {https://doi.org/10.1038/s41535-018-0128-x} {\bibfield  {journal} {\bibinfo  {journal} {npj Quantum Mater.}\ }\textbf {\bibinfo {volume} {3}},\ \bibinfo {pages} {54} (\bibinfo {year} {2018})}\BibitemShut {NoStop}%
\bibitem [{\citenamefont {Peters}\ \emph {et~al.}(2019)\citenamefont {Peters}, \citenamefont {van Loon}, \citenamefont {Rubtsov}, \citenamefont {Lichtenstein}, \citenamefont {Katsnelson},\ and\ \citenamefont {Stepanov}}]{PhysRevB.100.165128}%
  \BibitemOpen
  \bibfield  {author} {\bibinfo {author} {\bibfnamefont {L.}~\bibnamefont {Peters}}, \bibinfo {author} {\bibfnamefont {E.~G. C.~P.}\ \bibnamefont {van Loon}}, \bibinfo {author} {\bibfnamefont {A.~N.}\ \bibnamefont {Rubtsov}}, \bibinfo {author} {\bibfnamefont {A.~I.}\ \bibnamefont {Lichtenstein}}, \bibinfo {author} {\bibfnamefont {M.~I.}\ \bibnamefont {Katsnelson}},\ and\ \bibinfo {author} {\bibfnamefont {E.~A.}\ \bibnamefont {Stepanov}},\ }\bibfield  {title} {\bibinfo {title} {{Dual boson approach with instantaneous interaction}},\ }\href {https://doi.org/10.1103/PhysRevB.100.165128} {\bibfield  {journal} {\bibinfo  {journal} {Phys. Rev. B}\ }\textbf {\bibinfo {volume} {100}},\ \bibinfo {pages} {165128} (\bibinfo {year} {2019})}\BibitemShut {NoStop}%
\bibitem [{\citenamefont {Toschi}\ \emph {et~al.}(2007)\citenamefont {Toschi}, \citenamefont {Katanin},\ and\ \citenamefont {Held}}]{PhysRevB.75.045118}%
  \BibitemOpen
  \bibfield  {author} {\bibinfo {author} {\bibfnamefont {A.}~\bibnamefont {Toschi}}, \bibinfo {author} {\bibfnamefont {A.~A.}\ \bibnamefont {Katanin}},\ and\ \bibinfo {author} {\bibfnamefont {K.}~\bibnamefont {Held}},\ }\bibfield  {title} {\bibinfo {title} {{Dynamical vertex approximation: A step beyond dynamical mean-field theory}},\ }\href {https://doi.org/10.1103/PhysRevB.75.045118} {\bibfield  {journal} {\bibinfo  {journal} {Phys. Rev. B}\ }\textbf {\bibinfo {volume} {75}},\ \bibinfo {pages} {045118} (\bibinfo {year} {2007})}\BibitemShut {NoStop}%
\bibitem [{\citenamefont {Galler}\ \emph {et~al.}(2017)\citenamefont {Galler}, \citenamefont {Thunstr\"om}, \citenamefont {Gunacker}, \citenamefont {Tomczak},\ and\ \citenamefont {Held}}]{PhysRevB.95.115107}%
  \BibitemOpen
  \bibfield  {author} {\bibinfo {author} {\bibfnamefont {A.}~\bibnamefont {Galler}}, \bibinfo {author} {\bibfnamefont {P.}~\bibnamefont {Thunstr\"om}}, \bibinfo {author} {\bibfnamefont {P.}~\bibnamefont {Gunacker}}, \bibinfo {author} {\bibfnamefont {J.~M.}\ \bibnamefont {Tomczak}},\ and\ \bibinfo {author} {\bibfnamefont {K.}~\bibnamefont {Held}},\ }\bibfield  {title} {\bibinfo {title} {{Ab initio dynamical vertex approximation}},\ }\href {https://doi.org/10.1103/PhysRevB.95.115107} {\bibfield  {journal} {\bibinfo  {journal} {Phys. Rev. B}\ }\textbf {\bibinfo {volume} {95}},\ \bibinfo {pages} {115107} (\bibinfo {year} {2017})}\BibitemShut {NoStop}%
\bibitem [{\citenamefont {Galler}\ \emph {et~al.}(2018)\citenamefont {Galler}, \citenamefont {Kaufmann}, \citenamefont {Gunacker}, \citenamefont {Pickem}, \citenamefont {Thunstr\"om}, \citenamefont {Tomczak},\ and\ \citenamefont {Held}}]{doi:10.7566/JPSJ.87.041004}%
  \BibitemOpen
  \bibfield  {author} {\bibinfo {author} {\bibfnamefont {A.}~\bibnamefont {Galler}}, \bibinfo {author} {\bibfnamefont {J.}~\bibnamefont {Kaufmann}}, \bibinfo {author} {\bibfnamefont {P.}~\bibnamefont {Gunacker}}, \bibinfo {author} {\bibfnamefont {M.}~\bibnamefont {Pickem}}, \bibinfo {author} {\bibfnamefont {P.}~\bibnamefont {Thunstr\"om}}, \bibinfo {author} {\bibfnamefont {J.~M.}\ \bibnamefont {Tomczak}},\ and\ \bibinfo {author} {\bibfnamefont {K.}~\bibnamefont {Held}},\ }\bibfield  {title} {\bibinfo {title} {{Towards ab initio Calculations with the Dynamical Vertex Approximation}},\ }\href {https://doi.org/10.7566/JPSJ.87.041004} {\bibfield  {journal} {\bibinfo  {journal} {J. Phys. Soc. Jpn.}\ }\textbf {\bibinfo {volume} {87}},\ \bibinfo {pages} {041004} (\bibinfo {year} {2018})}\BibitemShut {NoStop}%
\bibitem [{\citenamefont {Kaufmann}\ \emph {et~al.}(2021)\citenamefont {Kaufmann}, \citenamefont {Eckhardt}, \citenamefont {Pickem}, \citenamefont {Kitatani}, \citenamefont {Kauch},\ and\ \citenamefont {Held}}]{PhysRevB.103.035120}%
  \BibitemOpen
  \bibfield  {author} {\bibinfo {author} {\bibfnamefont {J.}~\bibnamefont {Kaufmann}}, \bibinfo {author} {\bibfnamefont {C.}~\bibnamefont {Eckhardt}}, \bibinfo {author} {\bibfnamefont {M.}~\bibnamefont {Pickem}}, \bibinfo {author} {\bibfnamefont {M.}~\bibnamefont {Kitatani}}, \bibinfo {author} {\bibfnamefont {A.}~\bibnamefont {Kauch}},\ and\ \bibinfo {author} {\bibfnamefont {K.}~\bibnamefont {Held}},\ }\bibfield  {title} {\bibinfo {title} {{Self-consistent ladder dynamical vertex approximation}},\ }\href {https://doi.org/10.1103/PhysRevB.103.035120} {\bibfield  {journal} {\bibinfo  {journal} {Phys. Rev. B}\ }\textbf {\bibinfo {volume} {103}},\ \bibinfo {pages} {035120} (\bibinfo {year} {2021})}\BibitemShut {NoStop}%
\bibitem [{\citenamefont {Ayral}\ and\ \citenamefont {Parcollet}(2015)}]{PhysRevB.92.115109}%
  \BibitemOpen
  \bibfield  {author} {\bibinfo {author} {\bibfnamefont {T.}~\bibnamefont {Ayral}}\ and\ \bibinfo {author} {\bibfnamefont {O.}~\bibnamefont {Parcollet}},\ }\bibfield  {title} {\bibinfo {title} {{Mott physics and spin fluctuations: A unified framework}},\ }\href {https://doi.org/10.1103/PhysRevB.92.115109} {\bibfield  {journal} {\bibinfo  {journal} {Phys. Rev. B}\ }\textbf {\bibinfo {volume} {92}},\ \bibinfo {pages} {115109} (\bibinfo {year} {2015})}\BibitemShut {NoStop}%
\bibitem [{\citenamefont {Ayral}\ and\ \citenamefont {Parcollet}(2016)}]{PhysRevB.93.235124}%
  \BibitemOpen
  \bibfield  {author} {\bibinfo {author} {\bibfnamefont {T.}~\bibnamefont {Ayral}}\ and\ \bibinfo {author} {\bibfnamefont {O.}~\bibnamefont {Parcollet}},\ }\bibfield  {title} {\bibinfo {title} {{Mott physics and spin fluctuations: A functional viewpoint}},\ }\href {https://doi.org/10.1103/PhysRevB.93.235124} {\bibfield  {journal} {\bibinfo  {journal} {Phys. Rev. B}\ }\textbf {\bibinfo {volume} {93}},\ \bibinfo {pages} {235124} (\bibinfo {year} {2016})}\BibitemShut {NoStop}%
\bibitem [{\citenamefont {Vu\ifmmode \check{c}\else \v{c}\fi{}i\ifmmode \check{c}\else \v{c}\fi{}evi\ifmmode~\acute{c}\else \'{c}\fi{}}\ \emph {et~al.}(2017)\citenamefont {Vu\ifmmode \check{c}\else \v{c}\fi{}i\ifmmode \check{c}\else \v{c}\fi{}evi\ifmmode~\acute{c}\else \'{c}\fi{}}, \citenamefont {Ayral},\ and\ \citenamefont {Parcollet}}]{PhysRevB.96.104504}%
  \BibitemOpen
  \bibfield  {author} {\bibinfo {author} {\bibfnamefont {J.}~\bibnamefont {Vu\ifmmode \check{c}\else \v{c}\fi{}i\ifmmode \check{c}\else \v{c}\fi{}evi\ifmmode~\acute{c}\else \'{c}\fi{}}}, \bibinfo {author} {\bibfnamefont {T.}~\bibnamefont {Ayral}},\ and\ \bibinfo {author} {\bibfnamefont {O.}~\bibnamefont {Parcollet}},\ }\bibfield  {title} {\bibinfo {title} {{TRILEX and $GW$+EDMFT approach to $d$-wave superconductivity in the Hubbard model}},\ }\href {https://doi.org/10.1103/PhysRevB.96.104504} {\bibfield  {journal} {\bibinfo  {journal} {Phys. Rev. B}\ }\textbf {\bibinfo {volume} {96}},\ \bibinfo {pages} {104504} (\bibinfo {year} {2017})}\BibitemShut {NoStop}%
\bibitem [{\citenamefont {Stepanov}\ \emph {et~al.}(2019)\citenamefont {Stepanov}, \citenamefont {Harkov},\ and\ \citenamefont {Lichtenstein}}]{PhysRevB.100.205115}%
  \BibitemOpen
  \bibfield  {author} {\bibinfo {author} {\bibfnamefont {E.~A.}\ \bibnamefont {Stepanov}}, \bibinfo {author} {\bibfnamefont {V.}~\bibnamefont {Harkov}},\ and\ \bibinfo {author} {\bibfnamefont {A.~I.}\ \bibnamefont {Lichtenstein}},\ }\bibfield  {title} {\bibinfo {title} {{Consistent partial bosonization of the extended Hubbard model}},\ }\href {https://doi.org/10.1103/PhysRevB.100.205115} {\bibfield  {journal} {\bibinfo  {journal} {Phys. Rev. B}\ }\textbf {\bibinfo {volume} {100}},\ \bibinfo {pages} {205115} (\bibinfo {year} {2019})}\BibitemShut {NoStop}%
\bibitem [{\citenamefont {Harkov}\ \emph {et~al.}(2021)\citenamefont {Harkov}, \citenamefont {Vandelli}, \citenamefont {Brener}, \citenamefont {Lichtenstein},\ and\ \citenamefont {Stepanov}}]{PhysRevB.103.245123}%
  \BibitemOpen
  \bibfield  {author} {\bibinfo {author} {\bibfnamefont {V.}~\bibnamefont {Harkov}}, \bibinfo {author} {\bibfnamefont {M.}~\bibnamefont {Vandelli}}, \bibinfo {author} {\bibfnamefont {S.}~\bibnamefont {Brener}}, \bibinfo {author} {\bibfnamefont {A.~I.}\ \bibnamefont {Lichtenstein}},\ and\ \bibinfo {author} {\bibfnamefont {E.~A.}\ \bibnamefont {Stepanov}},\ }\bibfield  {title} {\bibinfo {title} {{Impact of partially bosonized collective fluctuations on electronic degrees of freedom}},\ }\href {https://doi.org/10.1103/PhysRevB.103.245123} {\bibfield  {journal} {\bibinfo  {journal} {Phys. Rev. B}\ }\textbf {\bibinfo {volume} {103}},\ \bibinfo {pages} {245123} (\bibinfo {year} {2021})}\BibitemShut {NoStop}%
\bibitem [{\citenamefont {Lenihan}\ \emph {et~al.}(2022)\citenamefont {Lenihan}, \citenamefont {Kim}, \citenamefont {\ifmmode~\check{S}\else \v{S}\fi{}imkovic},\ and\ \citenamefont {Kozik}}]{PhysRevLett.129.107202}%
  \BibitemOpen
  \bibfield  {author} {\bibinfo {author} {\bibfnamefont {C.}~\bibnamefont {Lenihan}}, \bibinfo {author} {\bibfnamefont {A.~J.}\ \bibnamefont {Kim}}, \bibinfo {author} {\bibfnamefont {F.}~\bibnamefont {\ifmmode~\check{S}\else \v{S}\fi{}imkovic}},\ and\ \bibinfo {author} {\bibfnamefont {E.}~\bibnamefont {Kozik}},\ }\bibfield  {title} {\bibinfo {title} {{Evaluating Second-Order Phase Transitions with Diagrammatic Monte Carlo: N\'eel Transition in the Doped Three-Dimensional Hubbard Model}},\ }\href {https://doi.org/10.1103/PhysRevLett.129.107202} {\bibfield  {journal} {\bibinfo  {journal} {Phys. Rev. Lett.}\ }\textbf {\bibinfo {volume} {129}},\ \bibinfo {pages} {107202} (\bibinfo {year} {2022})}\BibitemShut {NoStop}%
\bibitem [{\citenamefont {Rohringer}\ and\ \citenamefont {Toschi}(2016)}]{PhysRevB.94.125144}%
  \BibitemOpen
  \bibfield  {author} {\bibinfo {author} {\bibfnamefont {G.}~\bibnamefont {Rohringer}}\ and\ \bibinfo {author} {\bibfnamefont {A.}~\bibnamefont {Toschi}},\ }\bibfield  {title} {\bibinfo {title} {Impact of nonlocal correlations over different energy scales: A dynamical vertex approximation study},\ }\href {https://doi.org/10.1103/PhysRevB.94.125144} {\bibfield  {journal} {\bibinfo  {journal} {Phys. Rev. B}\ }\textbf {\bibinfo {volume} {94}},\ \bibinfo {pages} {125144} (\bibinfo {year} {2016})}\BibitemShut {NoStop}%
\bibitem [{\citenamefont {Sch\"afer}\ \emph {et~al.}(2021)\citenamefont {Sch\"afer}, \citenamefont {Wentzell}, \citenamefont {\ifmmode~\check{S}\else \v{S}\fi{}imkovic}, \citenamefont {He}, \citenamefont {Hille}, \citenamefont {Klett}, \citenamefont {Eckhardt}, \citenamefont {Arzhang}, \citenamefont {Harkov}, \citenamefont {Le~R\'egent}, \citenamefont {Kirsch}, \citenamefont {Wang}, \citenamefont {Kim}, \citenamefont {Kozik}, \citenamefont {Stepanov}, \citenamefont {Kauch}, \citenamefont {Andergassen}, \citenamefont {Hansmann}, \citenamefont {Rohe}, \citenamefont {Vilk}, \citenamefont {LeBlanc}, \citenamefont {Zhang}, \citenamefont {Tremblay}, \citenamefont {Ferrero}, \citenamefont {Parcollet},\ and\ \citenamefont {Georges}}]{PhysRevX.11.011058}%
  \BibitemOpen
  \bibfield  {author} {\bibinfo {author} {\bibfnamefont {T.}~\bibnamefont {Sch\"afer}}, \bibinfo {author} {\bibfnamefont {N.}~\bibnamefont {Wentzell}}, \bibinfo {author} {\bibfnamefont {F.}~\bibnamefont {\ifmmode~\check{S}\else \v{S}\fi{}imkovic}}, \bibinfo {author} {\bibfnamefont {Y.-Y.}\ \bibnamefont {He}}, \bibinfo {author} {\bibfnamefont {C.}~\bibnamefont {Hille}}, \bibinfo {author} {\bibfnamefont {M.}~\bibnamefont {Klett}}, \bibinfo {author} {\bibfnamefont {C.~J.}\ \bibnamefont {Eckhardt}}, \bibinfo {author} {\bibfnamefont {B.}~\bibnamefont {Arzhang}}, \bibinfo {author} {\bibfnamefont {V.}~\bibnamefont {Harkov}}, \bibinfo {author} {\bibfnamefont {F.-M.}\ \bibnamefont {Le~R\'egent}}, \bibinfo {author} {\bibfnamefont {A.}~\bibnamefont {Kirsch}}, \bibinfo {author} {\bibfnamefont {Y.}~\bibnamefont {Wang}}, \bibinfo {author} {\bibfnamefont {A.~J.}\ \bibnamefont {Kim}}, \bibinfo {author} {\bibfnamefont {E.}~\bibnamefont {Kozik}}, \bibinfo {author} {\bibfnamefont {E.~A.}\ \bibnamefont {Stepanov}}, \bibinfo
  {author} {\bibfnamefont {A.}~\bibnamefont {Kauch}}, \bibinfo {author} {\bibfnamefont {S.}~\bibnamefont {Andergassen}}, \bibinfo {author} {\bibfnamefont {P.}~\bibnamefont {Hansmann}}, \bibinfo {author} {\bibfnamefont {D.}~\bibnamefont {Rohe}}, \bibinfo {author} {\bibfnamefont {Y.~M.}\ \bibnamefont {Vilk}}, \bibinfo {author} {\bibfnamefont {J.~P.~F.}\ \bibnamefont {LeBlanc}}, \bibinfo {author} {\bibfnamefont {S.}~\bibnamefont {Zhang}}, \bibinfo {author} {\bibfnamefont {A.-M.~S.}\ \bibnamefont {Tremblay}}, \bibinfo {author} {\bibfnamefont {M.}~\bibnamefont {Ferrero}}, \bibinfo {author} {\bibfnamefont {O.}~\bibnamefont {Parcollet}},\ and\ \bibinfo {author} {\bibfnamefont {A.}~\bibnamefont {Georges}},\ }\bibfield  {title} {\bibinfo {title} {{Tracking the Footprints of Spin Fluctuations: A MultiMethod, MultiMessenger Study of the Two-Dimensional Hubbard Model}},\ }\href {https://doi.org/10.1103/PhysRevX.11.011058} {\bibfield  {journal} {\bibinfo  {journal} {Phys. Rev. X}\ }\textbf {\bibinfo {volume} {11}},\
  \bibinfo {pages} {011058} (\bibinfo {year} {2021})}\BibitemShut {NoStop}%
\bibitem [{\citenamefont {Vandelli}(2022)}]{vandelli2022quantum}%
  \BibitemOpen
  \bibfield  {author} {\bibinfo {author} {\bibfnamefont {M.}~\bibnamefont {Vandelli}},\ }\emph {\bibinfo {title} {Quantum embedding methods in dual space for strongly interacting electronic systems}},\ \href@noop {} {Ph.D. thesis},\ \bibinfo  {school} {Universit{\"a}t Hamburg Hamburg} (\bibinfo {year} {2022})\BibitemShut {NoStop}%
\bibitem [{\citenamefont {Sch\"afer}\ \emph {et~al.}(2015)\citenamefont {Sch\"afer}, \citenamefont {Geles}, \citenamefont {Rost}, \citenamefont {Rohringer}, \citenamefont {Arrigoni}, \citenamefont {Held}, \citenamefont {Bl\"umer}, \citenamefont {Aichhorn},\ and\ \citenamefont {Toschi}}]{PhysRevB.91.125109}%
  \BibitemOpen
  \bibfield  {author} {\bibinfo {author} {\bibfnamefont {T.}~\bibnamefont {Sch\"afer}}, \bibinfo {author} {\bibfnamefont {F.}~\bibnamefont {Geles}}, \bibinfo {author} {\bibfnamefont {D.}~\bibnamefont {Rost}}, \bibinfo {author} {\bibfnamefont {G.}~\bibnamefont {Rohringer}}, \bibinfo {author} {\bibfnamefont {E.}~\bibnamefont {Arrigoni}}, \bibinfo {author} {\bibfnamefont {K.}~\bibnamefont {Held}}, \bibinfo {author} {\bibfnamefont {N.}~\bibnamefont {Bl\"umer}}, \bibinfo {author} {\bibfnamefont {M.}~\bibnamefont {Aichhorn}},\ and\ \bibinfo {author} {\bibfnamefont {A.}~\bibnamefont {Toschi}},\ }\bibfield  {title} {\bibinfo {title} {{Fate of the false Mott-Hubbard transition in two dimensions}},\ }\href {https://doi.org/10.1103/PhysRevB.91.125109} {\bibfield  {journal} {\bibinfo  {journal} {Phys. Rev. B}\ }\textbf {\bibinfo {volume} {91}},\ \bibinfo {pages} {125109} (\bibinfo {year} {2015})}\BibitemShut {NoStop}%
\bibitem [{\citenamefont {\ifmmode~\check{S}\else \v{S}\fi{}imkovic}\ \emph {et~al.}(2020)\citenamefont {\ifmmode~\check{S}\else \v{S}\fi{}imkovic}, \citenamefont {LeBlanc}, \citenamefont {Kim}, \citenamefont {Deng}, \citenamefont {Prokof'ev}, \citenamefont {Svistunov},\ and\ \citenamefont {Kozik}}]{PhysRevLett.124.017003}%
  \BibitemOpen
  \bibfield  {author} {\bibinfo {author} {\bibfnamefont {F.}~\bibnamefont {\ifmmode~\check{S}\else \v{S}\fi{}imkovic}}, \bibinfo {author} {\bibfnamefont {J.~P.~F.}\ \bibnamefont {LeBlanc}}, \bibinfo {author} {\bibfnamefont {A.~J.}\ \bibnamefont {Kim}}, \bibinfo {author} {\bibfnamefont {Y.}~\bibnamefont {Deng}}, \bibinfo {author} {\bibfnamefont {N.~V.}\ \bibnamefont {Prokof'ev}}, \bibinfo {author} {\bibfnamefont {B.~V.}\ \bibnamefont {Svistunov}},\ and\ \bibinfo {author} {\bibfnamefont {E.}~\bibnamefont {Kozik}},\ }\bibfield  {title} {\bibinfo {title} {{Extended Crossover from a Fermi Liquid to a Quasiantiferromagnet in the Half-Filled 2D Hubbard Model}},\ }\href {https://doi.org/10.1103/PhysRevLett.124.017003} {\bibfield  {journal} {\bibinfo  {journal} {Phys. Rev. Lett.}\ }\textbf {\bibinfo {volume} {124}},\ \bibinfo {pages} {017003} (\bibinfo {year} {2020})}\BibitemShut {NoStop}%
\bibitem [{\citenamefont {Kim}\ \emph {et~al.}(2020)\citenamefont {Kim}, \citenamefont {Simkovic},\ and\ \citenamefont {Kozik}}]{PhysRevLett.124.117602}%
  \BibitemOpen
  \bibfield  {author} {\bibinfo {author} {\bibfnamefont {A.~J.}\ \bibnamefont {Kim}}, \bibinfo {author} {\bibfnamefont {F.}~\bibnamefont {Simkovic}},\ and\ \bibinfo {author} {\bibfnamefont {E.}~\bibnamefont {Kozik}},\ }\bibfield  {title} {\bibinfo {title} {{Spin and Charge Correlations across the Metal-to-Insulator Crossover in the Half-Filled 2D Hubbard Model}},\ }\href {https://doi.org/10.1103/PhysRevLett.124.117602} {\bibfield  {journal} {\bibinfo  {journal} {Phys. Rev. Lett.}\ }\textbf {\bibinfo {volume} {124}},\ \bibinfo {pages} {117602} (\bibinfo {year} {2020})}\BibitemShut {NoStop}%
\bibitem [{\citenamefont {{IV, Fedor \ifmmode \check{S}\else \v{S}\fi{}imkovic and Rossi, Riccardo and Ferrero, Michel}}(2022)}]{PhysRevResearch.4.043201}%
  \BibitemOpen
  \bibfield  {author} {\bibinfo {author} {\bibnamefont {{IV, Fedor \ifmmode \check{S}\else \v{S}\fi{}imkovic and Rossi, Riccardo and Ferrero, Michel}}},\ }\bibfield  {title} {\bibinfo {title} {{Two-dimensional Hubbard model at finite temperature: Weak, strong, and long correlation regimes}},\ }\href {https://doi.org/10.1103/PhysRevResearch.4.043201} {\bibfield  {journal} {\bibinfo  {journal} {Phys. Rev. Res.}\ }\textbf {\bibinfo {volume} {4}},\ \bibinfo {pages} {043201} (\bibinfo {year} {2022})}\BibitemShut {NoStop}%
\bibitem [{\citenamefont {Otsuki}\ \emph {et~al.}(2014)\citenamefont {Otsuki}, \citenamefont {Hafermann},\ and\ \citenamefont {Lichtenstein}}]{PhysRevB.90.235132}%
  \BibitemOpen
  \bibfield  {author} {\bibinfo {author} {\bibfnamefont {J.}~\bibnamefont {Otsuki}}, \bibinfo {author} {\bibfnamefont {H.}~\bibnamefont {Hafermann}},\ and\ \bibinfo {author} {\bibfnamefont {A.~I.}\ \bibnamefont {Lichtenstein}},\ }\bibfield  {title} {\bibinfo {title} {{Superconductivity, antiferromagnetism, and phase separation in the two-dimensional Hubbard model: A dual-fermion approach}},\ }\href {https://doi.org/10.1103/PhysRevB.90.235132} {\bibfield  {journal} {\bibinfo  {journal} {Phys. Rev. B}\ }\textbf {\bibinfo {volume} {90}},\ \bibinfo {pages} {235132} (\bibinfo {year} {2014})}\BibitemShut {NoStop}%
\bibitem [{\citenamefont {Kitatani}\ \emph {et~al.}(2015)\citenamefont {Kitatani}, \citenamefont {Tsuji},\ and\ \citenamefont {Aoki}}]{PhysRevB.92.085104}%
  \BibitemOpen
  \bibfield  {author} {\bibinfo {author} {\bibfnamefont {M.}~\bibnamefont {Kitatani}}, \bibinfo {author} {\bibfnamefont {N.}~\bibnamefont {Tsuji}},\ and\ \bibinfo {author} {\bibfnamefont {H.}~\bibnamefont {Aoki}},\ }\bibfield  {title} {\bibinfo {title} {{FLEX+DMFT approach to the $d$-wave superconducting phase diagram of the two-dimensional Hubbard model}},\ }\href {https://doi.org/10.1103/PhysRevB.92.085104} {\bibfield  {journal} {\bibinfo  {journal} {Phys. Rev. B}\ }\textbf {\bibinfo {volume} {92}},\ \bibinfo {pages} {085104} (\bibinfo {year} {2015})}\BibitemShut {NoStop}%
\bibitem [{\citenamefont {Kitatani}\ \emph {et~al.}(2019)\citenamefont {Kitatani}, \citenamefont {Sch\"afer}, \citenamefont {Aoki},\ and\ \citenamefont {Held}}]{PhysRevB.99.041115}%
  \BibitemOpen
  \bibfield  {author} {\bibinfo {author} {\bibfnamefont {M.}~\bibnamefont {Kitatani}}, \bibinfo {author} {\bibfnamefont {T.}~\bibnamefont {Sch\"afer}}, \bibinfo {author} {\bibfnamefont {H.}~\bibnamefont {Aoki}},\ and\ \bibinfo {author} {\bibfnamefont {K.}~\bibnamefont {Held}},\ }\bibfield  {title} {\bibinfo {title} {{Why the critical temperature of high-${T}_{c}$ cuprate superconductors is so low: The importance of the dynamical vertex structure}},\ }\href {https://doi.org/10.1103/PhysRevB.99.041115} {\bibfield  {journal} {\bibinfo  {journal} {Phys. Rev. B}\ }\textbf {\bibinfo {volume} {99}},\ \bibinfo {pages} {041115} (\bibinfo {year} {2019})}\BibitemShut {NoStop}%
\bibitem [{\citenamefont {Kitatani}\ \emph {et~al.}(2020)\citenamefont {Kitatani}, \citenamefont {Si}, \citenamefont {Janson}, \citenamefont {Arita}, \citenamefont {Zhong},\ and\ \citenamefont {Held}}]{kitatani2020nickelate}%
  \BibitemOpen
  \bibfield  {author} {\bibinfo {author} {\bibfnamefont {M.}~\bibnamefont {Kitatani}}, \bibinfo {author} {\bibfnamefont {L.}~\bibnamefont {Si}}, \bibinfo {author} {\bibfnamefont {O.}~\bibnamefont {Janson}}, \bibinfo {author} {\bibfnamefont {R.}~\bibnamefont {Arita}}, \bibinfo {author} {\bibfnamefont {Z.}~\bibnamefont {Zhong}},\ and\ \bibinfo {author} {\bibfnamefont {K.}~\bibnamefont {Held}},\ }\bibfield  {title} {\bibinfo {title} {{Nickelate superconductors — a renaissance of the one-band Hubbard model}},\ }\href {https://doi.org/10.1038/s41535-020-00260-y} {\bibfield  {journal} {\bibinfo  {journal} {npj Quantum Mater.}\ }\textbf {\bibinfo {volume} {5}},\ \bibinfo {pages} {59} (\bibinfo {year} {2020})}\BibitemShut {NoStop}%
\bibitem [{\citenamefont {Maier}\ \emph {et~al.}(2000)\citenamefont {Maier}, \citenamefont {Jarrell}, \citenamefont {Pruschke},\ and\ \citenamefont {Keller}}]{PhysRevLett.85.1524}%
  \BibitemOpen
  \bibfield  {author} {\bibinfo {author} {\bibfnamefont {T.}~\bibnamefont {Maier}}, \bibinfo {author} {\bibfnamefont {M.}~\bibnamefont {Jarrell}}, \bibinfo {author} {\bibfnamefont {T.}~\bibnamefont {Pruschke}},\ and\ \bibinfo {author} {\bibfnamefont {J.}~\bibnamefont {Keller}},\ }\bibfield  {title} {\bibinfo {title} {{$\mathit{d}$-Wave Superconductivity in the Hubbard Model}},\ }\href {https://doi.org/10.1103/PhysRevLett.85.1524} {\bibfield  {journal} {\bibinfo  {journal} {Phys. Rev. Lett.}\ }\textbf {\bibinfo {volume} {85}},\ \bibinfo {pages} {1524} (\bibinfo {year} {2000})}\BibitemShut {NoStop}%
\bibitem [{\citenamefont {Gull}\ \emph {et~al.}(2013)\citenamefont {Gull}, \citenamefont {Parcollet},\ and\ \citenamefont {Millis}}]{PhysRevLett.110.216405}%
  \BibitemOpen
  \bibfield  {author} {\bibinfo {author} {\bibfnamefont {E.}~\bibnamefont {Gull}}, \bibinfo {author} {\bibfnamefont {O.}~\bibnamefont {Parcollet}},\ and\ \bibinfo {author} {\bibfnamefont {A.~J.}\ \bibnamefont {Millis}},\ }\bibfield  {title} {\bibinfo {title} {{Superconductivity and the Pseudogap in the Two-Dimensional Hubbard Model}},\ }\href {https://doi.org/10.1103/PhysRevLett.110.216405} {\bibfield  {journal} {\bibinfo  {journal} {Phys. Rev. Lett.}\ }\textbf {\bibinfo {volume} {110}},\ \bibinfo {pages} {216405} (\bibinfo {year} {2013})}\BibitemShut {NoStop}%
\bibitem [{\citenamefont {Kamil}\ \emph {et~al.}(2018)\citenamefont {Kamil}, \citenamefont {Berges}, \citenamefont {Schönhoff}, \citenamefont {Rösner}, \citenamefont {Schüler}, \citenamefont {Sangiovanni},\ and\ \citenamefont {Wehling}}]{Kamil2018}%
  \BibitemOpen
  \bibfield  {author} {\bibinfo {author} {\bibfnamefont {E.}~\bibnamefont {Kamil}}, \bibinfo {author} {\bibfnamefont {J.}~\bibnamefont {Berges}}, \bibinfo {author} {\bibfnamefont {G.}~\bibnamefont {Schönhoff}}, \bibinfo {author} {\bibfnamefont {M.}~\bibnamefont {Rösner}}, \bibinfo {author} {\bibfnamefont {M.}~\bibnamefont {Schüler}}, \bibinfo {author} {\bibfnamefont {G.}~\bibnamefont {Sangiovanni}},\ and\ \bibinfo {author} {\bibfnamefont {T.}~\bibnamefont {Wehling}},\ }\bibfield  {title} {\bibinfo {title} {{Electronic structure of single layer 1T-NbSe$_2$: interplay of lattice distortions, non-local exchange, and Mott-Hubbard correlations}},\ }\href {https://doi.org/10.1088/1361-648X/aad215} {\bibfield  {journal} {\bibinfo  {journal} {J. Phys. Condens. Matter}\ }\textbf {\bibinfo {volume} {30}} (\bibinfo {year} {2018})}\BibitemShut {NoStop}%
\bibitem [{\citenamefont {Stepanov}\ \emph {et~al.}(2022)\citenamefont {Stepanov}, \citenamefont {Harkov}, \citenamefont {R\"osner}, \citenamefont {Lichtenstein}, \citenamefont {Katsnelson},\ and\ \citenamefont {Rudenko}}]{stepanov2021coexisting}%
  \BibitemOpen
  \bibfield  {author} {\bibinfo {author} {\bibfnamefont {E.~A.}\ \bibnamefont {Stepanov}}, \bibinfo {author} {\bibfnamefont {V.}~\bibnamefont {Harkov}}, \bibinfo {author} {\bibfnamefont {M.}~\bibnamefont {R\"osner}}, \bibinfo {author} {\bibfnamefont {A.~I.}\ \bibnamefont {Lichtenstein}}, \bibinfo {author} {\bibfnamefont {M.~I.}\ \bibnamefont {Katsnelson}},\ and\ \bibinfo {author} {\bibfnamefont {A.~N.}\ \bibnamefont {Rudenko}},\ }\bibfield  {title} {\bibinfo {title} {{Coexisting charge density wave and ferromagnetic instabilities in monolayer InSe}},\ }\href {https://doi.org/10.1038/s41524-022-00798-4} {\bibfield  {journal} {\bibinfo  {journal} {npj Comput. Mater.}\ }\textbf {\bibinfo {volume} {8}},\ \bibinfo {pages} {118} (\bibinfo {year} {2022})}\BibitemShut {NoStop}%
\bibitem [{\citenamefont {Stepanov}\ \emph {et~al.}(2024{\natexlab{a}})\citenamefont {Stepanov}, \citenamefont {Vandelli}, \citenamefont {Lichtenstein},\ and\ \citenamefont {Lechermann}}]{stepanov2024}%
  \BibitemOpen
  \bibfield  {author} {\bibinfo {author} {\bibfnamefont {E.~A.}\ \bibnamefont {Stepanov}}, \bibinfo {author} {\bibfnamefont {M.}~\bibnamefont {Vandelli}}, \bibinfo {author} {\bibfnamefont {A.~I.}\ \bibnamefont {Lichtenstein}},\ and\ \bibinfo {author} {\bibfnamefont {F.}~\bibnamefont {Lechermann}},\ }\bibfield  {title} {\bibinfo {title} {{Charge Density Wave Ordering in NdNiO$_2$: Effects of Multiorbital Nonlocal Correlations}},\ }\href {https://doi.org/10.1038/s41524-024-01298-3} {\bibfield  {journal} {\bibinfo  {journal} {npj Comput. Mater.}\ }\textbf {\bibinfo {volume} {10}},\ \bibinfo {pages} {108} (\bibinfo {year} {2024}{\natexlab{a}})}\BibitemShut {NoStop}%
\bibitem [{\citenamefont {Vilk}\ and\ \citenamefont {Tremblay}(1996)}]{vilk1996}%
  \BibitemOpen
  \bibfield  {author} {\bibinfo {author} {\bibfnamefont {Y.~M.}\ \bibnamefont {Vilk}}\ and\ \bibinfo {author} {\bibfnamefont {A.-M.~S.}\ \bibnamefont {Tremblay}},\ }\bibfield  {title} {\bibinfo {title} {{Destruction of {{Fermi-liquid}} Quasiparticles in Two Dimensions by Critical Fluctuations}},\ }\href {https://doi.org/10.1209/epl/i1996-00315-2} {\bibfield  {journal} {\bibinfo  {journal} {EPL}\ }\textbf {\bibinfo {volume} {33}},\ \bibinfo {pages} {159} (\bibinfo {year} {1996})}\BibitemShut {NoStop}%
\bibitem [{\citenamefont {van Loon}\ \emph {et~al.}(2014{\natexlab{b}})\citenamefont {van Loon}, \citenamefont {Hafermann}, \citenamefont {Lichtenstein}, \citenamefont {Rubtsov},\ and\ \citenamefont {Katsnelson}}]{PhysRevLett.113.246407}%
  \BibitemOpen
  \bibfield  {author} {\bibinfo {author} {\bibfnamefont {E.~G. C.~P.}\ \bibnamefont {van Loon}}, \bibinfo {author} {\bibfnamefont {H.}~\bibnamefont {Hafermann}}, \bibinfo {author} {\bibfnamefont {A.~I.}\ \bibnamefont {Lichtenstein}}, \bibinfo {author} {\bibfnamefont {A.~N.}\ \bibnamefont {Rubtsov}},\ and\ \bibinfo {author} {\bibfnamefont {M.~I.}\ \bibnamefont {Katsnelson}},\ }\bibfield  {title} {\bibinfo {title} {{Plasmons in Strongly Correlated Systems: Spectral Weight Transfer and Renormalized Dispersion}},\ }\href {https://doi.org/10.1103/PhysRevLett.113.246407} {\bibfield  {journal} {\bibinfo  {journal} {Phys. Rev. Lett.}\ }\textbf {\bibinfo {volume} {113}},\ \bibinfo {pages} {246407} (\bibinfo {year} {2014}{\natexlab{b}})}\BibitemShut {NoStop}%
\bibitem [{\citenamefont {Boehnke}\ \emph {et~al.}(2018)\citenamefont {Boehnke}, \citenamefont {Werner},\ and\ \citenamefont {Lechermann}}]{Boehnke_2018}%
  \BibitemOpen
  \bibfield  {author} {\bibinfo {author} {\bibfnamefont {L.}~\bibnamefont {Boehnke}}, \bibinfo {author} {\bibfnamefont {P.}~\bibnamefont {Werner}},\ and\ \bibinfo {author} {\bibfnamefont {F.}~\bibnamefont {Lechermann}},\ }\bibfield  {title} {\bibinfo {title} {{Multi-orbital nature of the spin fluctuations in Sr$_2$RuO$_4$}},\ }\href {https://doi.org/10.1209/0295-5075/122/57001} {\bibfield  {journal} {\bibinfo  {journal} {EPL}\ }\textbf {\bibinfo {volume} {122}},\ \bibinfo {pages} {57001} (\bibinfo {year} {2018})}\BibitemShut {NoStop}%
\bibitem [{\citenamefont {Acharya}\ \emph {et~al.}(2019)\citenamefont {Acharya}, \citenamefont {Pashov}, \citenamefont {Weber}, \citenamefont {Park}, \citenamefont {Sponza},\ and\ \citenamefont {Schilfgaarde}}]{acharya2019evening}%
  \BibitemOpen
  \bibfield  {author} {\bibinfo {author} {\bibfnamefont {S.}~\bibnamefont {Acharya}}, \bibinfo {author} {\bibfnamefont {D.}~\bibnamefont {Pashov}}, \bibinfo {author} {\bibfnamefont {C.}~\bibnamefont {Weber}}, \bibinfo {author} {\bibfnamefont {H.}~\bibnamefont {Park}}, \bibinfo {author} {\bibfnamefont {L.}~\bibnamefont {Sponza}},\ and\ \bibinfo {author} {\bibfnamefont {M.~V.}\ \bibnamefont {Schilfgaarde}},\ }\bibfield  {title} {\bibinfo {title} {{Evening out the spin and charge parity to increase $T_c$ in Sr$_2$RuO$_4$}},\ }\href {https://doi.org/10.1038/s42005-019-0254-1} {\bibfield  {journal} {\bibinfo  {journal} {Commun. Phys.}\ }\textbf {\bibinfo {volume} {2}},\ \bibinfo {pages} {163} (\bibinfo {year} {2019})}\BibitemShut {NoStop}%
\bibitem [{\citenamefont {Suzuki}\ \emph {et~al.}(2023)\citenamefont {Suzuki}, \citenamefont {Wang}, \citenamefont {Bertinshaw}, \citenamefont {Strand}, \citenamefont {K{\"a}ser}, \citenamefont {Krautloher}, \citenamefont {Yang}, \citenamefont {Wentzell}, \citenamefont {Parcollet}, \citenamefont {Jerzembeck}, \citenamefont {Kikugawa}, \citenamefont {Mackenzie}, \citenamefont {Georges}, \citenamefont {Hansmann}, \citenamefont {Gretarsson},\ and\ \citenamefont {Keimer}}]{suzuki2023distinct}%
  \BibitemOpen
  \bibfield  {author} {\bibinfo {author} {\bibfnamefont {H.}~\bibnamefont {Suzuki}}, \bibinfo {author} {\bibfnamefont {L.}~\bibnamefont {Wang}}, \bibinfo {author} {\bibfnamefont {J.}~\bibnamefont {Bertinshaw}}, \bibinfo {author} {\bibfnamefont {H.~U.~R.}\ \bibnamefont {Strand}}, \bibinfo {author} {\bibfnamefont {S.}~\bibnamefont {K{\"a}ser}}, \bibinfo {author} {\bibfnamefont {M.}~\bibnamefont {Krautloher}}, \bibinfo {author} {\bibfnamefont {Z.}~\bibnamefont {Yang}}, \bibinfo {author} {\bibfnamefont {N.}~\bibnamefont {Wentzell}}, \bibinfo {author} {\bibfnamefont {O.}~\bibnamefont {Parcollet}}, \bibinfo {author} {\bibfnamefont {F.}~\bibnamefont {Jerzembeck}}, \bibinfo {author} {\bibfnamefont {N.}~\bibnamefont {Kikugawa}}, \bibinfo {author} {\bibfnamefont {A.~P.}\ \bibnamefont {Mackenzie}}, \bibinfo {author} {\bibfnamefont {A.}~\bibnamefont {Georges}}, \bibinfo {author} {\bibfnamefont {P.}~\bibnamefont {Hansmann}}, \bibinfo {author} {\bibfnamefont {H.}~\bibnamefont {Gretarsson}},\ and\ \bibinfo {author}
  {\bibfnamefont {B.}~\bibnamefont {Keimer}},\ }\bibfield  {title} {\bibinfo {title} {{Distinct spin and orbital dynamics in Sr$_2$RuO$_4$}},\ }\href {https://doi.org/10.1038/s41467-023-42804-3} {\bibfield  {journal} {\bibinfo  {journal} {Nature Commun.}\ }\textbf {\bibinfo {volume} {14}},\ \bibinfo {pages} {7042} (\bibinfo {year} {2023})}\BibitemShut {NoStop}%
\bibitem [{\citenamefont {Vandelli}\ \emph {et~al.}(2024)\citenamefont {Vandelli}, \citenamefont {Galler}, \citenamefont {Rubio}, \citenamefont {Lichtenstein}, \citenamefont {Biermann},\ and\ \citenamefont {Stepanov}}]{Vandelli2024_PbSi}%
  \BibitemOpen
  \bibfield  {author} {\bibinfo {author} {\bibfnamefont {M.}~\bibnamefont {Vandelli}}, \bibinfo {author} {\bibfnamefont {A.}~\bibnamefont {Galler}}, \bibinfo {author} {\bibfnamefont {A.}~\bibnamefont {Rubio}}, \bibinfo {author} {\bibfnamefont {A.~I.}\ \bibnamefont {Lichtenstein}}, \bibinfo {author} {\bibfnamefont {S.}~\bibnamefont {Biermann}},\ and\ \bibinfo {author} {\bibfnamefont {E.~A.}\ \bibnamefont {Stepanov}},\ }\bibfield  {title} {\bibinfo {title} {{Doping-dependent Charge- and Spin-Density Wave Orderings in a Monolayer of {Pb} Adatoms on {Si}(111)}},\ }\href {https://doi.org/10.1038/s41535-024-00630-w} {\bibfield  {journal} {\bibinfo  {journal} {npj Quantum Mater.}\ }\textbf {\bibinfo {volume} {9}},\ \bibinfo {pages} {19} (\bibinfo {year} {2024})}\BibitemShut {NoStop}%
\bibitem [{\citenamefont {Stepanov}\ \emph {et~al.}(2024{\natexlab{b}})\citenamefont {Stepanov}, \citenamefont {Chatzieleftheriou}, \citenamefont {Wagner},\ and\ \citenamefont {Sangiovanni}}]{PhysRevB.110.L161106}%
  \BibitemOpen
  \bibfield  {author} {\bibinfo {author} {\bibfnamefont {E.~A.}\ \bibnamefont {Stepanov}}, \bibinfo {author} {\bibfnamefont {M.}~\bibnamefont {Chatzieleftheriou}}, \bibinfo {author} {\bibfnamefont {N.}~\bibnamefont {Wagner}},\ and\ \bibinfo {author} {\bibfnamefont {G.}~\bibnamefont {Sangiovanni}},\ }\bibfield  {title} {\bibinfo {title} {{Interconnected renormalization of Hubbard bands and Green's function zeros in Mott insulators induced by strong magnetic fluctuations}},\ }\href {https://doi.org/10.1103/PhysRevB.110.L161106} {\bibfield  {journal} {\bibinfo  {journal} {Phys. Rev. B}\ }\textbf {\bibinfo {volume} {110}},\ \bibinfo {pages} {L161106} (\bibinfo {year} {2024}{\natexlab{b}})}\BibitemShut {NoStop}%
\bibitem [{\citenamefont {Stepanov}(2025)}]{stepanov2024signatures}%
  \BibitemOpen
  \bibfield  {author} {\bibinfo {author} {\bibfnamefont {E.~A.}\ \bibnamefont {Stepanov}},\ }\bibfield  {title} {\bibinfo {title} {{Fingerprints of a charge ice state in the doped Mott insulator ${\mathrm{Nb}}_{3}{\mathrm{Cl}}_{8}$}},\ }\href {https://doi.org/10.1103/j6bj-gz7j} {\bibfield  {journal} {\bibinfo  {journal} {Phys. Rev. B}\ }\textbf {\bibinfo {volume} {112}},\ \bibinfo {pages} {045131} (\bibinfo {year} {2025})}\BibitemShut {NoStop}%
\bibitem [{\citenamefont {Stepanov}\ \emph {et~al.}(2021)\citenamefont {Stepanov}, \citenamefont {Nomura}, \citenamefont {Lichtenstein},\ and\ \citenamefont {Biermann}}]{PhysRevLett.127.207205}%
  \BibitemOpen
  \bibfield  {author} {\bibinfo {author} {\bibfnamefont {E.~A.}\ \bibnamefont {Stepanov}}, \bibinfo {author} {\bibfnamefont {Y.}~\bibnamefont {Nomura}}, \bibinfo {author} {\bibfnamefont {A.~I.}\ \bibnamefont {Lichtenstein}},\ and\ \bibinfo {author} {\bibfnamefont {S.}~\bibnamefont {Biermann}},\ }\bibfield  {title} {\bibinfo {title} {{Orbital Isotropy of Magnetic Fluctuations in Correlated Electron Materials Induced by Hund's Exchange Coupling}},\ }\href {https://doi.org/10.1103/PhysRevLett.127.207205} {\bibfield  {journal} {\bibinfo  {journal} {Phys. Rev. Lett.}\ }\textbf {\bibinfo {volume} {127}},\ \bibinfo {pages} {207205} (\bibinfo {year} {2021})}\BibitemShut {NoStop}%
\bibitem [{\citenamefont {Stepanov}(2022)}]{PhysRevLett.129.096404}%
  \BibitemOpen
  \bibfield  {author} {\bibinfo {author} {\bibfnamefont {E.~A.}\ \bibnamefont {Stepanov}},\ }\bibfield  {title} {\bibinfo {title} {{Eliminating Orbital Selectivity from the Metal-Insulator Transition by Strong Magnetic Fluctuations}},\ }\href {https://doi.org/10.1103/PhysRevLett.129.096404} {\bibfield  {journal} {\bibinfo  {journal} {Phys. Rev. Lett.}\ }\textbf {\bibinfo {volume} {129}},\ \bibinfo {pages} {096404} (\bibinfo {year} {2022})}\BibitemShut {NoStop}%
\bibitem [{\citenamefont {Vandelli}\ \emph {et~al.}(2023)\citenamefont {Vandelli}, \citenamefont {Kaufmann}, \citenamefont {Harkov}, \citenamefont {Lichtenstein}, \citenamefont {Held},\ and\ \citenamefont {Stepanov}}]{PhysRevResearch.5.L022016}%
  \BibitemOpen
  \bibfield  {author} {\bibinfo {author} {\bibfnamefont {M.}~\bibnamefont {Vandelli}}, \bibinfo {author} {\bibfnamefont {J.}~\bibnamefont {Kaufmann}}, \bibinfo {author} {\bibfnamefont {V.}~\bibnamefont {Harkov}}, \bibinfo {author} {\bibfnamefont {A.~I.}\ \bibnamefont {Lichtenstein}}, \bibinfo {author} {\bibfnamefont {K.}~\bibnamefont {Held}},\ and\ \bibinfo {author} {\bibfnamefont {E.~A.}\ \bibnamefont {Stepanov}},\ }\bibfield  {title} {\bibinfo {title} {{Extended regime of metastable metallic and insulating phases in a two-orbital electronic system}},\ }\href {https://doi.org/10.1103/PhysRevResearch.5.L022016} {\bibfield  {journal} {\bibinfo  {journal} {Phys. Rev. Res.}\ }\textbf {\bibinfo {volume} {5}},\ \bibinfo {pages} {L022016} (\bibinfo {year} {2023})}\BibitemShut {NoStop}%
\bibitem [{\citenamefont {Stepanov}\ and\ \citenamefont {Biermann}(2024)}]{PhysRevLett.132.226501}%
  \BibitemOpen
  \bibfield  {author} {\bibinfo {author} {\bibfnamefont {E.~A.}\ \bibnamefont {Stepanov}}\ and\ \bibinfo {author} {\bibfnamefont {S.}~\bibnamefont {Biermann}},\ }\bibfield  {title} {\bibinfo {title} {{Can Orbital-Selective N\'eel Transitions Survive Strong Nonlocal Electronic Correlations?}},\ }\href {https://doi.org/10.1103/PhysRevLett.132.226501} {\bibfield  {journal} {\bibinfo  {journal} {Phys. Rev. Lett.}\ }\textbf {\bibinfo {volume} {132}},\ \bibinfo {pages} {226501} (\bibinfo {year} {2024})}\BibitemShut {NoStop}%
\bibitem [{\citenamefont {Chatzieleftheriou}\ \emph {et~al.}(2025)\citenamefont {Chatzieleftheriou}, \citenamefont {Rudenko}, \citenamefont {Sidis}, \citenamefont {Biermann},\ and\ \citenamefont {Stepanov}}]{Ruthenates}%
  \BibitemOpen
  \bibfield  {author} {\bibinfo {author} {\bibfnamefont {M.}~\bibnamefont {Chatzieleftheriou}}, \bibinfo {author} {\bibfnamefont {A.~N.}\ \bibnamefont {Rudenko}}, \bibinfo {author} {\bibfnamefont {Y.}~\bibnamefont {Sidis}}, \bibinfo {author} {\bibfnamefont {S.}~\bibnamefont {Biermann}},\ and\ \bibinfo {author} {\bibfnamefont {E.~A.}\ \bibnamefont {Stepanov}},\ }\bibfield  {title} {\bibinfo {title} {{Nature of momentum- and orbital-dependent magnetic fluctuations in ${\mathrm{Sr}}_{2}{\mathrm{RuO}}_{4}$}},\ }\href {https://doi.org/10.1103/ts6y-zb6m} {\bibfield  {journal} {\bibinfo  {journal} {Phys. Rev. B}\ }\textbf {\bibinfo {volume} {112}},\ \bibinfo {pages} {195118} (\bibinfo {year} {2025})}\BibitemShut {NoStop}%
\bibitem [{\citenamefont {Dasari}\ \emph {et~al.}(2025)\citenamefont {Dasari}, \citenamefont {Strand}, \citenamefont {Eckstein}, \citenamefont {Lichtenstein},\ and\ \citenamefont {Stepanov}}]{vglv-2rmv}%
  \BibitemOpen
  \bibfield  {author} {\bibinfo {author} {\bibfnamefont {N.}~\bibnamefont {Dasari}}, \bibinfo {author} {\bibfnamefont {H.~U.~R.}\ \bibnamefont {Strand}}, \bibinfo {author} {\bibfnamefont {M.}~\bibnamefont {Eckstein}}, \bibinfo {author} {\bibfnamefont {A.~I.}\ \bibnamefont {Lichtenstein}},\ and\ \bibinfo {author} {\bibfnamefont {E.~A.}\ \bibnamefont {Stepanov}},\ }\bibfield  {title} {\bibinfo {title} {{Electron-magnon dynamics triggered by an ultrashort laser pulse: A real-time dual $GW$ study}},\ }\href {https://doi.org/10.1103/vglv-2rmv} {\bibfield  {journal} {\bibinfo  {journal} {Phys. Rev. B}\ }\textbf {\bibinfo {volume} {111}},\ \bibinfo {pages} {235129} (\bibinfo {year} {2025})}\BibitemShut {NoStop}%
\bibitem [{\citenamefont {Hafermann}\ \emph {et~al.}(2008)\citenamefont {Hafermann}, \citenamefont {Brener}, \citenamefont {Rubtsov}, \citenamefont {Katsnelson},\ and\ \citenamefont {Lichtenstein}}]{hafermann2008}%
  \BibitemOpen
  \bibfield  {author} {\bibinfo {author} {\bibfnamefont {H.}~\bibnamefont {Hafermann}}, \bibinfo {author} {\bibfnamefont {S.}~\bibnamefont {Brener}}, \bibinfo {author} {\bibfnamefont {A.~N.}\ \bibnamefont {Rubtsov}}, \bibinfo {author} {\bibfnamefont {M.~I.}\ \bibnamefont {Katsnelson}},\ and\ \bibinfo {author} {\bibfnamefont {A.~I.}\ \bibnamefont {Lichtenstein}},\ }\bibfield  {title} {\bibinfo {title} {{Cluster Dual Fermion Approach to Nonlocal Correlations}},\ }\href {https://doi.org/10.1134/S0021364007220134} {\bibfield  {journal} {\bibinfo  {journal} {JETP Letters}\ }\textbf {\bibinfo {volume} {86}},\ \bibinfo {pages} {677} (\bibinfo {year} {2008})}\BibitemShut {NoStop}%
\bibitem [{\citenamefont {Yang}\ \emph {et~al.}(2011)\citenamefont {Yang}, \citenamefont {Fotso}, \citenamefont {Hafermann}, \citenamefont {Tam}, \citenamefont {Moreno}, \citenamefont {Pruschke},\ and\ \citenamefont {Jarrell}}]{PhysRevB.84.155106}%
  \BibitemOpen
  \bibfield  {author} {\bibinfo {author} {\bibfnamefont {S.-X.}\ \bibnamefont {Yang}}, \bibinfo {author} {\bibfnamefont {H.}~\bibnamefont {Fotso}}, \bibinfo {author} {\bibfnamefont {H.}~\bibnamefont {Hafermann}}, \bibinfo {author} {\bibfnamefont {K.-M.}\ \bibnamefont {Tam}}, \bibinfo {author} {\bibfnamefont {J.}~\bibnamefont {Moreno}}, \bibinfo {author} {\bibfnamefont {T.}~\bibnamefont {Pruschke}},\ and\ \bibinfo {author} {\bibfnamefont {M.}~\bibnamefont {Jarrell}},\ }\bibfield  {title} {\bibinfo {title} {{Dual fermion dynamical cluster approach for strongly correlated systems}},\ }\href {https://doi.org/10.1103/PhysRevB.84.155106} {\bibfield  {journal} {\bibinfo  {journal} {Phys. Rev. B}\ }\textbf {\bibinfo {volume} {84}},\ \bibinfo {pages} {155106} (\bibinfo {year} {2011})}\BibitemShut {NoStop}%
\bibitem [{\citenamefont {Iskakov}\ \emph {et~al.}(2018)\citenamefont {Iskakov}, \citenamefont {Terletska},\ and\ \citenamefont {Gull}}]{PhysRevB.97.125114}%
  \BibitemOpen
  \bibfield  {author} {\bibinfo {author} {\bibfnamefont {S.}~\bibnamefont {Iskakov}}, \bibinfo {author} {\bibfnamefont {H.}~\bibnamefont {Terletska}},\ and\ \bibinfo {author} {\bibfnamefont {E.}~\bibnamefont {Gull}},\ }\bibfield  {title} {\bibinfo {title} {{Momentum-space cluster dual-fermion method}},\ }\href {https://doi.org/10.1103/PhysRevB.97.125114} {\bibfield  {journal} {\bibinfo  {journal} {Phys. Rev. B}\ }\textbf {\bibinfo {volume} {97}},\ \bibinfo {pages} {125114} (\bibinfo {year} {2018})}\BibitemShut {NoStop}%
\bibitem [{\citenamefont {Iskakov}\ \emph {et~al.}(2024)\citenamefont {Iskakov}, \citenamefont {Katsnelson},\ and\ \citenamefont {Lichtenstein}}]{DFQMC}%
  \BibitemOpen
  \bibfield  {author} {\bibinfo {author} {\bibfnamefont {S.}~\bibnamefont {Iskakov}}, \bibinfo {author} {\bibfnamefont {M.~I.}\ \bibnamefont {Katsnelson}},\ and\ \bibinfo {author} {\bibfnamefont {A.~I.}\ \bibnamefont {Lichtenstein}},\ }\bibfield  {title} {\bibinfo {title} {{Perturbative solution of fermionic sign problem in quantum Monte Carlo computations}},\ }\href {https://doi.org/10.1038/s41524-024-01221-w} {\bibfield  {journal} {\bibinfo  {journal} {npj Comput. Mater.}\ }\textbf {\bibinfo {volume} {10}},\ \bibinfo {pages} {36} (\bibinfo {year} {2024})}\BibitemShut {NoStop}%
\bibitem [{\citenamefont {Ayral}\ \emph {et~al.}(2017{\natexlab{b}})\citenamefont {Ayral}, \citenamefont {Vu\ifmmode \check{c}\else \v{c}\fi{}i\ifmmode \check{c}\else \v{c}\fi{}evi\ifmmode~\acute{c}\else \'{c}\fi{}},\ and\ \citenamefont {Parcollet}}]{PhysRevLett.119.166401}%
  \BibitemOpen
  \bibfield  {author} {\bibinfo {author} {\bibfnamefont {T.}~\bibnamefont {Ayral}}, \bibinfo {author} {\bibfnamefont {J.}~\bibnamefont {Vu\ifmmode \check{c}\else \v{c}\fi{}i\ifmmode \check{c}\else \v{c}\fi{}evi\ifmmode~\acute{c}\else \'{c}\fi{}}},\ and\ \bibinfo {author} {\bibfnamefont {O.}~\bibnamefont {Parcollet}},\ }\bibfield  {title} {\bibinfo {title} {{Fierz Convergence Criterion: A Controlled Approach to Strongly Interacting Systems with Small Embedded Clusters}},\ }\href {https://doi.org/10.1103/PhysRevLett.119.166401} {\bibfield  {journal} {\bibinfo  {journal} {Phys. Rev. Lett.}\ }\textbf {\bibinfo {volume} {119}},\ \bibinfo {pages} {166401} (\bibinfo {year} {2017}{\natexlab{b}})}\BibitemShut {NoStop}%
\bibitem [{\citenamefont {S\'emon}\ and\ \citenamefont {Tremblay}(2012)}]{PhysRevB.85.201101}%
  \BibitemOpen
  \bibfield  {author} {\bibinfo {author} {\bibfnamefont {P.}~\bibnamefont {S\'emon}}\ and\ \bibinfo {author} {\bibfnamefont {A.-M.~S.}\ \bibnamefont {Tremblay}},\ }\bibfield  {title} {\bibinfo {title} {{Importance of subleading corrections for the Mott critical point}},\ }\href {https://doi.org/10.1103/PhysRevB.85.201101} {\bibfield  {journal} {\bibinfo  {journal} {Phys. Rev. B}\ }\textbf {\bibinfo {volume} {85}},\ \bibinfo {pages} {201101} (\bibinfo {year} {2012})}\BibitemShut {NoStop}%
\bibitem [{\citenamefont {Shinaoka}\ \emph {et~al.}(2015)\citenamefont {Shinaoka}, \citenamefont {Nomura}, \citenamefont {Biermann}, \citenamefont {Troyer},\ and\ \citenamefont {Werner}}]{PhysRevB.92.195126}%
  \BibitemOpen
  \bibfield  {author} {\bibinfo {author} {\bibfnamefont {H.}~\bibnamefont {Shinaoka}}, \bibinfo {author} {\bibfnamefont {Y.}~\bibnamefont {Nomura}}, \bibinfo {author} {\bibfnamefont {S.}~\bibnamefont {Biermann}}, \bibinfo {author} {\bibfnamefont {M.}~\bibnamefont {Troyer}},\ and\ \bibinfo {author} {\bibfnamefont {P.}~\bibnamefont {Werner}},\ }\bibfield  {title} {\bibinfo {title} {{Negative sign problem in continuous-time quantum Monte Carlo: Optimal choice of single-particle basis for impurity problems}},\ }\href {https://doi.org/10.1103/PhysRevB.92.195126} {\bibfield  {journal} {\bibinfo  {journal} {Phys. Rev. B}\ }\textbf {\bibinfo {volume} {92}},\ \bibinfo {pages} {195126} (\bibinfo {year} {2015})}\BibitemShut {NoStop}%
\bibitem [{\citenamefont {Kanamori}(1963)}]{10.1143/PTP.30.275}%
  \BibitemOpen
  \bibfield  {author} {\bibinfo {author} {\bibfnamefont {J.}~\bibnamefont {Kanamori}},\ }\bibfield  {title} {\bibinfo {title} {{Electron Correlation and Ferromagnetism of Transition Metals}},\ }\href {https://doi.org/10.1143/PTP.30.275} {\bibfield  {journal} {\bibinfo  {journal} {Prog. Theor. Phys.}\ }\textbf {\bibinfo {volume} {30}},\ \bibinfo {pages} {275} (\bibinfo {year} {1963})}\BibitemShut {NoStop}%
\bibitem [{\citenamefont {Wallerberger}\ \emph {et~al.}(2019)\citenamefont {Wallerberger}, \citenamefont {Hausoel}, \citenamefont {Gunacker}, \citenamefont {Kowalski}, \citenamefont {Parragh}, \citenamefont {Goth}, \citenamefont {Held},\ and\ \citenamefont {Sangiovanni}}]{wallerberger2019}%
  \BibitemOpen
  \bibfield  {author} {\bibinfo {author} {\bibfnamefont {M.}~\bibnamefont {Wallerberger}}, \bibinfo {author} {\bibfnamefont {A.}~\bibnamefont {Hausoel}}, \bibinfo {author} {\bibfnamefont {P.}~\bibnamefont {Gunacker}}, \bibinfo {author} {\bibfnamefont {A.}~\bibnamefont {Kowalski}}, \bibinfo {author} {\bibfnamefont {N.}~\bibnamefont {Parragh}}, \bibinfo {author} {\bibfnamefont {F.}~\bibnamefont {Goth}}, \bibinfo {author} {\bibfnamefont {K.}~\bibnamefont {Held}},\ and\ \bibinfo {author} {\bibfnamefont {G.}~\bibnamefont {Sangiovanni}},\ }\bibfield  {title} {\bibinfo {title} {{w2dynamics: Local one- and two-particle quantities from dynamical mean field theory}},\ }\href {https://doi.org/https://doi.org/10.1016/j.cpc.2018.09.007} {\bibfield  {journal} {\bibinfo  {journal} {Comput. Phys. Commun.}\ }\textbf {\bibinfo {volume} {235}},\ \bibinfo {pages} {388} (\bibinfo {year} {2019})}\BibitemShut {NoStop}%
\bibitem [{\citenamefont {Backes}\ \emph {et~al.}(2022)\citenamefont {Backes}, \citenamefont {Sim},\ and\ \citenamefont {Biermann}}]{PhysRevB.105.245115}%
  \BibitemOpen
  \bibfield  {author} {\bibinfo {author} {\bibfnamefont {S.}~\bibnamefont {Backes}}, \bibinfo {author} {\bibfnamefont {J.-H.}\ \bibnamefont {Sim}},\ and\ \bibinfo {author} {\bibfnamefont {S.}~\bibnamefont {Biermann}},\ }\bibfield  {title} {\bibinfo {title} {Nonlocal correlation effects in fermionic many-body systems: Overcoming the noncausality problem},\ }\href {https://doi.org/10.1103/PhysRevB.105.245115} {\bibfield  {journal} {\bibinfo  {journal} {Phys. Rev. B}\ }\textbf {\bibinfo {volume} {105}},\ \bibinfo {pages} {245115} (\bibinfo {year} {2022})}\BibitemShut {NoStop}%
\bibitem [{\citenamefont {Vu\ifmmode \check{c}\else \v{c}\fi{}i\ifmmode \check{c}\else \v{c}\fi{}evi\ifmmode~\acute{c}\else \'{c}\fi{}}\ \emph {et~al.}(2018)\citenamefont {Vu\ifmmode \check{c}\else \v{c}\fi{}i\ifmmode \check{c}\else \v{c}\fi{}evi\ifmmode~\acute{c}\else \'{c}\fi{}}, \citenamefont {Wentzell}, \citenamefont {Ferrero},\ and\ \citenamefont {Parcollet}}]{PhysRevB.97.125141}%
  \BibitemOpen
  \bibfield  {author} {\bibinfo {author} {\bibfnamefont {J.}~\bibnamefont {Vu\ifmmode \check{c}\else \v{c}\fi{}i\ifmmode \check{c}\else \v{c}\fi{}evi\ifmmode~\acute{c}\else \'{c}\fi{}}}, \bibinfo {author} {\bibfnamefont {N.}~\bibnamefont {Wentzell}}, \bibinfo {author} {\bibfnamefont {M.}~\bibnamefont {Ferrero}},\ and\ \bibinfo {author} {\bibfnamefont {O.}~\bibnamefont {Parcollet}},\ }\bibfield  {title} {\bibinfo {title} {{Practical consequences of the Luttinger-Ward functional multivaluedness for cluster DMFT methods}},\ }\href {https://doi.org/10.1103/PhysRevB.97.125141} {\bibfield  {journal} {\bibinfo  {journal} {Phys. Rev. B}\ }\textbf {\bibinfo {volume} {97}},\ \bibinfo {pages} {125141} (\bibinfo {year} {2018})}\BibitemShut {NoStop}%
\bibitem [{\citenamefont {Lee}\ and\ \citenamefont {Haule}(2017)}]{PhysRevB.95.155104}%
  \BibitemOpen
  \bibfield  {author} {\bibinfo {author} {\bibfnamefont {J.}~\bibnamefont {Lee}}\ and\ \bibinfo {author} {\bibfnamefont {K.}~\bibnamefont {Haule}},\ }\bibfield  {title} {\bibinfo {title} {{Diatomic molecule as a testbed for combining DMFT with electronic structure methods such as $GW$ and DFT}},\ }\href {https://doi.org/10.1103/PhysRevB.95.155104} {\bibfield  {journal} {\bibinfo  {journal} {Phys. Rev. B}\ }\textbf {\bibinfo {volume} {95}},\ \bibinfo {pages} {155104} (\bibinfo {year} {2017})}\BibitemShut {NoStop}%
\bibitem [{\citenamefont {Petocchi}\ \emph {et~al.}(2021)\citenamefont {Petocchi}, \citenamefont {Christiansson},\ and\ \citenamefont {Werner}}]{PhysRevB.104.195146}%
  \BibitemOpen
  \bibfield  {author} {\bibinfo {author} {\bibfnamefont {F.}~\bibnamefont {Petocchi}}, \bibinfo {author} {\bibfnamefont {V.}~\bibnamefont {Christiansson}},\ and\ \bibinfo {author} {\bibfnamefont {P.}~\bibnamefont {Werner}},\ }\bibfield  {title} {\bibinfo {title} {{Fully ab initio electronic structure of ${\mathrm{Ca}}_{2}{\mathrm{RuO}}_{4}$}},\ }\href {https://doi.org/10.1103/PhysRevB.104.195146} {\bibfield  {journal} {\bibinfo  {journal} {Phys. Rev. B}\ }\textbf {\bibinfo {volume} {104}},\ \bibinfo {pages} {195146} (\bibinfo {year} {2021})}\BibitemShut {NoStop}%
\bibitem [{\citenamefont {Nilsson}\ \emph {et~al.}(2017)\citenamefont {Nilsson}, \citenamefont {Boehnke}, \citenamefont {Werner},\ and\ \citenamefont {Aryasetiawan}}]{PhysRevMaterials.1.043803}%
  \BibitemOpen
  \bibfield  {author} {\bibinfo {author} {\bibfnamefont {F.}~\bibnamefont {Nilsson}}, \bibinfo {author} {\bibfnamefont {L.}~\bibnamefont {Boehnke}}, \bibinfo {author} {\bibfnamefont {P.}~\bibnamefont {Werner}},\ and\ \bibinfo {author} {\bibfnamefont {F.}~\bibnamefont {Aryasetiawan}},\ }\bibfield  {title} {\bibinfo {title} {{Multitier self-consistent $GW+\text{EDMFT}$}},\ }\href {https://doi.org/10.1103/PhysRevMaterials.1.043803} {\bibfield  {journal} {\bibinfo  {journal} {Phys. Rev. Mater.}\ }\textbf {\bibinfo {volume} {1}},\ \bibinfo {pages} {043803} (\bibinfo {year} {2017})}\BibitemShut {NoStop}%
\bibitem [{\citenamefont {Chen}\ \emph {et~al.}(2022)\citenamefont {Chen}, \citenamefont {Petocchi},\ and\ \citenamefont {Werner}}]{PhysRevB.105.085102}%
  \BibitemOpen
  \bibfield  {author} {\bibinfo {author} {\bibfnamefont {J.}~\bibnamefont {Chen}}, \bibinfo {author} {\bibfnamefont {F.}~\bibnamefont {Petocchi}},\ and\ \bibinfo {author} {\bibfnamefont {P.}~\bibnamefont {Werner}},\ }\bibfield  {title} {\bibinfo {title} {{Causal versus local $GW+\mathrm{EDMFT}$ scheme and application to the triangular-lattice extended Hubbard model}},\ }\href {https://doi.org/10.1103/PhysRevB.105.085102} {\bibfield  {journal} {\bibinfo  {journal} {Phys. Rev. B}\ }\textbf {\bibinfo {volume} {105}},\ \bibinfo {pages} {085102} (\bibinfo {year} {2022})}\BibitemShut {NoStop}%
\bibitem [{\citenamefont {Titvinidze}\ \emph {et~al.}(2025)\citenamefont {Titvinidze}, \citenamefont {Stobbe}, \citenamefont {Leusch},\ and\ \citenamefont {Rohringer}}]{Titvinidze_2025}%
  \BibitemOpen
  \bibfield  {author} {\bibinfo {author} {\bibfnamefont {I.}~\bibnamefont {Titvinidze}}, \bibinfo {author} {\bibfnamefont {J.}~\bibnamefont {Stobbe}}, \bibinfo {author} {\bibfnamefont {M.}~\bibnamefont {Leusch}},\ and\ \bibinfo {author} {\bibfnamefont {G.}~\bibnamefont {Rohringer}},\ }\bibfield  {title} {\bibinfo {title} {{Suppression of the charge fluctuations by nonlocal correlations close to the Mott transition: Insights from the ladder dynamical vertex approximation}},\ }\href {https://doi.org/10.1103/n2g3-nv58} {\bibfield  {journal} {\bibinfo  {journal} {Phys. Rev. B}\ }\textbf {\bibinfo {volume} {112}},\ \bibinfo {pages} {155125} (\bibinfo {year} {2025})}\BibitemShut {NoStop}%
\end{thebibliography}%

\end{document}